\documentclass[prx,aps,amsart,amssymb,epsfig,showpacs,twocolumn,eqsecnum,superscriptaddress]{revtex4-1}

\usepackage{amsmath,amssymb,amsfonts,graphicx,multirow,color,bm,textcomp,mathtools}
\usepackage{paralist}
\usepackage{srcltx}
\usepackage[all,cmtip]{xy}
\usepackage{mathrsfs}
\usepackage{srcltx,soul}
\usepackage{tabularx} 
\usepackage{threeparttable,dsfont,bbm}

\usepackage{hyperref}
\usepackage{cleveref}

\usepackage[caption=false,subrefformat=parens,labelformat=parens]{subfig}

\crefname{equation}{Eq.}{Eqs.}
\newcommand{\vk}{{\bf k}}

\newcommand{\ket}[1]{| #1 \rangle}
\newcommand{\braket}[2]{\left \langle #1 | #2 \right\rangle}
\newcommand{\braketnoresize}[2]{\langle #1 | #2 \rangle }

\newcommand{\bee}{\begin{eqnarray}}
\newcommand{\ee}{\end{eqnarray}}
\newcommand{\bma}{\begin{pmatrix}}
\newcommand{\ema}{\end{pmatrix}}
\newcommand{\balig}{\begin{align}}
\newcommand{\ealig}{\end{align}}

\newcommand{\ba}{\begin{align}}
\newcommand{\ea}{\end{align}}

\newcommand{\ignore}[1]{}

\usepackage{dcolumn}
\newcolumntype{C}[1]{>{\centering\let\newline\\\arraybackslash\hspace{0pt}}m{#1}}

\begin{document}

\title{Symmetry-enforced band crossings in tetragonal materials: \\
Dirac and Weyl degeneracies on points, lines, and planes}

\author{Moritz~M.~Hirschmann}
\affiliation{Max-Planck-Institut f\"ur Festk\"orperforschung, Heisenbergstrasse 1, D-70569 Stuttgart, Germany} 

\author{Andreas Leonhardt}
\affiliation{Max-Planck-Institut f\"ur Festk\"orperforschung, Heisenbergstrasse 1, D-70569 Stuttgart, Germany} 

\author{Berkay Kilic}
\affiliation{Max-Planck-Institut f\"ur Festk\"orperforschung, Heisenbergstrasse 1, D-70569 Stuttgart, Germany} 
\affiliation{Institute for Functional Materials and Quantum Technologies, University of Stuttgart, Pfaffenwaldring 57, D-70550 Stuttgart, Germany}

\author{Douglas H. Fabini}
\affiliation{Max-Planck-Institut f\"ur Festk\"orperforschung, Heisenbergstrasse 1, D-70569 Stuttgart, Germany} 

\author{Andreas P. Schnyder} 
\email{a.schnyder@fkf.mpg.de}
\affiliation{Max-Planck-Institut f\"ur Festk\"orperforschung, Heisenbergstrasse 1, D-70569 Stuttgart, Germany}

\begin{abstract}
We study the occurrence of symmetry-enforced topological band
crossings in tetragonal crystals with strong spin-orbit coupling.
By computing the momentum dependence of the symmetry eigenvalues
and the global band topology in the entire Brillouin zone, we
determine all symmetry-enforced band crossings in tetragonal
space groups. In particular, we classify all Dirac and Weyl
degeneracies on points, lines, and planes, and find a rich variety
of topological degeneracies. This includes, among others, double
Weyl points, fourfold-double Weyl points, fourfold-quadruple Weyl
points, Weyl and Dirac nodal lines, as well as topological nodal
planes. For the space groups with symmetry-enforced Weyl points,
we determine the minimal number  of Weyl points for a given band
pair and, remarkably, find that materials in space groups 119 and 120 can have 
band pairs with only \emph{two} Weyl points in the entire Brillouin zone. This simplifies
the topological responses, which would be useful for device applications.
Using the classification of  symmetry-enforced band crossings, we perform an extensive database
search for candidate materials with tetragonal space groups.
Notably, we find that Ba$_5$In$_4$Bi$_5$ and NaSn$_5$ exhibit
twofold and fourfold Weyl nodal lines, respectively, which cross
the Fermi energy.
Hf$_3$Sb and Cs$_2$Tl$_3$ have band pairs with few number of Weyl points
near the Fermi energy. 
Furthermore, we show that Ba$_3$Sn$_2$ has Weyl points with an
accordion dispersion and topological nodal planes, while AuBr and
Tl$_4$PbSe$_3$ possess Dirac points with hourglass dispersions.
For each of these candidate materials we present the
\emph{ab-initio} band structures and discuss possible experimental
signatures of the nontrivial band topology.
\end{abstract}

\date{\rm\today}

\maketitle

\section{Introduction}
Topological semimetals exhibit protected band crossings near the
Fermi energy, which carry nonzero topological
charges~\cite{chiu_RMP_16,volovikLectNotes13,armitage_mele_vishwanath_review,burkov_review_Weyl,yang_ali_review_ndoal_line,ann_review_transport_top_semimetals_19,burkov_review_nat_mat_16,wang_zhi_min_Review_ACS_nano}.
The existence of such band crossings has been recognized early on
in the development of solid-state physics~\cite{herring_PR_37},
but their importance was appreciated only recently.  Indeed, over
the last few years it was shown that topological band crossings
lead to a number of interesting phenomena, such as, unusual
magnetotransport~\cite{bzdusek_soluyanov_nature_16}, intrinsic
anomalous Hall effects~\cite{wang_hechang_nat_comm_intrinsic_AHE},
large thermopower~\cite{Skinnereaat2621}, exotic surface
states~\cite{WanVishwanathSavrasovPRB11,BurkovBalentsPRB11,xu_hasan_Na3Bi_fermi_arc_science_15,huang_hasan_TaAs_Fermi_Arc_nat_comm_15,nodal_line_Yang,bian_hasan_PbTaSe2_drumhead_nat_comm_16,bian_hasan_TlTaSe2_drumhead_PRB_16},
and various responses related to quantum
anomalies~\cite{fukushima_warringa_PRD_08,vazifeh_franz_PRL_13,goswami_sarma_axial_anomaly,huang_genfu_chiral_anomaly_PRX}.
Due to these unusual properties, topological semimetals hold great
potential for novel device
applications~\cite{pengzi_cha_nat_reviews_2019}.  For example, the
helical nature of the electronic states can be utilized for
low-dissipation transport~\cite{burkov_review_nat_mat_16}.  Using
the spin-momentum locking of the surface states, low-consumption
spintronic
devices~\cite{spin_pol_current_BTS_sci_rep_15,tian_spin_pol_TI_sci_adv_17}
and magnetic memory
devices~\cite{peng_nat_commun_spin_orbit_torque} can be
constructed.  It may also be possible to build topological field
effect transistors, by controlling the phase transitions of
topological semimetals~\cite{Kim723_tunable_band_gap_science_15}.
The high photosensitivity of topological semimetals may provide a
possibility for building ultrafast broadband
photodetectors~\cite{wang_photodetector_nano_letters_17,yang_photodetector_ACS_photonics_18,chumeng_photosensitivity_Adv_Mat_18}.
Besides, many topological semimetals exhibit large thermoelectric
responses, which makes them promising for high-efficiency energy
converters and thermal
detectors~\cite{lundgren_greg_fiete_PRB_14}. 

Despite these extensive research activities, there is still a lack
of suitable materials for device applications. In this paper, we
use the principle of symmetry-enforced band crossings to look for
new topological materials. That is, we investigate under which
circumstances nonsymmorphic symmetries lead to enforced band
crossings on high-symmetry points, lines, or
planes~\cite{bzdusek_soluyanov_nature_16,young_kane_rappe_Dirac_3D_PRL_12,schoop_zrsis,zhao_schnyder_PRB_16,young_kane_non_symmorphic_PRL_15,furusaki_non_symmorphic_17,ryo_murakami_PRB_17,yang_furusaki_PRB_2017,tsirkin_vanderbilt_PRB_17,fang_kee_fu_off_center_PRB_15,malard_johannesson_18,zhang_hexagonals_PRMAT_18,chan_trigonal_PRMAT_19,Orthorhombic_to_be_published}.
We focus on space groups (SGs) in the tetragonal crystal system, which exhibit fourfold
(screw) rotations about the $z$-axis. Previously, we have applied
this strategy to classify symmetry-enforced band crossings in
materials with hexagonal and trigonal SG
symmetries~\cite{zhang_hexagonals_PRMAT_18,chan_trigonal_PRMAT_19}.
A future work will be concerned with orthorhombic
SGs~\cite{Orthorhombic_to_be_published}.

We summarize our results in Tables~\ref{mTab1} and~\ref{mTab2},
which classify all possible symmetry-enforced band crossings in
tetragonal materials with strong spin-orbit coupling.  We find a
large variety of different types of topological band degeneracies
on points, lines, and planes.  These include Dirac and Weyl
points, Dirac nodal lines, twofold and fourfold Weyl nodal lines,
as well as topological nodal planes.  Regarding the
symmetry-enforced Weyl points, we find four different variants,
namely, single Weyl points with a linear band crossing and Chern
number $| \mathcal{C} | = 1$, double Weyl points with a quadratic
band crossing along two directions and $| \mathcal{C} | =2$,
fourfold double Weyl points with a fourfold degeneracy and $|
\mathcal{C} | =2$, and fourfold quadruple Weyl points with a
fourfold degeneracy and $| \mathcal{C} | =4$
(Sec.~\ref{Sec_NonSymWeyl}).  The degeneracies along nodal lines
also come in different varieties: Dirac nodal lines, fourfold Weyl
nodal lines, and twofold Weyl nodal lines, which can form nodal
chains or ``armillary spheres"
(Sec.~\ref{sec_two_fold_weyl_lines}).  Additionally, there are
``almost movable nodal lines", which are pinned to high-symmetry points but
otherwise freely movable (Sec.~\ref{sec_movableTRIMlines})
\footnote{We note that the existence of these almost movable nodal lines cannot be derived
solely form the compatibility relations of irreps.}.  Finally, for the nodal planes,
we find that they occur on the Brillouin zone (BZ) boundaries and are enforced by
the combination of screw rotation with time-reversal symmetry.
The nontrivial topology of these nodal planes follows from the
Nielsen-Ninomiya fermion doubling
theorem~\cite{nielsen_no_go,NIELSEN198120} together with the
global band topology (i.e. the topological charge of all band
crossings of a given band pair), see Sec.~\ref{weyl_planes}.

With these classification results in hand, we proceed to identify
materials that exhibit the aforementioned symmetry-enforced band
crossings.  For that purpose, we perform extensive database
searches for materials with strong spin-orbit coupling and the
relevant space group symmetries (Sec.~\ref{example_materials} and
Fig.~\ref{heat_map}).  Our search yields seven candidate
materials, whose band structures are studied in detail using
density functional theory (DFT) calculations.  In particular, we find that
Ba$_5$In$_4$Bi$_5$ possesses Weyl nodal lines close to the Fermi
energy, which form nodal chains (Sec.~\ref{sec_Ba5In4Bi5}).
NaSn$_5$ exhibits fourfold nodal lines crossing the Fermi energy
(Sec.~\ref{sec_NaSn5}).  Ba$_3$Sn$_2$ has single and double Weyl
points with accordion dispersions and topological nodal planes
(Secs.~\ref{Ba3Sn2_chapter} and~\ref{Ba3Sn2_chapter_zwei}), while
Hf$_3$Sb and Cs$_2$Tl$_3$ exhibit a small number of Weyl points close to the
Fermi energy (Secs.~\ref{Hf3Sb_chapter} and~\ref{Cs2Tl3_chapter}).
The centrosymmetric compounds Tl$_4$PbSe$_3$ and AuBr possess
movable Dirac points forming  hourglass dispersions
(Secs.~\ref{Tl4PbSe3_chapter} and~\ref{sec_AuBr}).

The remainder of this paper is organized as follows. In
Sec.~\ref{mSec2} we introduce our naming conventions for the
different topological band degeneracies and explain our notation
for the symmetry operators and high-symmetry points
(Sec.~\ref{mSec2A}).  Moreover, we explain the details of our
database search for candidate materials
(Sec.~\ref{example_materials}).  We find candidate materials for
seven of the tetragonal SGs. Their band structures and topological
features are presented in those sections, where the corresponding
SGs are analyzed.  In Sec.~\ref{sec_application_kramers_theorem}
we study band degeneracies that are enforced by Kramers theorem.
We show that Kramers theorem leads to Weyl points at time-reversal
invariant momenta (TRIMs)
(Sec.~\ref{sec_application_kramers_theorem_A}), as well as to
point and line degeneracies on other high-symmetry points and
lines, respectively
(Sec.~\ref{sec_application_kramers_theorem_B}).
Section~\ref{Sec_NonSymWeyl} is concerned with hourglass and
accordion dispersions (Sec.~\ref{Sec_NonSymWeyl_A}), as well as
Weyl points that are symmetry enforced by screw rotations.
Depending on the tetragonal SG, these Weyl points are of the
following types: single Weyl points (Sec.~\ref{Sec_NonSymWeyl_B}),
double Weyl points (Sec.~\ref{Sec_NonSymWeyl_C}), fourfold double
Weyl points (Sec.~\ref{Sec_NonSymWeyl_D}), and fourfold quadruple
Weyl points (Sec.~\ref{Sec_NonSymWeyl_E}).  In
Sec.~\ref{sec_theo_Dirac_points} we discuss movable fourfold
points and Dirac points, whose existence is enforced by screw
rotations combined with glide reflections or other multiple
(non)symmorphic symmetries.
Sections~\ref{sec_two_fold_weyl_lines}
and~\ref{sec_four_fold_weyl_lines} are devoted to the study of
twofold and fourfold Weyl nodal lines, respectively.  In
particular, we investigate nodal lines forming chains of connected
rings (Sec.~\ref{sec_nodal_chain_metals}) and armillary spheres
(Sec.~\ref{section_on_SG110}).  In Sec.~\ref{weyl_planes} we study
nodal planes, i.e. band degeneracies on two-dimensional planes at
the BZ boundary, which are enforced by screw rotations together
with time-reversal symmetry. The nontrivial topological charge of
these nodal planes is inferred from the global band topology of
all band crossings in the BZ.  Finally, in
Sec.~\ref{Sec_AccBandCross_OffCenter} we analyze band crossings
protected by off-entered symmetries, i.e. by the combination of
screw rotation (or glide mirror) with inversion.  The conclusions
and outlook of our work are given in Sec.~\ref{sec_conclusion}.
Additional band structure calculations  are presented in
Appendix~\ref{appendix_band_structure}.  In
Appendix~\ref{generic_models} we discuss tight-binding models and
their topological surface states for some tetragonal SGs. In
Appendix~\ref{appendix_low_energy_ham} we derive effective
Hamiltonians describing the low-energy physics near different
types of topological band crossings.

\renewcommand{\arraystretch}{1.03}
\begin{table*}[htbp!]
 \resizebox{\textwidth}{!}{
\begin{tabular}{| r@{ }l || c | c | c | c | c | c | }
   \hline
\multicolumn{2}{|c||}{\text{SG}} &  \text{movable Weyl points} &  \text{movable Weyl lines} &  \text{fourfold points} & \# \text{ Weyl} & \text{nodal planes} & \text{notable features} \\
 \hline 
 75 & ( $P4$ ) & & & & 8 & & \\
 \hline 
 76 & ( $P4_1$ ) & $\Gamma$-Z(8), M-A(8), X-R(4) & & & 4,8 & $k_z = \pi$ & double Weyl \\
 \hline 
 77 & ( $P4_2$ ) & $\Gamma$-Z(4), M-A(4) & & & 4,8 & & \text{double Weyl} \\
 \hline 
 78 & ( $P4_3$ ) &  $\Gamma$-Z(8), M-A(8), X-R(4) & & & 4,8 & $k_z = \pi$ & double Weyl \\
 \hline 
 79 & ( $I4$ ) & & & & 10 & & Weyl at P \\
 \hline
 80 & ( $I4_1$ ) & $\Gamma$-Z-M(4) & & & 4,8 & & double Weyl, Weyl at P(*) \\
 \hline 
 81 & ( $P\bar{4}$ ) & & & & 4 & & \\
 \hline 
 82 & ( $I\bar{4}$ ) & & & & 6 & & \\
 \hline 
 89 & ( $P422$ ) & & & & 8 & & \\
 \hline
 90 & ( $P42_12$ ) & $\Gamma$-X(4), Z-R(4) & & A, M & 2,10 & $k_x,k_y = \pi$ & fourfold Weyl \\
 \hline 
 91 & ( $P4_122$ )&   $\Gamma$-Z(8), M-A(8), X-R(4) & & & 4,8 & $k_z =\pi $ & double Weyl \\
 \hline 
 92 & ( $P4_12_12$ )& $\Gamma$-Z(8), $\Gamma$-X(4) & & M, R, A(*) & 1,3,9 & $k_x,k_y,k_z = \pi$ & 
\parbox{4cm}{ top. nodal plane, twofold/fourfold (double/quadruple) Weyl }
\\
 \hline 
 93 & ( $P4_222$ )& $\Gamma$-Z(4), M-A(4) & & & 4,8 & & double Weyl \\
 \hline 
 94 & ( $P4_22_12$ )& $\Gamma$-Z(4), $\Gamma$-X(4), Z-R(4) & & A, M & 2,12 & $k_x,k_y = \pi$ & \parbox{4cm}{top. nodal plane, double/fourfold Weyl} \\
 \hline
 95 & ( $P4_322$ )& $\Gamma$-Z(8), M-A(8), X-R(4) & & & 4,8 & $k_z = \pi$ & double Weyl \\
 \hline 
 96 & ( $P4_3 2_1 2$ )&  $\Gamma$-Z(8), $\Gamma$-X(4) & & M, R, A(*) & 1,3,9 & $k_x,k_y,k_z = \pi$ & see SG 92 \\
 \hline 
 97 & ( $I422$ ) & & & & 10 & & Weyl at P \\
 \hline 
 98 & ( $I4_122$ ) & $\Gamma$-Z-M(4) & & & 4,8 & & Weyl at P(*) \\
 \hline 
 99 & ( $P4mm$ ) & & & & & & \\
 \hline 
 100 & ( $P4bm$ ) & & ($\Gamma$-Z ; X, R)(4) & M, A & & & \\
 \hline
 101 & ( $P4_2cm$ ) & & & Z, A, R & & & \\
 \hline 
 102 & ( $P4_2nm$ ) & & ($\Gamma$-$Z$-R ; X)(4), (X-$M$-A ; R)(4) & Z, M & & & nodal chain metal \\
 \hline 
 103 & ( $P4cc$ ) & & & Z, R, A & & & \\
 \hline 
 104 & ( $P4nc$ ) &  & ($\Gamma$-$Z$-R ; X)(4), (X-$M$-$A$ ; R)(4) & Z, M, A & & & nodal chain metal \\
 \hline
 105 & ( $P4_2mc$ ) & & & Z, A & & & \\
 \hline 
 106 & ( $P4_2 bc$ ) &  & ($\Gamma$-$Z$ ; X, R)(4) &  Z, A, M, M-A(8) & & & movable fourfold point \\
 \hline 
 107 & ( $I4mm$ ) & & (N, $\Gamma$-Z, M-Z$_1$ ; $-$)(2) & & & & \\
 \hline 
 108 & ( $I4cm$ ) & & ($\Gamma$-Z, M-Z$_1$ ; N)(4) & P & & & fourfold at P \\
 \hline 
 109 & ( $I4_1md$ ) & & ($\Gamma$-Z ; X)(4), (N, $\Gamma$-Z, M-Z$_1$ ; $-$)(2) & M & & & P(*), nodal chain metal\\ 
  \hline
   110 & ( $I4_1cd $ )&  & \parbox{5.0cm}{ ($\Gamma$-Z, P ; X)(4), ($\Gamma$-Z, X ; P )(4), (X-$M$ ; P)(4), ($\Gamma$-Z, $M$-Z$_1$ ; N)(4), (P, $\Gamma$-Z; $-$)(2), (P, X-M-Z$_1$ ; $-$)(2) } & M  &  & & 
\parbox{2.5cm}{$8\mathbb{N}$ bands,\\ in-gap nodal lines}\\
  \hline 
 111 & ( $P\bar{4}2m $ )& & & & 4 & & \\ 
 \hline 
 112 & ( $P\bar{4}2c $ )& & & Z, A  & 4 & & \\
 \hline 
 113 & ( $P\bar{4}2_1m $ )& $\Gamma$-X(4), Z-R(4) & & $\overline{\text{MA}}$ & 8 & $k_x,k_y = \pi$ & fourfold line \\
 \hline 
 114 & ( $P\bar{4}2_1c $ )&  $\Gamma$-X(4) & & Z, $\overline{\text{MA}}$ & 4 & $k_x,k_y = \pi$ & fourfold line \\ 
 \hline 
 115 & ( $P\bar{4}m2 $ ) & & & & & & \\
 \hline 
 116 & ( $P\bar{4}c2 $ ) & & & Z, A, R & & & \\
 \hline 
 117 & ( $P\bar{4}b2 $ ) &  & ($\Gamma$-Z ; X, R)(4) & A, M & & & \\
 \hline 
 118 & ( $P\bar{4}n2 $ ) & A-R(4), R-X(4) & \parbox{3cm}{ ($\Gamma$-$Z$-R ; X)(4) (X-$M$-A ; R)(4) }& Z, M & & & nodal chain metal \\
 \hline 
 119 & ( $I\bar{4}m2 $ ) & & (N, $\Gamma$-Z, M-Z$_1$ ; $-$)(2) & & 2 & & \\
 \hline 
 120 & ( $I\bar{4}c2 $ ) & X-P(4) & ($\Gamma$-Z, M-Z$_1$ ; N)(4) & & 2 & & \\
 \hline 
 121 & ( $I\bar{4}2m $ ) & & & & 4 & & \\
 \hline 
 122 & ( $I\bar{4}2d $ ) & & ($\Gamma$-Z ; X)(4) & M & 4 &  & nodal chain metal  \\
 \hline 
\end{tabular}
 }
\caption{
Classification of band crossings in \emph{non-centrosymmetric}
tetragonal crystals with strong spin-orbit coupling. The first
column lists the tetragonal space groups (SG) that lack inversion
symmetry by their number and symbol.  The second and third columns
list the positions of Weyl points, see Sec.~\ref{accordion}, and
Weyl nodal lines, see Sec.~\ref{weyl_lines}, which are movable on
the indicated rotation axis or mirror plane. For Weyl
lines we group the high-symmetry points, e.g. ``X, R,'' or lines,
e.g. ``$\Gamma$-$Z$-R,'' into sets with the same mirror eigenvalue
pairing.  The sets of points for different eigenvalues are
separated by semicolons.  Note fourfold degeneracies fall into neither
category, which is indicated by italicized type, e.g. $Z$ in $\Gamma$-$Z$-R.  On a path between any two points with different
eigenvalues nodal lines are found. 
If a point (instead of a line) is given in the first eigenvalue set an almost nodal line passes through it (see Sec.~\ref{sec_movableTRIMlines}). 
For example, (P, $\Gamma$-Z; $-$)(2) denotes a nodal line passing through P situated in the mirror plane containing P and $\Gamma$-Z.  
The number in brackets
corresponds to the number of bands connected by the given feature.
The fourth column denotes points in reciprocal space with a
fourfold degeneracy. The fifth column gives the lowest possible
number of point crossings which carry a non-zero chirality,
thereby including fourfold degeneracies but excluding nodal
planes. Several numbers are given if the minimal number of Weyl
points depends on the band index. The second to last column lists
conditions on $\mathbf{k}$ for which all bands are twofold
degenerate (see the discussion of nodal planes in
Sec.~\ref{weyl_planes}). The last column contains a selection of
noteworthy features of the respective SG. Nodal points allowing
for irreducible representations of different dimensions are marked
by (*). 
\label{mTab1}
}
\end{table*}
\renewcommand{\arraystretch}{1}
\begin{table*}[htb!]
\resizebox{\textwidth}{!}{
\begin{tabular}{| r@{ }l || c | c | c | c | c | c | }
   \hline
\multicolumn{2}{|c||}{\text{SG}} &  \text{Dirac points} & \text{Dirac lines} & \text{accidental points} & \text{accidental lines} & \text{notable features} \\
\hline 
 83 & ( $P4/m$ ) & & & & & \\
 \hline 
 84 & ( $P4_2/m$ ) & Z, A & & & & \\
 \hline 
 85 & ( $P4/n$ ) & X, A, M, R & & X-R & & \\
 \hline 
 86 & ( $P4_2/n$ ) & X, Z, M, R & &  X-R & & \\
 \hline 
 87 & ( $I4/m$ ) & & & & &  \\
 \hline 
 88 & ( $I4_1/a$ ) & X, M & & X-P & &  \\
 \hline 
 123 & ( $P4/mmm$ ) & & & & & \\
 \hline 
 124 & ( $P4/mcc$ ) & A, Z, R  & & Z-R, Z-A, R-A & & \\
 \hline 
 125 & ( $P4/nbm$ ) & M, A, X, R  & & X-R, X-M, R-A & & \\
 \hline 
 126 & ( $P4/nnc$ ) & X, Z, A, R, M & & X-R, X-M, Z-R, Z-A & & \\
 \hline 
 127 & ( $P4/mbm$ ) & & $\overline{\text{MX}}$, $\overline{\text{AM}}$, $\overline{\text{AR}}$ &  & $k_x, k_y = \pi$ & \\
 \hline 
 128 & ( $P4/mnc$ ) & Z & $\overline{\text{AM}}$, $\overline{\text{MX}}$ & Z-R, Z-A, R-A & $k_x, k_y = \pi$  & \\
 \hline 
 129 & ( $P4/nmm$ ) & & $\overline{\text{RX}}$, $\overline{\text{AM}}$ & & $k_x, k_y = \pi$ & \\
 \hline 
 130 & ( $P4/ncc$ ) & Z, R-Z(4) & $\overline{\text{RX}}$, $\overline{\text{AM}}$, $\overline{\text{AR}}$ & Z-R, Z-A & $k_x, k_y = \pi$ & movable Dirac; eightfold A \\
 \hline 
 131 & ( $P4_2/mmc$ ) & Z, A & & Z-A & & \\
 \hline 
 132 & ( $P4_2/mcm$ ) & Z, R, A & & Z-R, R-A & & \\
 \hline 
 133 & ( $P4_2/nbc$ ) & X, Z, A, R, M, A-M(4) & & X-R, X-M, Z-A,  R-A & & movable Dirac \\
 \hline 
 134 & ( $P4_2/nnm$ ) & X, Z, M, R & & X-R, X-M, Z-R  & & \\
 \hline 
 135 & ( $P4_2/mbc$ ) & Z & $\overline{\text{MX}}$, $\overline{\text{AM}}$, $\overline{\text{AR}}$ & Z-A & $k_x, k_y = \pi$  & eightfold A\\
 \hline 
 136 & ( $P4_2/mnm$ ) & Z & $\overline{\text{MX}}$, $\overline{\text{AM}}$ & Z-R & $k_x, k_y = \pi $ & \\
 \hline 
 137 & ( $P4_2/nmc$ ) & Z & $\overline{\text{AM}}$, $\overline{\text{RX}}$ & Z-A &  $k_x, k_y = \pi$ & \\
 \hline 
 138 & ( $P4_2/ncm$ ) & Z, R-Z(4) & $\overline{\text{AR}}$, $\overline{\text{AM}}$, $\overline{\text{RX}}$  & Z-R &  $k_x, k_y = \pi$ & movable Dirac \\
 \hline 
 139 & ( $I4/mmm$ ) & & & & & \\
 \hline 
 140 & ( $I4/mcm$ ) & N, P & & N-P & & \\
 \hline 
 141 & ( $I4/amd$ ) & X, M & & X-P, X-M & & \\
 \hline
 142 & ( $I4_1/acd$ ) & X, M, N, P(*) & & X-P, X-M, N-P & & Dirac at P(*) \\
 \hline  
\end{tabular}
}
\caption{ \label{mTab2}
Classification of enforced band crossings in
\emph{centrosymmetric} tetragonal crystals with strong spin-orbit
coupling.  The first column lists the tetragonal space groups
including inversion symmetry by their number and symbol. The
second and third column list the positions of Dirac points and
Dirac lines.  Here, R-Z(4) refers to a movable Dirac point on the
line connecting R and Z in an hourglass structure comprising 4
spin degenerate bands (see Sec.~\ref{maintext_130} and
\ref{maintext_138}).  The fourth and fifth column list lines and
planes on which accidental points and lines are possible,
respectively. Here, the fourfold rotation axis $\Gamma$-Z and
M-A are omitted, because they always allow accidental crossings
except when they host a Dirac line (Sec.~\ref{Sec_AccBandCross_OffCenter}). The last column points to
noteworthy features. Nodal points allowing for irreducible
representations of different dimensions are marked by (*) and are discussed in Sec.~\ref{maintext_142}.
}
\end{table*}

\section{Preliminaries}
\label{mSec2} 

Before presenting the results of our classification, we start in
this section by introducing our conventions for the naming of
topological band degeneracies and by explaining the notations for
the symmetry operators (Sec.~\ref{mSec2A}). We also give a brief
discussion on how the database search for candidate materials was
performed (Sec.~\ref{example_materials}).

There are in total 68 tetragonal space groups, of which 49 have primitive
lattices (P-tetragonal), and the remaining 19 have body-centered
lattices (I-tetragonal). All tetragonal space groups posses a
fourfold rotation symmetry around the $z$-axis. The BZs for the tetragonal space groups are shown in
Fig.~\ref{fig_bulk_BZs}. Tetragonal lattices contain two
independent lattice constants $a$ and $c$, which correspond to the
breadth and height of the conventional unit cell along the $x$
and $z$ directions, respectively. Depending on the cell metrics,
the Brillouin zone for body-centered tetragonal crystals is one
of two distinct polyhedra: an elongated dodecahedron (BCT$_1$, for 
$c<a$) or a truncated square bipyramid (BCT$_2$, for
$c>a$)~\cite{setyawan2010high}.

\subsection{Conventions}
\label{mSec2A} 

The discussion of enforced band features necessitates the frequent
use of space group (SG) symmetries. For the tetragonal SGs the
relevant symmetries are two- and fourfold rotations and mirror
symmetries. We define the abbreviations for symmetries as
$N_{xyz}(abc)$, where $N=2,4$ refers to an $N$-fold rotation
around the axis given by the vector $(x,y,z)$ followed by a
(fractional) lattice translation $(a,b,c)$. For mirror symmetries
the notation $M_{xyz}(abc)$ declares $(x,y,z)$ to be the normal
direction of the mirror plane and $(a,b,c)$ again the translation.
All symbols implicitly contain also the action on the electron
spin. Combined with the spatial action the symmetries can be
exemplarily defined as
\begin{align}
&2_{001}(a,b,c) : \nonumber
\\
&(x,y,z) \to (-x+ a, -y +b, z +c) \otimes \mathrm{i} \sigma_z,
 \\
&4_{001}(a,b,c):  \nonumber   
\\
&(x,y,z) \to (-y+ a,x+b ,z+c) \otimes \frac{\sigma_0 + \mathrm{i} \sigma_z}{\sqrt{2}},
 \label{symmetries_4} \\
&M_{001}(a,b,c):\nonumber
\\
&(x,y,z) \to (x+ a, y+b, -z +c ) \otimes \mathrm{i} \sigma_z,
\label{symmetry_definitions}
\end{align}
where $\sigma_x, \sigma_y, \sigma_z$ refers to the Pauli matrices
and $\sigma_0$ is the 2x2 unit matrix. We denote pure translations by a vector $(a,b,c)$ with $t(a,b,c)$.

Additionally, we will make frequent use of time-reversal symmetry
$\mathcal{T}=\mathds{1}\otimes \mathrm{i}\sigma_y \mathcal{K}$,
which consists of an unitary part acting on spin space and complex
conjugation $\mathcal{K}$, whose action in momentum space is
$\mathcal{T}\vk = -\vk$.

It is sufficient to study points, line segments and planes of one
eighth of the full BZ, because all space groups we consider in
this work contain time-reversal $\mathcal{T}$ and the fourfold
rotation symmetry $4_{001}$. Therefore the band structure and any
topological features are always related between different parts of
the BZ and one octant of the full BZ suffices.
Without loss of generality, we set the lattice constants $a=c=1$.

\begin{figure*}[th]
\centering
\includegraphics[width=\textwidth]{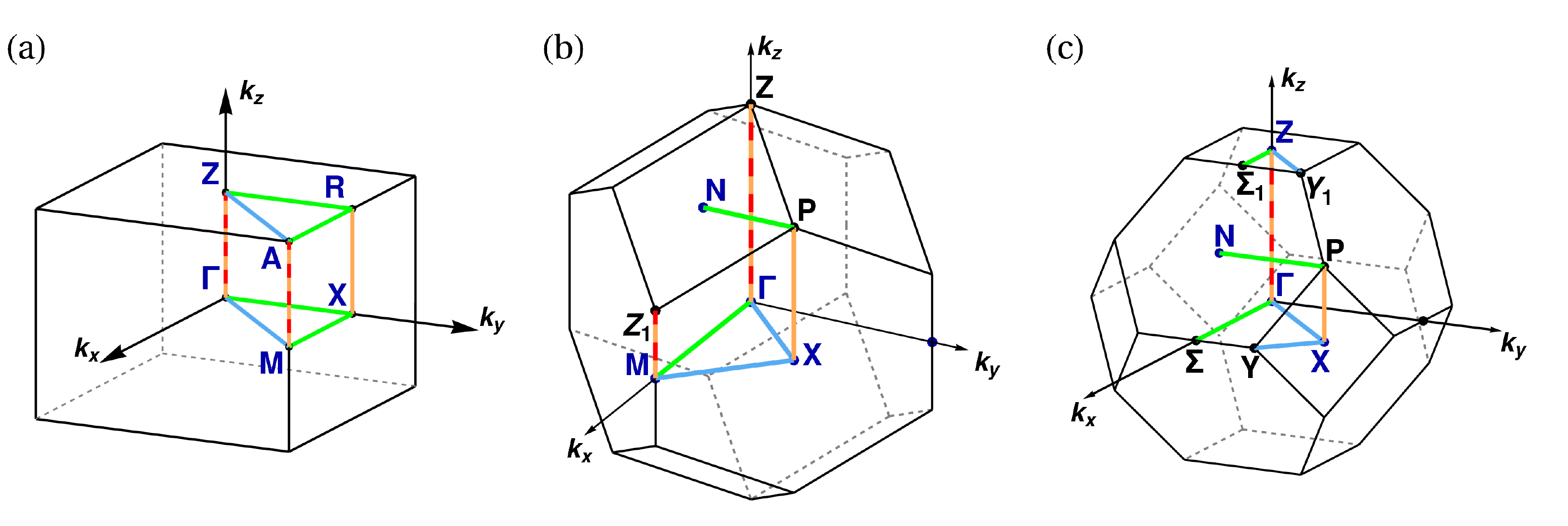}
\caption{
Brillouin zones of the tetragonal crystal system. TRIMs are labeled
in dark blue. Lines of high-symmetry, i.e. sections of rotation
axes, are highlighted for the fourfold rotation (red) along
$[001]$, and the twofold rotations along $\langle 100 \rangle
$ (green), $\langle 110 \rangle $ (light blue), and $[001]$
(orange).
(a) Primitive BZ.
(b) Body-centered BZ for $c<a$,
BCT$_1$.
(c) Body-centered BZ for $c>a$, BCT$_2$.
}
\label{fig_bulk_BZs} 
\end{figure*}

A time-reversal invariant momentum (TRIM) is an important point of
interest for any space group symmetry analysis. It is defined as a
point $\bm{k}_\text{TRIM}$ of the BZ, which is left invariant by
the action of time-reversal, i.e. $\mathcal{T} \bm{k}_\text{TRIM}
= \bm{k}_\text{TRIM} + \bm{K}$, where $\bm{K}$ is a reciprocal
lattice vector. By describing the reciprocal lattice with
coordinates given in its primitive basis, it is evident that there
are always 8 TRIMs in the Brillouin zone. To be specific in the
primitive basis the TRIMs are 
$(0,0,0)$, 
$( \frac{1}{2},0,0)$,
$(0, \frac{1}{2},0)$, 
$(0,0, \frac{1}{2})$, 
$( \frac{1}{2}, \frac{1}{2},0)$,
$( \frac{1}{2},0, \frac{1}{2})$, 
$( \frac{1}{2}, \frac{1}{2},0)$, 
$( \frac{1}{2}, \frac{1}{2}, \frac{1}{2})$ 
for any unit cell type, i.e, primitive as well as body-centered
cells. Keeping this result in mind we describe the Brillouin zone
in Cartesian coordinates, which makes the possible symmetries more
obvious. Hereby, the primitive and body-centered Brillouin zones
must be distinguished, see Fig.~\ref{fig_bulk_BZs}. Adopting this
description, the eight TRIMs for the primitive cell are
$\Gamma(0,0,0)$, X$(0,\pi,0)$ (2), Z$(0,0,\pi)$, R$(0,\pi,\pi)$
(2), M$(\pi,0,\pi)$, A$(\pi,\pi,\pi)$, where $(2)$ denotes TRIMs
that appear in 2 distinct copies related by symmetry. For the
body-centered Brillouin zone BCT$_1$ the TRIMs are
$\Gamma(0,0,0)$, X$(\pi,\pi,0)$(2), M$(2\pi,0,0)$,
N$(\pi,0,\pi)$(4) and for BCT$_2$ M$(2\pi,0,0)$ is replaced by
Z$(0,0,2\pi)$. Without loss of generality we use the labels of
BCT$_1$ unless a material realization requires BCT$_2$.

The point P=$(\pi,\pi,\pi)$ is not a TRIM of the body-centered BZ.
Yet, it is invariant under the combination of time-reversal
symmetry $\mathcal{T}$ with fourfold rotation $4_{001}$. As we
will discuss below in Sec.~\ref{eigval_Kramers}, there is a version
of Kramers theorem with this combined symmetry leading to nodal
points at P. Hereby a prerequisite of Kramers theorem can become
dependent on the eigenvalues at P for the twofold rotation.

To denote segments of high-symmetry lines we use two or more
points on the line connected by a hyphen. The shortest connection
between the points defines the line. The notation with a hyphen is
used, when any point on the high-symmetry line may be the locus of
a feature of interest and also when the whole line exhibits a
property, e.g. $\Gamma$-Z in SG 76 contains movable Weyl points
and forms accordion states. For clarity we denote the fourfold
rotation axis $(0,0,k_z)$ of the body-centered cell as
$\Gamma$-Z-M. If a feature appears for every single point on a
line, we use the same points but with an overline, e.g.
$\overline{\text{MA}}$ for SG~113 is a line of fourfold degenerate
points. 

The focus of this work lies on enforced features of the band
structure. We distinguish symmetry enforced properties, which must
occur based on symmetry arguments for any realization of a space
group with spin degrees of freedom, from accidental features that
may exist but depend on details of the system. For example
consider for now $\Gamma$-X to be a twofold screw rotation axis.
There all bands can be labeled by one out of two symmetry
eigenvalues. We will show that each band must exchange its
eigenvalue at least once forming an enforced crossing on the line
$\Gamma$-X, cf.  Sec.~\ref{accordion}. Furthermore, any crossing
of bands with different eigenvalues must be gapless, because any
term introduced to gap the crossing necessarily breaks the
symmetry. It is possible that the bands exchange several times
leading to what we refer to as accidental band crossings.
Nevertheless they are symmetry protected like the enforced
crossings. Note, if both occur it is generally not possible to
label one of them as enforced. This notion of accidental crossings
includes non-guaranteed band crossings that are protected by other
means than symmetry eigenvalues, e.g. Weyl points situated away
from high-symmetry points or lines, which are protected only by
translation symmetry and their inherent non-zero Chern number. In
other words, we regard enforced features to be stable under
arbitrary symmetry-preserving perturbations. They are unaffected
by the details of a respective realization.

When naming point-like topological band crossings, we call
fourfold crossings Dirac points only in the presence of inversion
symmetry.  We regard a point crossing as a species of Weyl point
if and only if it carries a non-zero chirality. By default, such a Weyl point
is twofold degenerate with a chirality $\mathcal{C} = 1$, double
Weyl points are twofold degenerate with a chirality of 2.  We
refer to fourfold double Weyl points, when discussing two
superimposed regular Weyl points with a total chirality of
$\mathcal{C} = 2$, which is also called a doubled spin-1/2 fermion
in the literature~\cite{PhysRevB.98.155145}.  We further discuss
the fourfold quadruple Weyl point, which consists of two
superimposed regular double Weyl points and carries a total charge
of $\mathcal{C} = 4$, see Sec.~\ref{fourfold_quadruple_Weyl}.

\subsection{Systematic search for example materials}
\label{example_materials}

 \begin{figure}[t!]
 \includegraphics[width = 0.89\columnwidth]{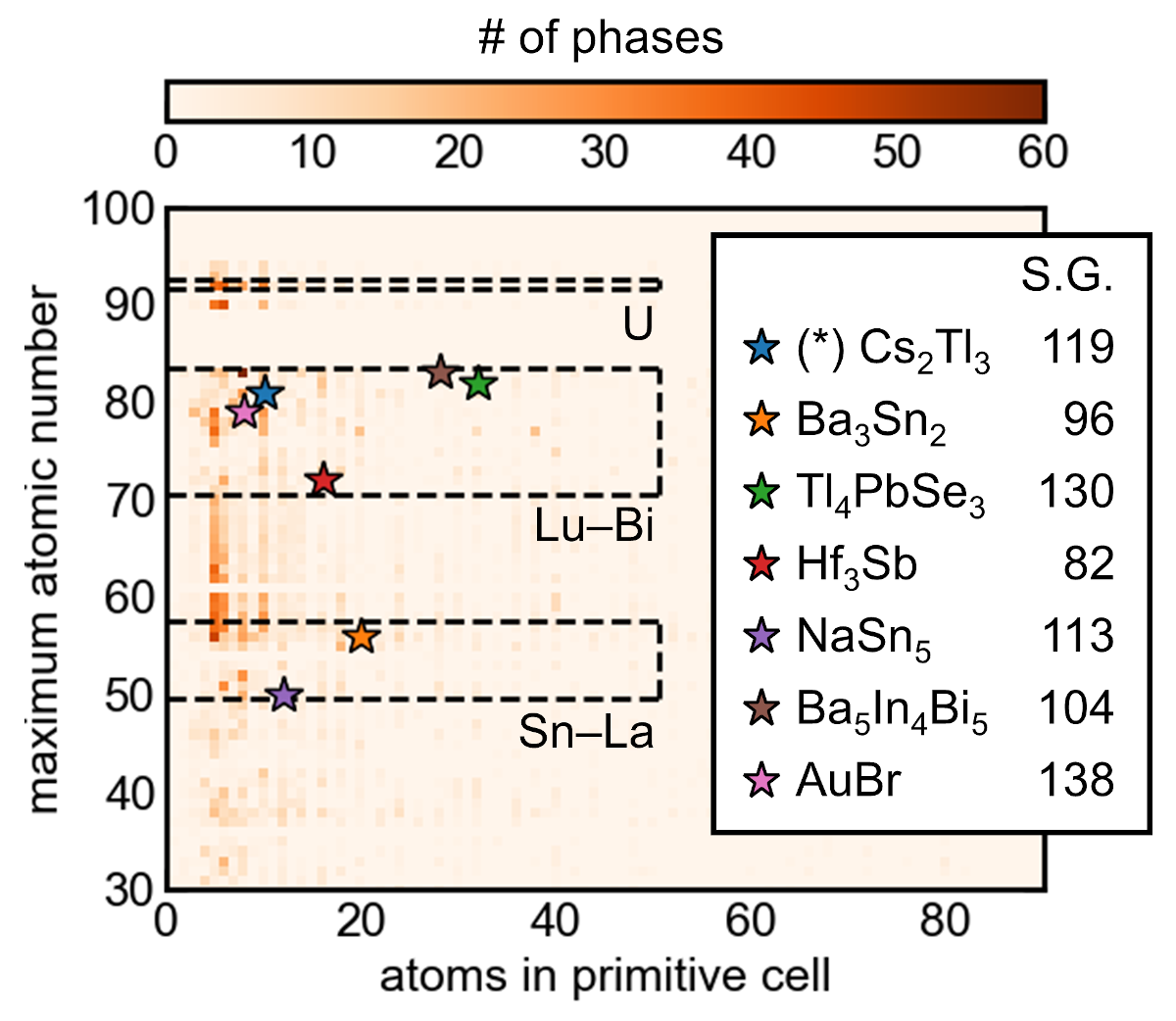}
\caption{
\label{heat_map}
Distribution of ordered, tetragonal phases in the Materials
Project and ICSD (MP $\cap$ ICSD, $N$~=~7039) by atomic number of
the heaviest element and the number of atoms in the primitive unit
cell. Seven example phases discussed in the text are highlighted.
The regions bounded by black dashed lines indicate two of the
automated screening criteria explained in the text. (*)
Cs$_2$Tl$_3$ is a hypothetical phase (see text).
}
\end{figure}
 
In the following sections, we will identify and discuss the
tetragonal SGs, whose non-symmorphic symmetries
enforce various topological and trivial degeneracies on points,
lines and planes. For each of those SGs, we then perform a
database search for material examples among the subset of
ordered, inorganic, crystalline phases in the Materials Project
database~\cite{materials_project_citation,materials_project_web_link}
that correspond to entries in the Inorganic Crystal Structure
Database (ICSD)~\cite{ICSD_link}. The distribution of these
$\sim$7000 tetragonal phases by maximum atomic number and number
of atoms in the primitive cell is given in Fig.~\ref{heat_map}.
To identify materials with strong spin-orbit coupling, we screen
for phases containing elements Sn$-$La, Lu$-$Bi, or U (we exclude
most rare-earths as $f$ electrons are poorly described by single
particle DFT, but compounds of La$^{3+}$, Lu$^{3+}$, or U$^{6+}$
have empty or filled $f$-shells). We apply further screens based
on the number of atoms in the unit cell (as larger cells have more
bands and typically more limited band widths), thermodynamic
stability (formation energy $\leq$50~meV~atom$^{-1}$ off the
convex hull), and bandgap ($\leq$1~eV by a generalized gradient approximation functional). These
criteria yield between zero and several dozen phases for each of
the space groups of greatest interest. Seven example materials,
annotated in Fig.~\ref{heat_map}, are discussed in detail in the
corresponding sections. AuBr and Cs$_2$Tl$_3$ were added manually
to this group, as AuBr has a bandgap wider than the screening
criterion, and Cs$_2$Tl$_3$ is a hypothetical heavier analogue of
the known Cs$_2$In$_3$.

For the materials of interest we perform DFT band structure
calculations with the VASP code~\cite{Kresse1996,Kresse1996-2},
using the projector augmented wave (PAW)
method~\cite{Bloechl1994,Kresse1999} and the PBE~\cite{PBE_PRL_96}
exchange-correlation functional. The relaxed structures stored in
the Materials Project are used directly for all phases except
hypothetical Cs$_2$Tl$_3$, whose structure is determined by
relaxing Tl-substituted Cs$_2$In$_3$.

In the main text we will focus on the band dispersion around the
features of interest. The complete band structures along the
high-symmetry lines of the full BZ are presented in
Appendix~\ref{appendix_band_structure}.

\section{Applications of Kramers theorem} 
\label{sec_application_kramers_theorem}

Kramers theorem states that an antiunitary operation
$\mathcal{T}$, e.g. time-reversal, with $\mathcal{T}^2 = -1$
leads to twofold degenerate bands at each invariant
momentum~\cite{wigner_Kramers_1932}. These Kramers-Weyl points
were discussed for chiral space groups
\cite{chang2018topological}. We extend the argument to groups
containing mirror and rotoinversion symmetries and find a minimal
number of paired Weyl points in SG~119 and 120. It is favorable to
obtain simple realizations of Weyl semimetals comprising few Weyl
points with a large separation to simplify further analysis.
In such systems the signatures of Weyl points are more prominent,
e.g. in transport properties and as a less complex arrangement of
the surface Fermi arcs \cite{minimumNumberArcs_MoTe_2016_prl,
minimumWeyl_TaIrTe_2016_prb, belopolski2017signatures}.

Beyond this well known application of Kramers theorem at TRIMs, it
also leads to other topological features and can be adapted to
convey insights into the tabulated irreducible group
representations~\cite{elcoro_aroyo_JAC_17, miller_love_irreps}.
Among these features there are nodal planes as a result of twofold
screw rotations, which we discuss separately in
Sec.~\ref{weyl_planes}, as well as nodal lines on high-symmetry
axes following an analogous argument with glide mirror symmetries
and rotoinversions.

We first discuss Kramers-Weyl points at TRIMs and then elaborate
on how Kramers pairing occurs away from TRIMs. For the second part
the combination of time-reversal with (non-)symmorphic symmetries
results in line and point degeneracies.

\subsection{Kramers-Weyl points} 
\label{sec_application_kramers_theorem_A}

Weyl points are, in their basic form, twofold degenerate crossings
acting as sources and sinks of Berry curvature, which is measured
by their topological charge, the chirality $\mathcal{C}$. It corresponds to
the Chern number calculated on an enclosing two-dimensional
manifold, e.g. a sphere. 
Since the Berry
curvature transforms as a pseudo vector under reflections, one
derives that the chirality of any nodal feature on a mirror plane must be equal to
its negative and thereby vanish. By excluding point groups with
mirror and inversion symmetry it is possible to conclude that any
given TRIM carries twofold degenerate nodal points with non-zero
chirality dubbed as Kramers-Weyl
points~\cite{chang2018topological}.

Yet, it is insufficient to consider only chiral crystals, meaning
crystals with rotations and translations only, because a mirror
symmetry can coexist with a Weyl point as long as the
Weyl point is not situated on the mirror plane.
In that case, the mirror symmetry relates two Weyl points of
opposite chirality.
The tetragonal SGs 111-114 and 119-122 contain such
enforced topological crossings. To aid the search of new
materials with few Weyl points we have listed the multiplicities
of enforced Weyl points in the column "\# Weyl" in
Table~\ref{mTab1}. Depending on the considered band index
different multiplicities appear and are listed in ascending order.
Note that this column contains all charged nodal points, not just
\mbox{(Kramers-)Weyl} points, independently of the magnitude of
their charge. The SGs 99-110, 115-118 each contain a version of
$M_{100}$, which has mirror planes intersecting all TRIMs, and
also do not contain Weyl points enforced by other means thus
leaving the entry in column "\# Weyl" empty. Although each space
group without mirror symmetries contains in principle a set of 8
Kramers-Weyl points, nodal lines or planes may lead to extended
instead of point-like degeneracies. Consider SG 81 with its
fourfold rotoinversion, where the lines $\Gamma$-Z and M-A exhibit
only twofold representations, such that just X and R remain as
Kramers-Weyl points yielding a total multiplicity of 4.

Finally, we find the minimal number of one Weyl point pair in
SG~119 by choosing a space group with a specific mirror symmetry
and a body-centered unit cell. Consider SG~82 as starting point,
a body-centered version of SG~81, which is generated by a fourfold
rotoinversion. Analogous to SG~81 the twofold degenerate line
$\Gamma$-Z-M reduces the number of TRIMs with Kramers-Weyl points,
leaving six Weyl points at X and N. The rotoinversion relates Weyl
points of opposite chirality, i.e. the nodal points $(\pi,\pi,0)$
and $(\pi,-\pi,0)$ both labeled X, and likewise for the crossings
at N. If the mirror symmetry $M_{100}$ is added to remove the Weyl
points at N as well, the resulting group is SG~119. For SG~119 the
twofold degeneracies at N 
are part of almost movable nodal lines, see
Sec.~\ref{sec_movableTRIMlines}, and carry no chirality. Hence,
SG~119 yields two enforced Weyl points at X related by an improper
rotation and a mirror symmetry, which achieves the minimal number
of one pair of Weyl points on a
lattice~\cite{nielsen_no_go,NIELSEN198120}. For SG~120 a single
pair of Kramers-Weyl points follows from an identical argument. 

This presence of only two Weyl points in SGs~119 and 120 per band
pair is at odds with the common notion that four Weyl points are
the minimal number achievable in a system with time-reversal
symmetry \cite{minimumNumberArcs_MoTe_2016_prl,
minimumWeyl_TaIrTe_2016_prb, belopolski2017signatures}. The
argument assumes that time reversal relates distinct Weyl points
with the same chirality and concludes by the Nielsen-Ninomiya
theorem that two additional Weyl points of opposite chirality must
exist \cite{nielsen_no_go,NIELSEN198120}. Although for a Weyl
point at a TRIM the first assumption is false, the minimal number
of four seems unchallenged, since there are eight TRIMs. But as
we have noted above, due the mirror symmetry $M_{100}$ in the
body-centered SGs~119 and 120 all but two nodal points have
vanishing chirality. Note however that it is possible to obtain a
single Weyl point if the nodal manifold of opposite chirality
appears in the shape of a topological nodal plane instead of a
Weyl point, see Sec.~\ref{weyl_planes}.

To further study SG~119 we have created a minimal model, confirmed
that the chiralities of Weyl points at X are indeed $+1$ and $-1$,
and determined the surface states, see
Appendix~\ref{Appendix_119}. Here we want to highlight one
property of the surface states for SG~119. 
We find that each surface exhibits two
Fermi arcs connecting the projections of the two X points in the
surface BZ. The arcs are related by time-reversal symmetry and
appear in disjoint bulk band gaps, see Fig.~\ref{Surface_SG119}.

Nevertheless it is unusual that two arcs appear if there is only
one pair of singly charged Weyl points. To understand the
relation of Weyl points and Fermi arcs, the standard approach is
to consider gapped planar subsystems on which one may calculate
the Chern number. It is a peculiarity of the body-centered BZ,
see Fig.~\ref{fig_bulk_BZs}(b,c), that an oriented plane normal to
(1,1,0) that passes for example the point
$(\tfrac{\pi}{2},\tfrac{\pi}{2},0)$ will also include the point
$(-\tfrac{\pi}{2},-\tfrac{\pi}{2},0)$. Hence, considering the
orientation of the normal vector at the two exemplary points the
plane does not capture the chirality of one of the Kramers-Weyl
points at X. The plane has Chern number zero. If then a slab is
made, i.e. the periodic boundary conditions are lifted, such that
this plane is truncated, the subsystem of the plane appears as two
parallel lines in the surface BZ. We find indeed that the surface
states pierce this line twice but in opposing directions, which
agrees with the vanishing Chern number of the corresponding
subsystem.

One concludes that considering a gapped and planar subsystem is
not sufficient to discern the nontrivial topology of SG~119 and
analogously SG~120. Rather, it is the chiralities of the
Weyl points alone that indicate the presence of topological surface
states and the connectivity of the body-centered BZ allows for two
Fermi arcs even for singly charged Weyl points.

\subsubsection{Material example: Hf$_3$Sb}
\label{Hf3Sb_chapter}

As an example for SG 82 ($I\bar{4}$), which enforces only six Kramers-Weyl points per pair of bands, we present the band structure of Hf$_3$Sb in Fig.~\ref{Hf3Sb_Cs2Tl3_bands}(a). 
It crystallizes with
$c<a$~\cite{willerstrom1981128,schubert1964einige}, i.e. its BZ
is of type BCT$_1$. 
Kramers-Weyl points appear at N and X for all
bands, whereas the points $\Gamma$ and M are part of a nodal line.
Accidental band crossings occur for several bands along the twofold rotation axis X-P. 

\subsubsection{Material example: Cs$_2$Tl$_3$}
\label{Cs2Tl3_chapter}

\begin{figure}
\centering
\includegraphics[width=\linewidth]{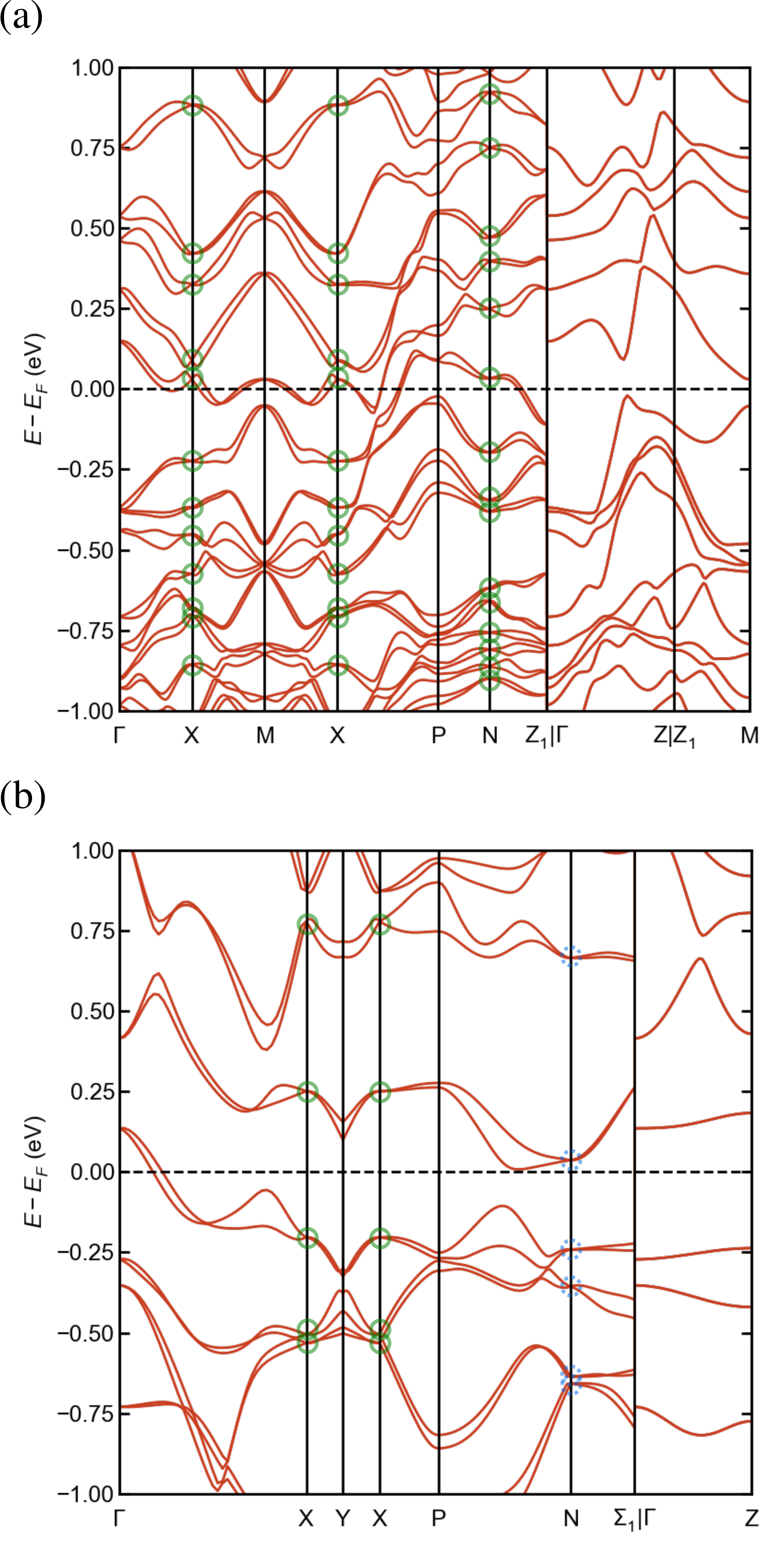}
\caption{ \label{Hf3Sb_Cs2Tl3_bands}
Kramers-Weyl realization. Green circles mark the Weyl points at
TRIMs. 
(a) Band structure of Hf$_3$Sb (SG~82). Each pair of bands
exhibits 6 independent Weyl points.
(b) Band structure of Cs$_2$Tl$_3$ (SG~119). Only 2 Weyl points
per pair of bands are enforced. Dashed blue circles highlight
crossings at N which are part of almost movable nodal lines in mirror planes
and contribute no chirality, see Sec.~\ref{sec_movableTRIMlines}.
}
\end{figure}

Cs$_2$Tl$_3$ is a hypothetical, heavier analogue to Cs$_2$In$_3$
in SG 119 ($I\bar4m2$) with $c>a$~\cite{yatsenko1985339}. We find
the formation energy of Cs$_2$Tl$_3$ with respect to the elements
is favorable, but the compound is not among the reported binaries
in the Cs$-$Tl system~\cite{dong_CsTl, dong19958, dong_Cs15Tl27}.
The band structure shown in Fig.~\ref{Hf3Sb_Cs2Tl3_bands}(b) shows
the two distinct Kramers-Weyl points at X.
The nodal crossings at N are also enforced by Kramers theorem, but
they are part of nodal lines within mirror planes and thus without
chirality. Similarly, the TRIMs $\Gamma$ and Z do not exhibit a
nodal point because they are part of a nodal line.

\subsection{Kramers theorem beyond TRIMs}
\label{sec_application_kramers_theorem_B}
\label{eigval_Kramers}

In this section we discuss band degeneracies due to
(non-)symmorphic, antiunitary symmetries. It is instructive to
consider for a start symmorphic symmetries combined with the
time-reversal operation $\mathcal{T}$ in the context of Kramers
theorem. The best known case is inversion symmetry, $\mathcal{P}$.
The combination $\mathcal{PT}$ enforces twofold degeneracies at
all momenta, because $(\mathcal{PT})^2 = -1$ and all $k$ are
invariant momenta. Due to this ubiquitous twofold degeneracy in
centrosymmetric groups, we have split the results of this paper
into the cases without and with inversion symmetry, see
Tables~\ref{mTab1} and~\ref{mTab2}. 

The combination of time reversal with non-symmorphic symmetries
can lead to nodal lines and planes, which occur at invariant $k$
points where the compound symmetry squares to any number different from 1, see
Sec.~\ref{weyl_planes}. In the following paragraphs we describe
variations of this argument, which lead to topological band
crossings. First we discuss symmorphic antiunitary compound
symmetries, which unlike $\mathcal{PT}$ do not square to the $-1$ but still yield nodal
lines or points and then continue to non-symmorphic operations
responsible for Kramers pairing restricted to certain rotation
eigenvalues. Both arguments enforce Weyl points, which are not
situated at TRIMs but pinned to the point P.

\subsubsection{Symmorphic antiunitary symmetries}

Consider the non-centrosymmetric SGs~81, 82 and SG~119
from the previous section, which exhibit a twofold degenerate
nodal line along $\Gamma$-Z and $\Gamma$-Z-A, respectively. This
degeneracy can be understood with the symmorphic operation
$\overline{4}_{001} \mathcal{T}$, where the rotoinversion
$\overline{4}_{001}$ acts on the fourfold rotation axes like 
inversion by itself. There is an analogy to the case of $\mathcal{PT}$
symmetry even though $(\overline{4}_{001} \mathcal{T})^2 \neq -1$.
As $\overline{4}_{001} \mathcal{T}$ leaves $\mathbf{k}$ invariant,
for the moment one may assume that $\overline{4}_{001} \mathcal{T}
$ creates a linearly dependent state $\overline{4}_{001}
\mathcal{T} \ket{\psi_{\mathbf{k}}} = \exp(\mathrm{i} \varphi)
\ket{\psi_{\mathbf{k}}}$, when acting on the eigenstate
$\ket{\psi_{\mathbf{k}}}$ of a Hamiltonian by producing only a
phase factor $\exp(\mathrm{i} \varphi)$. 
By substituting this equation in the form $\ket{\psi_{\mathbf{k}}} = \exp(-\mathrm{i} \varphi) \overline{4}_{001} \mathcal{T} \ket{\psi_{\mathbf{k}}}$ four times into itself, a contradiction $ \ket{\psi_{\mathbf{k}}} = (\overline{4}_{001} \mathcal{T})^4
\ket{\psi_{\mathbf{k}}} = -\ket{\psi_{\mathbf{k}}}$ is found by
using $ \mathcal{T}^2 = -1$ with $\overline{4}_{001}^4 = -1$ for the last step. 
Hence, the assumption is wrong and $\overline{4}_{001}
\mathcal{T}$ must relate the eigenstate $\ket{\psi_{\mathbf{k}}}$
to a linearly independent state $\overline{4}_{001} \mathcal{T}
\ket{\psi_{\mathbf{k}}}$ of the same energy.

The same argument can be used as well at TRIMs for the operation $4_{001} \mathcal{T}$ comprising the
symmorphic proper rotation $4_{001}$. 
But more intriguingly for body-centered space groups the symmetry $4_{001} \mathcal{T}$ leaves another 
momentum invariant, the point P. 
By the same argument as above P will always be degenerate. 
We highlight Weyl points at P in the last column of 
Table~\ref{mTab1}, because they add to the number of Kramers-Weyl
points without being at a TRIM.

\subsubsection{Eigenvalue dependent Kramers theorem}
\label{eigenvalue_dependent_Kramers_theorem}
Here we show a variant of Kramers theorem that only applies to
states with certain symmetry eigenvalues, which we refer to as the
eigenvalue dependent Kramers theorem. To illustrate this term,
note that typically Kramers theorem pairs bands differently
depending on whether symmetry eigenvalues are complex or real. In
that case a band with a real eigenvalue is paired to a band with
the same eigenvalue, whereas a complex eigenvalue is paired to its
complex conjugate. But for the eigenvalue dependent
Kramers theorem only some symmetry eigenvalues are paired, whereas
for others the theorem does not hold at all and no pairing is
obtained. It is a subtle and common feature of many band
structures that all bands at a given high-symmetry point of
reciprocal space have the same degeneracy. But this is not always
true. The tetragonal SGs 80, 92, 96, 98, 109, and 142 have
irreducible representations of different dimension at
high-symmetry points. This can be understood with the eigenvalue
dependent Kramers theorem. To do so, we revisit the above
observation that $(\overline{4}_{001} \mathcal{T})^2 \neq -1$ for
non-symmorphic rotations as a starting point for the theorem.

The presence of irreducible representations of unequal dimensions
at the same point can be explained by considering a time-reversal-containing
symmetry, which fulfills Kramers theorem only for
certain eigenvalues. To illustrate this, consider the
proof~\cite{wigner_Kramers_1932} of Kramers theorem for a state
$\ket{\psi}$ and an antiunitary operation $\tilde{\mathcal{T}}$
fulfilling $\tilde{\mathcal{T}}^2 \ket{\psi} = a \ket{\psi}$ with
$a \in \mathbb{C}$ :

\begin{align}
\braket{\psi}{\tilde{\mathcal{T}} \psi}
=
\braket{\tilde{\mathcal{T}}^2\psi}{\tilde{\mathcal{T}} \psi}
= 
a^* \braket{\psi}{\tilde{\mathcal{T}} \psi},
\label{Kramers_def}
\end{align}
where we used the property $\braket{\psi}{\phi} =
\braketnoresize{\tilde{\mathcal{T}} \phi}{\tilde{\mathcal{T}}
\psi}$ of antiunitary operators. $\tilde{\mathcal{T}} \ket{\psi}$
is an orthogonal state to $\ket{\psi}$ if $a \neq 1$. The
conventional time-reversal yields $a = -1$ leaving no room for a
dependence on eigenvalues of $\psi$. If we consider the
antiunitary symmetry $\tilde{\mathcal{T}} = \mathcal{T}
4_{001}(a,b,c)$ at an invariant k point, then $a = -
\alpha_{2_{001}}$ depends on the rotation eigenvalues
$\alpha_{2_{001}}$ of the twofold rotation symmetry resulting from
$\left(4_{001}(a,b,c)\right)^2$. For $c = \frac{1}{4}$ or
$\frac{3}{4}$ the two eigenvalues differ at $k_z = \pi$ by a
factor of $-1$. Given that, we can conclude that one but not the
other eigenstate of $2_{001}(a-b,b+a,2c)$ will have a Kramers
partner. In other words, only a band with twofold rotation
eigenvalue $+1$ but not $-1$ has a Kramers partner.

The SGs 80, 92, 96, 98, 109, and 142 contain high-symmetry points
hosting irreducible representations of different dimensions due to
eigenvalue dependent pairing. We highlight these features in
Tables~\ref{mTab1} and \ref{mTab2} with the symbol "(*)" next to
the label of the high-symmetry point, which can be found either in
the fourth or the last column. 

This eigenvalue dependent Kramers theorem makes a difference at
points in the BZ with $k_z = \pi$ that are invariant under
$\tilde{\mathcal{T}}$, i.e. where one eigenvalue of the twofold
screw rotation is $\alpha_{2_{001}} = -1$. These are the points Z
and A in the primitive unit cell and P for the body-centered case.
Note that at TRIMs the degeneracy will not always be increased
because time-reversal $\mathcal{T}$ and $\tilde{\mathcal{T}}$ can
relate the same symmetry eigenvalues. For them to act differently
on $2_{001}$ eigenvalues it is necessary that the fourfold
rotation contains a partial lattice translation perpendicular to
the rotation axis, as is the case in SGs 92 and 96 (see the
point A in the example in Sec.~\ref{Ba3Sn2_chapter}). Similarly,
additional symmetries can equalize the number of degenerate bands.
Compare the unequal dimension of representations at P of SG 109
with the uniformly twofold bands at P for SG~110. The latter is
due to a line of Kramers degeneracies caused by
$M_{010}(0,0,\frac{1}{2}) \mathcal{T}$, cf.
Fig.~\ref{Fig:SG110_BZ_NodalLines}(b) along the path N-P.

Beyond the tetragonal SGs this argument can be extended to a
few hexagonal and cubic cases, which is left for future work.


\section{Non-symmorphic Weyl points}
\label{Sec_NonSymWeyl}

In this section we discuss how screw-rotation symmetries in
tetragonal SGs lead to protected Weyl points. That is, we show how
the momentum dependence of the symmetry eigenvalues enforces band
crossings along a line in the BZ. Weyl points in tetragonal SGs
come in four different varieties: single Weyl points, double Weyl
points, fourfold Weyl points, and fourfold double Weyl points.

\subsection{Hourglass and accordion states}
\label{accordion}
\label{Sec_NonSymWeyl_A}

\begin{figure}[htbp]
\begin{minipage}[b]{0.40\linewidth}
\subfloat[$\,2_{001}(a,b,\frac12)$]{
\includegraphics[width=\linewidth]{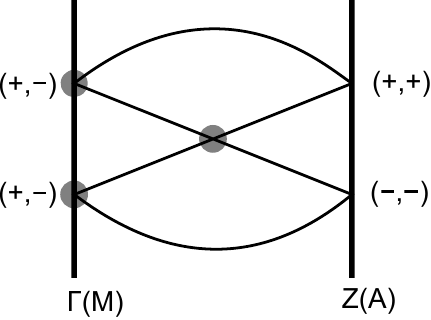}
\label{hourglass_figure}
}
 
\subfloat[$\,4_{001}(a,b,\frac12)$]{
\includegraphics[width=\linewidth]{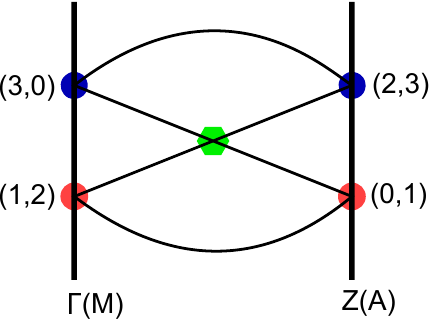}
\label{accordion_even}
}
\end{minipage}
\subfloat[$\,4_{001}(a,b,\frac14)$]{
\includegraphics[width=.40\linewidth]{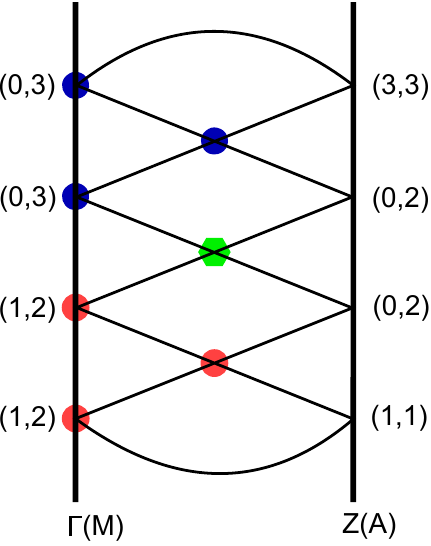}
\label{accordion_odd}
}
\caption{Connectivity diagrams with minimal band crossings for the
          screw rotations found in tetragonal space groups.
            The case for a fourfold screw rotation with $n=3$ looks
            like $n=1$ with redefined labels. 
            The colored symbols indicate the chiral charge of each
            crossing. Blue and red dots mark chirality +1 and -1,
            gray dots a chirality of $|\mathcal{C}|=1$ with undetermined
            sign and green hexagons indicate a double Weyl point
            with $\mathcal{C}=\pm2$.
        } 
\label{accordion_figure}
\end{figure}

A twofold screw rotation consists of a twofold rotation followed
by half a lattice translation along the rotation axis, e.g.
$2_{001}(a,b,\frac12)$ for arbitrary lateral translation
components $a$ and $b$. These symmetries square to a full lattice
translation and acquire a negative sign from the spin component,
which we denote as $-t(0,0,1)$ for the example above. 
Translation eigenvalues are given by a phase factor, defined
through the crystal momentum $\mathbf{k}$.
The eigenvalues of the screw rotation are therefore restricted to
the two square roots $\pm\sqrt{-\mathrm{e}^{\mathrm{i}k_z}}$ in
the example above, where the rotation axis is in $z$-direction
and we will label them with their sign.
Evaluated at the time-reversal invariant momenta on the rotation
axis, these are $\pm\mathrm{i}$ and $\pm1$ for $k_z=0$ and $\pi$,
respectively, corresponding to $\Gamma$ (M) and Z (A). Time
reversal symmetry ensures a Kramers partner at the same energy
with the complex conjugate eigenvalue, i.e. it creates pairs
\mbox{$(+,-)$} at $k_z=0$ and two pairs \mbox{$(+,+)$} and
\mbox{$(-,-)$} at $k_z=\pi$. The bands connecting these
degenerate points must necessarily cross at some point on the axis
and this band crossing is protected by the different symmetry
eigenvalues. On the axes left invariant by the rotation, the total
dispersion of the four bands involved creates an hourglass
shape~\cite{wang2016hourglass} as shown in Fig.~\ref{accordion_figure}(a). 

Combining a twofold screw rotation with time-reversal symmetry
results in an additional antiunitary symmetry squaring to a full
lattice translation. Invariant $k$-points are restricted to two
planes perpendicular to the rotation axis $\hat{\mathbf{e}}_i$,
with $\vk \hat{\mathbf{e}}_i=0$ and $\pi$. On the latter plane,
this symmetry squares to a lattice translation with eigenvalue 
$\mathrm{e}^{\mathrm{i}\vk\hat{\mathbf{e}}_i}=-1$ and enforces
Kramers degenerate states in the whole plane. Hence, the
degeneracies at Z (A) in the band connectivity diagrams are part of
these nodal planes and do not form Weyl points. For the topology
of such band degeneracies, see the section about topological nodal
planes (Sec.~\ref{weyl_planes}).


In a similar manner, we can construct the band
connectivity for fourfold non-symmorphic rotation symmetries
$4_{001}(a,b,\frac{n}{4})$ as defined in Eq.~\eqref{symmetries_4}.
The two invariant axes in primitive lattices are $\Gamma$-Z and
M-A. Together with translations, they generate SGs 76-78 with
$n=1,2,3$, respectively.
Where additional symmetries do not impose further degeneracies,
these arguments also hold for their supergroups, specifically for SGs 91 and 95 on both invariant lines and for SGs 92 and 96 on $\Gamma$-Z.
Applying such a symmetry four times results in a full lattice
translation and a minus sign from the spin component, 
$-t\,(0,0,n)$.
The eigenvalues are therefore given by the fourth roots of
$-\exp(\mathrm{i}n k_z)$, which we label with the integer 
$p \in \{0,1,2,3\}$,
defined through the parameterization
\begin{equation}
\alpha_p = \exp\left((\mathrm{i}\tfrac{\pi}{4}(2p+1)\right)
           \exp\left(\mathrm{i}\tfrac{n}{4}k_z\right).
\label{4fold_eigenvalues}
\end{equation}
At the two TRIMs on any invariant axis, time-reversal symmetry
$\mathcal{T}$ pairs bands with complex conjugate eigenvalues.
For $k_z=0$, i.e. at $\Gamma$ and M, we find
${\alpha_0}^*=\alpha_3$ and ${\alpha_1}^*=\alpha_2$, 
and we label the degeneracy with 
$(p,p^\prime)=(0,3)$ and (1,2), respectively.
For $k_z=\pi$, i.e. at Z and A, we have to distinguish
the three different possible fractional translations. 

With $n=1$, we find ${\alpha_0}^*=\alpha_2=-\mathrm{i}$, leading to
the pairing (0,2). The two real eigenvalues $\alpha_1=-1$ and
$\alpha_3=1$ get a Kramers partner with the same eigenvalue
each, leading to the pairing (3,3) and (1,1).
Connecting these pairs creates a minimum of three band crossings
on the invariant axis. This pattern is called an accordion 
state~\cite{zhang_hexagonals_PRMAT_18, gatti_ARPES_accordion_2020} 
and is shown in Fig.~\ref{accordion_figure}(c). Each of these
crossings is protected by different rotation symmetry
eigenvalues.
For a screw rotation with $n=3$ the same pattern is found, but
with interchanged labels 
\mbox{$p\rightarrow (p-1)\mod 4 $}.

For $n=2$ there are no real eigenvalues at either TRIM. Whereas
the pairing at
$\Gamma$ (M) remains unchanged, the pairing at Z (A) is now
(0,1) and (2,3). This allows a simpler band connectivity made up
of 4 bands only and one band crossing along the path,
see Fig.~\ref{accordion_figure}(b).

The chirality $C$ of these Weyl points can be inferred from the
ratio of eigenvalues involved in a
crossing~\cite{tsirkin_vanderbilt_PRB_17}. Twofold screw-rotations
can create only single Weyl points of chirality $\pm1$, whereas
fourfold screw-rotations lead to at least one double Weyl point
with chirality $\pm2$. Further details of each type of crossing
will be discussed below.

\subsection{Single Weyl points}
\label{Sec_NonSymWeyl_B}

These are conventional Weyl points with a linear dispersion in
each direction and a topological charge of $\mathcal{C}=\pm 1$~\cite{tsirkin_vanderbilt_PRB_17}.

For an hourglass dispersion from a twofold screw rotation, all
movable crossings and the ones at $\Gamma$ are of this type.
In the accordion dispersion, single Weyl points are found whenever
the ratio of rotation eigenvalues in a band crossing is purely
complex. In that case, the chirality is completely determined by
symmetry eigenvalues,
\begin{equation}
\frac{\alpha_p}{\alpha_{p^\prime}} = \pm\mathrm{i}
\Rightarrow \mathcal{C} = \mp 1.
\end{equation}
Here, $\alpha_p$, as defined in Eq.~\eqref{4fold_eigenvalues}, is
the symmetry eigenvalue of the upper band when moving to larger
$k_z$ and $\alpha_{p^\prime}$ the eigenvalue of the lower band.
This implies a sign change when rearranging the order of bands.
This condition is always met for the band crossings at $\Gamma$ and for
two of the four movable crossings in the accordion state. In both
cases, the order in energy might be exchanged, but the sum of all
chiralities vanishes.

\subsection{Double Weyl points}
\label{Sec_NonSymWeyl_C}

\begin{figure*}
\centering
\subfloat[Double Weyl point]{
\includegraphics[width=.30\linewidth]{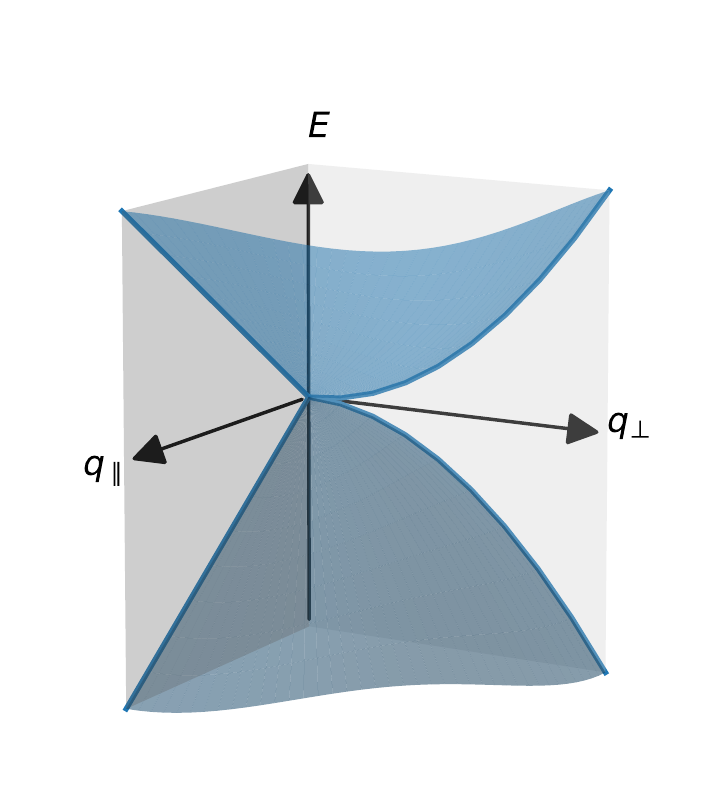}
\label{double_weyl_fig}
}
\subfloat[Fourfold double Weyl point]{
\includegraphics[width=.29\linewidth]{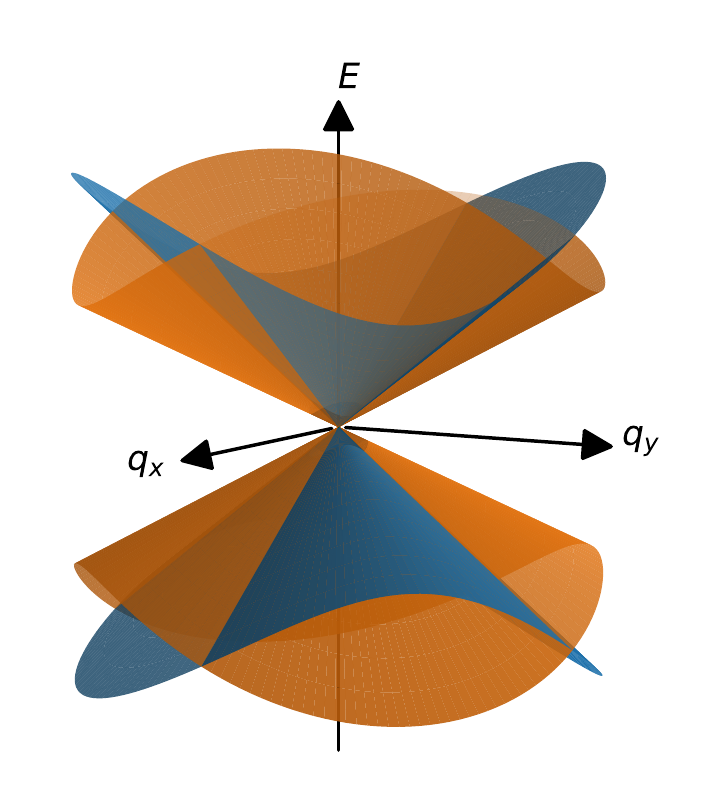}
\label{4fold_weyl_fig}
}
\subfloat[Fourfold quadruple Weyl point]{
\includegraphics[width=.29\linewidth]{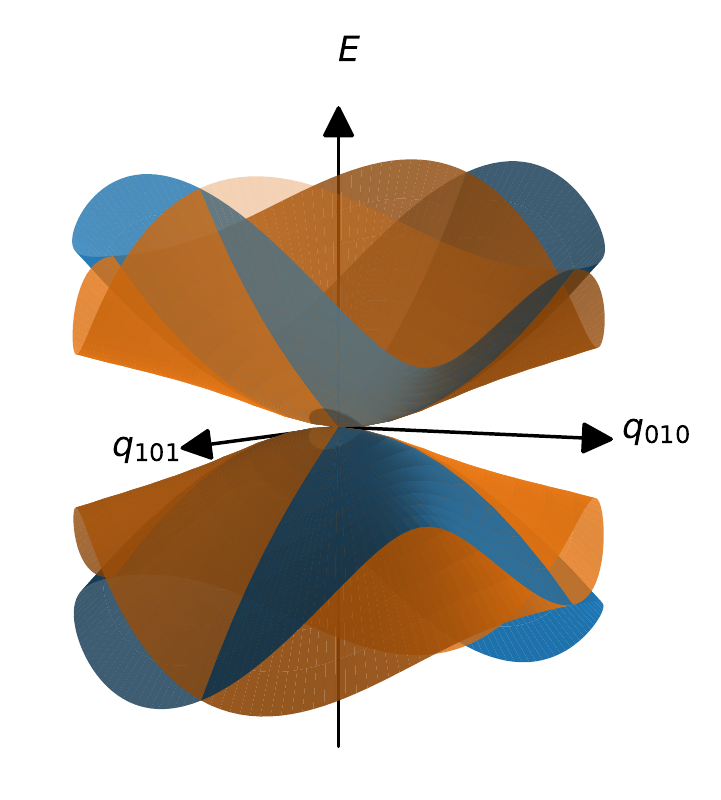}
\label{4fold_double_weyl_fig}
}\caption{
    Dispersion of symmetry enforced Weyl nodes with 
    Chern number $>1$. The coordinates $q$ in reciprocal space are given relative to the respective nodal point.
    (a) Double Weyl nodes are twofold degenerate,
    with linear dispersion along the fourfold rotation axis, $q_\parallel$,
    and quadratic dispersion perpendicular to it,  $q_\perp$, 
    $C_\mathrm{dW}=\pm2$.
    (b) A fourfold double Weyl point is a fourfold degeneracy with linear
    dispersion in all directions. Along
    high-symmetry lines a twofold degeneracy is enforced. 
    The colors mark the two symmetry-related Weyl points.
    The Chern number is the sum of these two Weyl points,
    $C_\mathrm{fW}=2C_\mathrm{Weyl}=\pm2$.
    (c)
    The fourfold double Weyl is built from two symmetry-related
    double Weyl points of equal chirality, the total Chern number
    is $C_\mathrm{fdW}=2 C_\mathrm{dW}=\pm4$.
    Twofold-degenerate lines in (b) and (c) are part of nodal
    planes (see Sec.\ref{weyl_planes}).
    The dispersion perpendicular to a fourfold Weyl line of
    Sec.~\ref{4fold_lines} is of the same type as in (b).
    }
\label{higher_weyl_fig}
\end{figure*}

Double Weyl points are twofold band crossings, where the
dispersion to lowest order is linear along one axis and quadratic
in directions perpendicular to it \cite{huang_hasan_double_Weyl_PNAS_16}, shown in
Fig.~\ref{higher_weyl_fig}(a). The Chern number of a manifold
enclosing such a node has absolute value $|\mathcal{C}|=2$.
They can be found among the crossings enforced by fourfold screw
rotations, where the ratio of rotation eigenvalues of the crossing
bands is $\frac{\alpha_p}{\alpha_{p^\prime}}=-1$. With the
parameterization of Eq.~\eqref{4fold_eigenvalues}, this is the
case for the pairs $(p,p^\prime)=(0,2)$ and (1,3).
For all of these, the direction of linear dispersion is along the
screw axis and the sign of their topological charge depends on
details of the Hamiltonian and is not determined from the order of
eigenvalues alone.

\subsubsection{Material example: Ba$_3$Sn$_2$}
\label{Ba3Sn2_chapter}

Here, we present an example material with Weyl nodal points, whose
existence is enforced by screw rotation symmetries. These Weyl
points are part of an hourglass or accordion dispersion, as
discussed in the previous sections. 

\begin{figure}
\centering
\includegraphics[width=.99\linewidth]{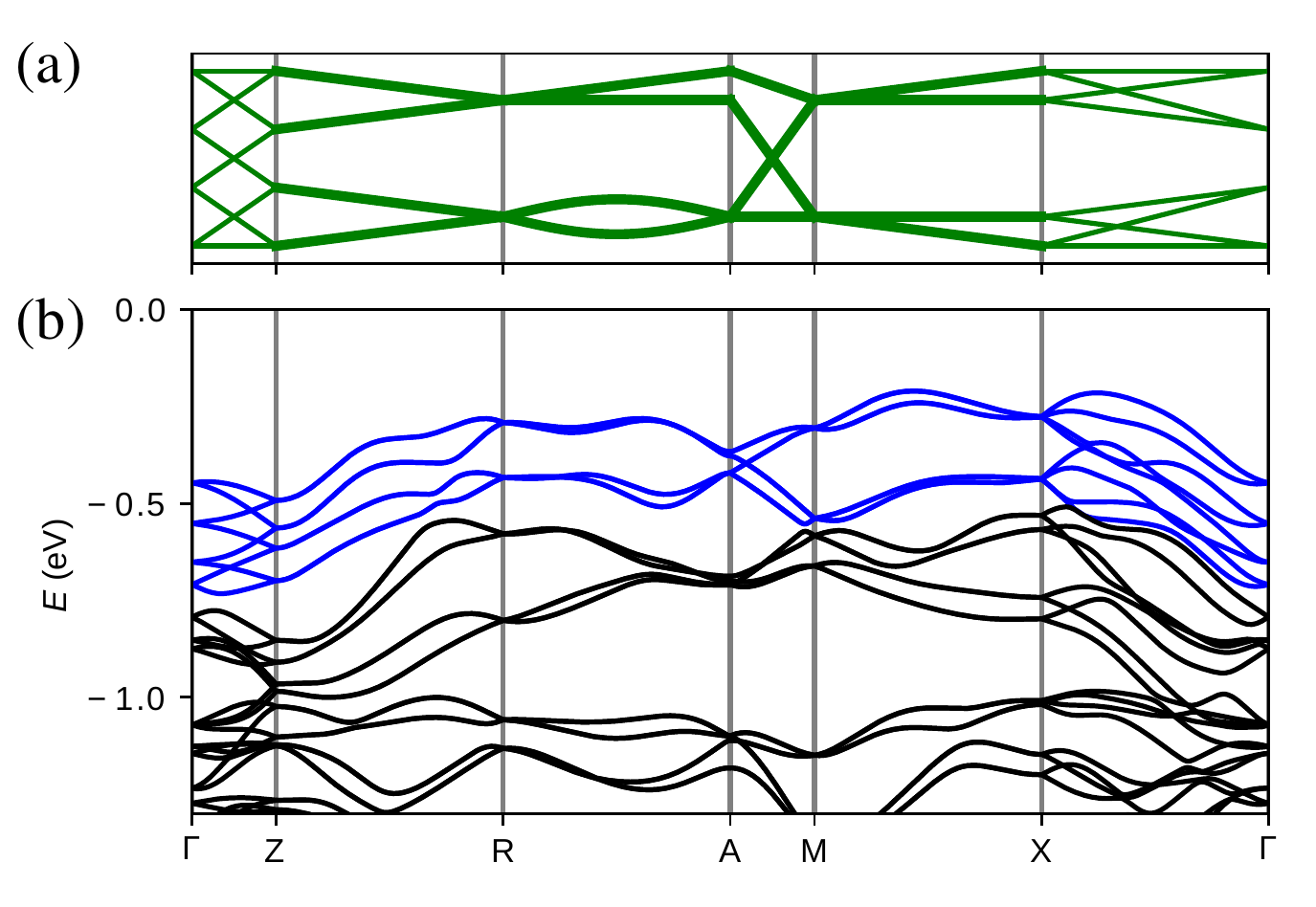} 
\caption{
        Band structures for SG 96:
        (a) Schematic band structure based on the 
        connectivity of irreducible representations. Thick lines
        indicate twofold degenerate bands within nodal planes.
        (b) Band
        structure of Ba$_3$Sn$_2$ calculated from first principles. 
        All Bands show the same connectivity as in the above sketch.
        The accordion states along $\Gamma$-Z can be seen most
        clearly in the uppermost eight bands (blue),
        whereas the hourglass states are only resolved in some
        bands lower in energy.
        }
\label{SG96_bands}
\end{figure}

The binary compound Ba$_3$Sn$_2$, which crystallizes in SG No.~96
($P4_32_12$)~\cite{Ba3Sn2_icsd_paper_12}, is an example of a
material with accordion states along the $\Gamma$-Z line. This
material is an electron-precise (insulating) Zintl phase, with a
narrow bandgap [Fig.~\ref{full_band_structures}(f)] and a metallic
luster. The [Sn$_2$]$^{6-}$ Zintl ion is unlikely to tolerate
heavy doping before decomposition, but light heterovalent doping
may be possible.
The accordion states are shown in Fig.~\ref{SG96_bands} alongside
a connectivity diagram for this SG and contains single and double
Weyl points as indicated in the accordion states in
Fig.~\ref{accordion_figure}(c).

Furthermore, this SG has twofold screw rotations with axes along
the 100 and 010 direction. As explained above, this enforces an
hourglass dispersion along $\Gamma$-X.
In the uppermost occupied bands, the movable crossings of the
hourglass dispersion are to close to X to be resolved, but they
are visible in the bands below.

An additional feature is the fourfold degenerate band crossing
along A-M. This feature is protected by symmetry eigenvalues, but
not mandated by band connectivity. It can in principle be removed
by exchanging the bands at A such that the fourfold degenerate
representation falls between the two twofold representations.
In either case, the fourfold degeneracy at half filling has to
carry a Chern number of $\mathcal{C}=\mp2$ to cancel the topological charges
of the double Weyl point with $\mathcal{C}=\pm2$ in the accordion state (see
Sec.~\ref{4fold_Weyl_chapter} about fourfold Weyl nodes below).
This SG and example material will also serve as an example
in the context of topological nodal planes in
Sec.\ref{weyl_planes}. 

\subsection{Fourfold double Weyl points}
\label{4fold_Weyl_chapter}
\label{Sec_NonSymWeyl_D}

A fourfold double Weyl point is a fourfold, point-like degeneracy with
linear dispersion, that splits into four non-degenerate bands in
all directions excluding the twofold-degenerate nodal planes,
which are covered in more detail in Sec.~\ref{weyl_planes}.
From a topological point of view these degeneracies are two
symmetry-related Weyl points of identical chirality on top of each
other, leading to a total Chern number of $\pm2$.
An exemplary dispersion is shown in Fig.~\ref{higher_weyl_fig}(b).
Such a feature is enforced by a combination of time reversal and a
spatial symmetry which enforces a different Kramers partner than
time reversal symmetry alone.

These conditions are met in the tetragonal crystal system by SGs 90 and 94 at M
and A and in SGs 92 and 96 at M. These TRIMs are on an axis left
invariant by the fourfold rotation, therefore we can label the
states with their eigenvalues,
\begin{equation}
4_{001}(\tfrac12,\tfrac12,0)\ket{p} 
    = \mathrm{e}^{\mathrm{i}(2p+1)\tfrac{\pi}{4}} \ket{p},
\end{equation}
indexed by $p=0,1,2,3$.
Time reversal symmetry pairs states with complex conjugate
eigenvalues, i.e. $(0,3)$ and $(1,2)$.

Consider now the additional antiunitary symmetry from combining
a twofold screw rotation with time reversal symmetry,
$2_{010}(\frac12,\frac12,0) \mathcal{T}$.
Its invariant points are restricted to the two planes $k_y=0$ and
$\pi$ where it squares to 1 and -1, respectively. In the latter
case Kramers pairs are enforced, i.e. every state in the plane is
twofold degenerate and we call it a nodal plane.

To find the eigenvalue of the Kramers partner, the commutation
relation of the fourfold rotation and the anti\-unitary symmetry
is needed. After some algebra, we find
\begin{align}
\label{eq:fourfoldWeylEW}
&4_{001}(\tfrac12,\tfrac12,0) \ \ 
    2_{010}(\tfrac12,\tfrac12,0) \mathcal{T}
\nonumber
\\
&=
2_{010}(\tfrac12,\tfrac12,0) \mathcal{T}
 \ \ t(1,0,0) \left(-4_{001}(\tfrac12,\tfrac12,0)\right)^3.
\end{align}
Therefore, the Kramers partner
$2_{010}(\frac12,\frac12,0)\mathcal{T}\ket{p}$
has the eigenvalue 
$\alpha_{p^\prime}=(\alpha_p^3)^* 
    = \exp( \mathrm{i}(2(p+2)+1)\frac{\pi}4 )$,
i.e. $p^\prime = p+2$ and thus the pairing is $(0,2)$ and
$(1,3)$. In combination with the pairings $(0,3)$ and $(1,2)$ from
time reversal symmetry, this establishes the fourfold degeneracy.
In other space groups, the translational part of these symmetries
might differ, but Eq.~\eqref{eq:fourfoldWeylEW} holds with a
modified translation. For all TRIMs, where the eigenvalue of the
translation is $-1$, the same pairing is found.
Alternatively, the presented results can be understood by
considering the eigenvalues $\pm \mathrm{i}$ of the twofold rotation
$2_{001}$. Time-reversal $\mathcal{T}$ pairs opposite signs at the
TRIMs, whereas $2_{010}(\frac12,\frac12,0) \mathcal{T}$
anticommutes with $2_{001}$ on the line M-A and therefore pairs
identical eigenvalues.
 
The fourfold degeneracy splits into non-degenerate bands, except
for the $k_x=\pi$ and $k_y=\pi$ nodal planes. The spectrum is
linear to lowest order in
$\mathbf{q}=\vk-\mathbf{K}_\mathrm{TRIM}$ and is made from two
Weyl cones, that are related by a $\frac{\pi}{2}$-rotation, see
Fig.~\ref{higher_weyl_fig}(b). A minimal, linearized Hamiltonian
for this band crossing has four bands and can always be brought
into block diagonal form, consisting of two
Weyl-Hamiltonians~\cite{PhysRevB.98.155145}. Furthermore,
$2_{010}(\frac12,\frac12,0)\mathcal{T}$ relates the coefficients
of the Kramers-Weyl points such that they have the same
chirality. Therefore, the total chirality of the fourfold Weyl
point for two occupied bands has to add up to $\pm2$. Hence, they
are also referred to as double-spin-1/2 Weyl points \cite{PhysRevB.98.155145}.

SG~92 (and 96) has another fourfold crossing at the R point, which
is not left invariant by fourfold rotations. Again, we use the
eigenvalues of the twofold rotation $2_{001}(0,0,\frac12)$, but
this time they are $\pm1$ because of the translational part.
Therefore, $2_{010}(\frac12,\frac12,\frac14)\mathcal{T}$ pairs
different eigenvalues on the line X-R, whereas $\mathcal{T}$ pairs
identical eigenvalues at R. From lattice and low-energy models we
find that the total chirality is $\mathcal{C} = \pm 2$ as for
previously discussed fourfold Weyl points (see
Appendix~\ref{Appendix_RinSG96}).

\subsection{Fourfold quadruple Weyl points}
\label{fourfold_quadruple_Weyl}
\label{Sec_NonSymWeyl_E}

In the previous chapter we have shown how a fourfold double Weyl
point can be made out of two symmetry related single Weyl points
on top of each other. The $\vk$- point A in SGs 92 and 96 also has
a fourfold degeneracy, but in contrast to M, it is made up from
two double instead of single Weyl points. Again, they have to
have the same Chern number of absolute value 2 and thus its total
Chern number is $|\mathcal{C}|=4$~\cite{higherorderDirac}.

The irreducible representations at A are two- and four-dimensional.
In principle the eigenvalue dependent Kramers theorem
applies as introduced in Sec.~\ref{eigval_Kramers}, i.e. the
combined operation $4_{001}(\frac12, \frac12, \frac14)
\mathcal{T}$ pairs only some of the rotation eigenstates. But the
regular Kramers theorem using $\mathcal{T}$ already pairs the same
$4_{001}(\frac12, \frac12, \frac14)$ eigenvalues 
and it is always
applicable. Thus the following argument utilizes the 
Kramers theorem based on time reversal symmetry alone. 

Time-reversal symmetry $\mathcal{T}$ pairs the bands as (0,2),
(1,1), and (3,3), when labeled with the parameter $p$
corresponding to the fourfold rotation eigenvalue
as defined in Eq.\eqref{4fold_eigenvalues}.
Analogously to the previous chapter, the combined operation
$2_{010}(\frac12, \frac12, \frac14) \mathcal{T}$ pairs bands into
nodal planes. Hereby, the non-symmorphic fourfold rotation in
SGs~92 and 96 modifies Eq.~\eqref{eq:fourfoldWeylEW}, where the
translation on the right hand side is now $t(1,0,-1)$ with
eigenvalue 1 at A. Together with the $k_z$ dependent eigenvalues
the resulting pairing due to $2_{010}(\frac12, \frac12, \frac14)
\mathcal{T}$ is (0,2) and (1,3). 

In conclusion, there is a twofold degeneracy (0,2) as well as a
fourfold degeneracy (1,1,3,3), where each state in the tuple is
orthogonal to the others by either different eigenvalues, i.e.
different $p$, or due to Kramers theorem. We interpret the
fourfold crossing as two copies of double Weyl points with
$|\mathcal{C}| = 2$. This chirality is determined from the
symmetry eigenvalues (1,3)~\cite{tsirkin_vanderbilt_PRB_17}.
Their chiralities must be equal, because the double Weyl points
are related by time-reversal symmetry.

To see this in detail, we provide a low-energy Hamiltonian of this
crossing. Since most terms linear in
$\mathbf{q}=\mathbf{k}-(\pi,\pi,\pi)$ vanish, terms up to
quadratic order need to be considered. Up to unitary
transformations, such a 4x4 Hamiltonian is restricted to a
block-diagonal form with the two double Weyl Hamiltonians
$H_\mathrm{dW}^\pm$ making up the blocks,
\begin{align}
H_\mathrm{dW}^{\pm} =& 
     \pm (v_x q_x q_y +\lambda_z q_z) \sigma_x
\nonumber  \\ &
    + v_y (q_y^2 - q_x^2) \sigma_y 
\label{Ham_4fold_quadruple}
\\ &
    \pm (v_z q_z +\lambda_x q_x q_y) \sigma_z.
\nonumber
\end{align}
See Appendix~\ref{appendix_4fold_quadruple} for a detailed
derivation. The Chern number is equal in both blocks and the
different signs only show in the dispersion when all three
components of $\mathbf{q}$ are non-zero. Only then, the bands
are non-degenerate. 
Otherwise the
eigenvalues of both Weyl points are identical, which ensures the
twofold degenerate planes.
Because of the splitting into non-degenerate bands and the non-zero
Chern number we do not use the name Dirac point, which has been used
in a prior report~\cite{higherorderDirac}.

\section{Non-symmorphic Dirac points}
\label{sec_theo_Dirac_points}

Enforced Dirac points pinned to TRIMs are a common and readily
accessible result of space group symmetries. For completeness
these crossings are listed in the second column of
Table~\ref{mTab2}. Here we want to give another perspective by
focusing on movable enforced fourfold crossings
in the presence of inversion or mirror symmetries.
 SGs 106, 130, 133, and 138 host either movable fourfold crossings
without chirality or, in the presence of inversion symmetry, movable Dirac points. The
following study of their symmetry eigenvalues highlights
similarities and differences. 
Finally, we discuss the pinned fourfold crossings enforced in SG~108 and SG~142 at the point P.

\subsection{SGs 106 and 133}
 
The symmetry operations in SGs 106 and 133 enforce a movable fourfold crossing on the
line M-A, which is part of a fourfold rotation axis. First we
focus on SG~133, a centrosymmetric group, for which the movable
crossing is a Dirac point. We explain its existence in terms of
its symmetry eigenvalues and their connectivity within the BZ. 

SG~133 contains the fourfold rotation $4_{001}(\frac12, 0,
\frac12)$ and a glide mirror symmetry $M_{010}(\frac12, 0, 0)$.
Along \mbox{M-A} the bands can be labeled by $p \in \{ 0, 1, 2, 3
\}$ referring to the eigenvalues of $4_{001}(\frac12, 0,
\frac12)$, namely $\exp(\mathrm{i} (2p+1)\frac{\pi}{4} ) \exp(\mathrm{i}
\frac{k_z}{2})$. An explicit calculation shows that the fourfold
rotation is related by $M_{010}(\frac{1}{2}, 0, 0)$ to its cube
\begin{align} &4_{001}(\tfrac{1}{2}, 0, \tfrac{1}{2})
M_{010}(\tfrac{1}{2}, 0, 0) \nonumber \\ &= -
M_{010}(\tfrac{1}{2}, 0, 0)\, (4_{001}(\tfrac{1}{2}, 0,
\tfrac{1}{2}))^3\, t(1,0,-1), \end{align} where the factor of $-1$
stems from the spin sector. On the line M-A $(\pi,\pi,k_z)$ this
relation pairs a band described by $p$ to the band with $p' = 3p
+ 1 \mod 4$. If we denote paired bands with $p$ and $p'$ as $(p,
p')$ then the only two possibilities on the M-A axis are
$(0,1)(2,3)$. 

The second part of the argument considers the Kramers theorem at
the endpoints of the line M-A. Time reversal forms band pairs
$(0,3)(1,2)$ at M, $k_z = 0$, and $(0,1)(2,3)$ at A, $k_z = \pi$.
Finally, one obtains the degeneracy at A by the combined operation
$\mathcal{T}' = \mathcal{T} M_{010}(\frac{1}{2}, 0, 0)$, which
fulfills $\mathcal{T}'^2=-1$ at M and A. It pairs at $k_z = 0$ the
eigenvalues into the tuples $(0,2)(1,3)$ and at $k_z = \pi$
$(0,0)(1,1)(2,2)(3,3)$. 

\begin{figure}[th]
\centering
\includegraphics[width=0.45\textwidth]{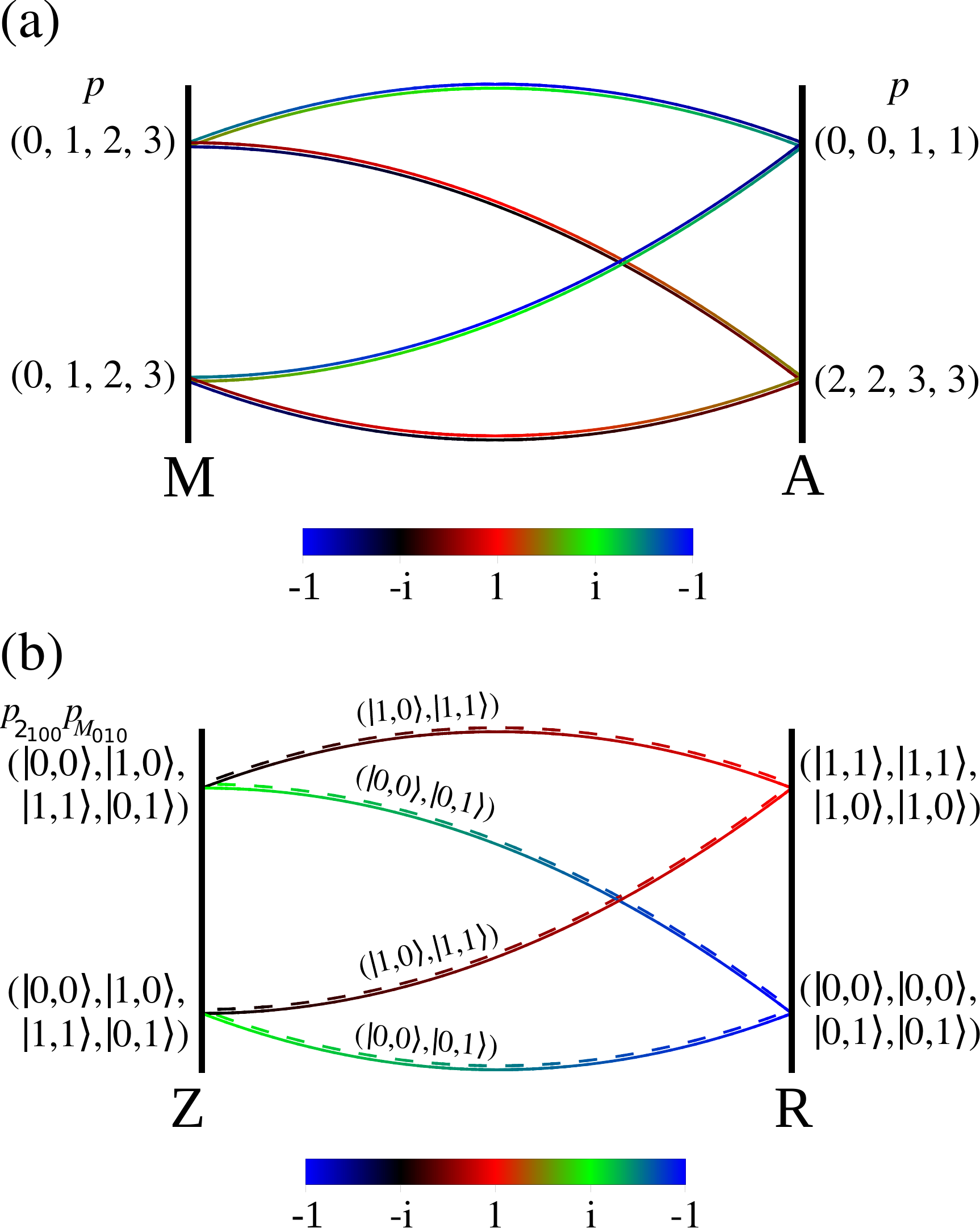}
\caption{ \label{movable_Dirac_points}
Movable fourfold crossings.
(a) SG 106 and 133: The symmetry eigenvalues $\alpha_{4_{001}}$ on
the M-A line are represented by the line color. For SG~133 the
M-A line hosts a Dirac point, whereas for SG~106 the crossing
corresponds to a fourfold point with zero chirality. 
(b) SG 130 and 138: The color corresponds to the symmetry
eigenvalues $\alpha_{2_{100}}$, whereas the line style (dashed)
distinguishes the values of $p_{M_{010}}$. A Dirac point is
enforced on the Z-R line.
} 
\end{figure}

We conclude that at M there is only one representation containing
all $4_{001}(\frac12, 0, \frac12)$ eigenstates once: $(0,1,2,3)$,
whereas at A two representations are possible, for which either
$(0,0,1,1)$ or $(2,2,3,3)$ correspond to distinct sets of
degenerate eigenstates. The twofold degenerate bands on the M-A
line with rotation eigenstates $(0,1)$ or $(2,3)$ must interpolate
between M and A. They exchange an odd number of times and thus
lead to an enforced fourfold crossing as presented in
Fig.~\ref{movable_Dirac_points}(a). 

So far we have not used the inversion symmetry at all. Its
presence together with time reversal specifies that the bands away
from M-A are twofold degenerate, thus the movable crossing is a
Dirac point. Consequentially, this argument can be applied
in the absence of inversion symmetry. Since SG~106 contains
nearly identical fourfold rotation and mirror symmetries as
SG~133, one may repeat the arguments above and find again the
band structure on M-A as shown in
Fig.~\ref{movable_Dirac_points}(a). Yet, SG~106 is special,
because there is no $\mathcal{PT}$ symmetry present and therefore
the bands split into four non-degenerate ones away from the
movable crossing. The movable crossing of SG~106 is a fourfold
crossing on a mirror plane and thus has a vanishing Chern number.

The minimal number of connected bands for SG 106 and 133 is eight
(or four when excluding spin). Thereby, only the movable crossings
on the M-A line are necessary to connect two sets of four bands.
One concludes that SG 106 and 133 can lead to semimetals with
movable fourfold crossings at a filling of $4 + 8\mathbb{N}$
electrons per unit cell. They can be thought of as two
superimposed Weyl semimetals with vanishing total chirality. As
such they show two sets of surface states which can hybridize with
each other unlike the Fermi arcs for a Weyl semimetal on its own.
Thus, van-Hove singularities appear for the surface spectrum, see
Appendix~\ref{Appendix_106_133}.

\subsection{SGs 130}
\label{maintext_130}

In SG 130 movable Dirac points appear and they require the
presence of inversion symmetry unlike the fourfold crossings of
SGs 106 and 133. They appear on the line Z-R, a twofold rotation
axis. A movable Dirac point cannot be understood by a single
twofold symmetry alone. It would require two different ways to
pair eigenvalues at the TRIMs, i.e. endpoints of the line and a
third point along the line itself, which is impossible with only two
eigenvalues. 

Therefore we must consider two symmetries at once. Indeed, SG 130
contains the off-centered screw rotation $2_{100}(\frac{1}{2}, 0,
\frac{1}{2})$ and the mirror symmetry $M_{010}(0, \frac{1}{2},
\frac{1}{2})$. On the BZ path Z-R, e.g. $(k_x, 0, \pi )$, these
symmetries commute and we label the bands and states by
$\ket{p_{2_{100}}, p_{M_{010}} }$, where the eigenvalues
$\alpha_{2_{100}} = \exp(\mathrm{i} (2p_{2_{100}}+1) \frac{\pi}{2} ) \exp(\mathrm{i}
\frac{k_x}{2})$ and $ \alpha_{M_{010}} = \exp(\mathrm{i} (2p_{M_{010}}+1)
\frac{\pi}{2} ) \exp(\mathrm{i} \frac{k_z}{2})$ are referred to by their
respective value of $p \in \{0, 1\}$. As before round brackets are
used to denote the pairing of irreducible representations denoted
by the different $p$. 

Inversion with time-reversal symmetry, $\mathcal{PT}$, pairs
identical eigenvalues for off-centered symmetries on certain
planes or lines~\cite{yang_furusaki_PRB_2017}. On Z-R, identical
$2_{100}(\frac{1}{2}, 0, \frac{1}{2})$ eigenvalues are paired, but
different eigenvalues for $M_{010}(0, \frac{1}{2}, \frac{1}{2})$.
Therefore, the pairing there is
$(\ket{0,0},\ket{0,1})$ and $(\ket{1,0},\ket{1,1})$. At the TRIMs
time reversal also pairs the symmetry eigenvalues. Whereas the
phase of $\alpha_{2_{100}}$ depends on $k_x$, $\alpha_{M_{010}}$
is real at Z and R, resulting in the pairing
$(\ket{0,0},\ket{1,0})$ and $(\ket{0,1},\ket{1,1})$ at $k_x = 0$
and $(\ket{0,0},\ket{0,0})$, $(\ket{0,1},\ket{0,1})$,
$(\ket{1,0},\ket{1,0})$, and $(\ket{1,1},\ket{1,1})$ at $k_x =
\pi$. Finally, inversion $\mathcal{P}$ anticommutes (commutes)
with $2_{100}(\frac{1}{2}, 0, \frac{1}{2})$ at Z (at R) and
anticommutes with $M_{010}(0, \frac{1}{2}, \frac{1}{2})$ at both Z
and R. States related by inversion have the same energy but are
not necessarily orthogonal unless at least one of their
eigenvalues differs.

With this preparation we can infer all degenerate states at the
TRIMs Z and R. We start from $\ket{0,0}$ and apply to it
$\mathcal{T}$, $\mathcal{P}$, or $\mathcal{PT}$ and determine
which eigenvalue tuple the consecutive results have. The resulting
states after the application of above symmetries are at Z
$(\ket{0,0}, \ket{1,0}, \ket{1,1},\ket{0,1})$, whereas at R one
obtains $(\ket{0,0}, \ket{0,0}', \ket{0,1},\ket{0,1}')$ and
$(\ket{1,1}, \ket{1,1}', \ket{1,0}, \ket{1,0}')$. Here, states
orthogonal by Kramers theorem are primed for clarity. To
interpolate between the fourfold degeneracies at Z and R a movable
Dirac point must appear, see Fig.~\ref{movable_Dirac_points}(b).

Since the mirror symmetry $M_{001}(\frac{1}{2}, \frac{1}{2}, 0)$
is not off-centered, there can be no Dirac nodal line on $k_z =
\pi$. Therefore the identified crossing is not part of a fourfold
nodal line, which concludes our derivation of the movable Dirac
points on Z-R in SG 130.

The movable Dirac point of SG~130 has been considered in
\cite{2016_wieder_kane_prl_doubleDirac} alongside the eightfold
degeneracy at the A point appearing in SG~130 as well as 135
\cite{Bradlynaaf5037}. These double Dirac points at the A point
are linear band crossings and lie at the boundary between
topological insulating phases, which can be reached by breaking
spatial symmetries for example with strain
\cite{2016_wieder_kane_prl_doubleDirac}. Since the eightfold point
is at the boundary one of the achievable phases must be
topological but the details depend on the values of the mass terms
introduced by the symmetry breaking perturbations.

\subsubsection{Material example: Tl$_4$PbSe$_3$}
\label{Tl4PbSe3_chapter}

Tl$_4$PbSe$_3$~\cite{malakhovska2009crystal} is an air-stable
narrow-gap semiconductor which crystallizes in SG 130 (along with
several other thallium tetrel chalcogenides and thallium
chalcohalides), and has been studied recently for possible
thermoelectric applications~\cite{Reshak2015}. Its band structure
is shown in Fig.~\ref{Dirac_points_AuBr_Tl4PbSe3}(b) with the
movable Dirac points close to the R point and the full band
structure is given in Fig.~\ref{full_band_structures}(e).

\subsection{SG 138}
\label{maintext_138}

SG~138 
hosts movable Dirac points on the line Z-R, which can be shown by
identical arguments as in SG 130, see Sec.~\ref{maintext_130}.
As the Dirac points have been discussed above, we focus here only
on the distinction between both groups. Unlike SG~130, SG~138
exhibits a nontrivial topology described by a $\mathbb{Z}_2$
invariant, which is enforced by the connectivity of bands. Whereas
the nontrivial value of the invariant is fixed for any band order,
the robustness is found to be dependent on details of the system.

Since materials in SG~138 are Dirac semimetals at half filling (gapless systems with
Dirac points on the line Z-R), the invariants for
topological insulators do not apply. Nevertheless, we can
consider two-dimensional gapped subsystems for which the existence
of surface states can be inferred from changes in the
time-reversal polarization~$\pi_a$, where we use the results and
notation of \cite{2007_FuKane_Z2Inversion}.

To determine the time-reversal polarization $\pi_a$ we can employ
the quantities $\delta_{\Gamma_i} = \prod_{m=1}^N
\xi_{2m}(\Gamma_i)$, where $\xi_{2m}(\Gamma_i)$ denotes the
eigenvalue of inversion $\mathcal{P}$ for the band $2m$ at the
TRIM $\Gamma_i \in \{ \Gamma, X, Z, M, R, A \}$. We consider the
number of occupied bands $N$ to be $4 + 8\mathbb{N}$ for SG~138,
i.e. the band index of the movable Dirac points. From the
quantities $\delta_{\Gamma_i}$ one can obtain the polarization
$\pi_a = \delta_{a1} \delta_{a2}$ \cite{2007_FuKane_Z2Inversion}.
The TRIMs labeled by $a1$ and $a2$ will fall on top of each other,
once we terminate the system in real space. To evaluate the
time-reversal polarization $\pi_a$ for SG~138 in a meaningful way
we consider subsystems with a band gap, e.g. the plane containing
the TRIMs XMAR. Edge states of said planes appear as lines in the
full surface BZ.
To be more specific, the plane XMAR projects to the path
$\bar{\text{X}}$-$\bar{\text{M}}$ in the two-dimensional surface
BZ for a (001) termination of the full three dimensional system
and thus its edge states appear as lines on this path.

In the following we give the inversion eigenvalues at the TRIMs
for SG~138, which can be determined to a large extent from the
commutation relations between the symmetries. Due to
time-reversal symmetry $\mathcal{T}$ each inversion eigenvalue
always appears twice, connected as a Kramers pair. This is
already considered by taking only even band indices $2m$ for
$\xi_{2m}(\Gamma_i)$ in the expression for $\delta_{\Gamma_i}$
\cite{2007_FuKane_Z2Inversion}. Furthermore, for SG 138 all TRIMs
except $\Gamma$ are fourfold degenerate. The non-symmorphic
mirror symmetries relate opposite inversion eigenvalues at the
points X, R, M and Z. At A, an eigenstate can be labeled by
inversion and mirror eigenvalues simultaneously. The application of
$\mathcal{T}$ and the twofold screw rotation yields that four
bands with the same inversion eigenvalue are degenerate at A. We
can thus explicitly give the values of $\delta_{\Gamma_i}$ for the
TRIMs $\delta_{\text{X}} = \delta_{\text{R}} = \delta_{\text{M}} =
\delta_{\text{Z}} = \xi_{2}(\text{X}) \xi_{4}(\text{X}) = -1 $ and
$\delta_{\text{A}}= \xi_{2}(\text{A}) \xi_{4}(\text{A}) = +1 $,
whereas $\delta_{\Gamma} = \pm 1$ is not determined from symmetry
alone. Note that $\delta_{\text{A}} = +1 $ is independent of the
band order, i.e. the value of $\xi_{2}(\text{A}) =
\xi_{4}(\text{A}) = \pm 1$.

Below we discuss the surface states for a slab with (001)
termination, which we compare to the explicit calculation for a
generic model, see Appendix~\ref{Appendix_130_138}. For the plane
XMAR there are two time-reversal polarizations $\pi_1 =
\delta_{\text{X}} \delta_{\text{R}} = 1$ and $\pi_2 =
\delta_{\text{M}} \delta_{\text{A}} = -1$. This difference
between $\pi_1$ and $\pi_2$ leads to surface states on the line
$\bar{\text{X}}$-$\bar{\text{M}}$, which is confirmed by the
surface spectrum, see Fig.~\ref{Surface_SG130_138}(c). The number
of surface states depends on the details of the system, but by the
above argument SG~138 ensures that the surface states cross the
gap connecting valence and conduction bands
an odd number of times. 
Computing the product $\pi_1 \pi_2 = (-1)^{\nu_1}$ analogously to
the weak invariant \cite{2007_FuKane_Z2Inversion} yields 
$\nu_1 = -1$.

To support this approach we compare SG~138 to SG~130. Although
the mirror operations of SG 130 and 138 are identical and the
enforced inversion eigenvalues are in principle the same, no
topological Dirac surface state appears for SG 130 on the line
$\bar{\text{X}}$-$\bar{\text{M}}$. The difference between SG 130
and 138 is the band touching at the point A. Due to this
eightfold crossing in SG~130 the plane XMAR is gapless and the
same number of positive and negative inversion eigenvalues are
present at A. Also note that the movable Dirac points, which
appear for both space groups, are trivial because
$(\mathcal{PT})^2=
-1$~\cite{WeylDiracCharges_furusaki,zhao_PT_PRL_16}. Both remarks
support the interpretation that the time-reversal polarizations
$\pi_a$ capture the topology (see the further analysis in
Appendix~\ref{Appendix_130_138}).

A robust topological phase does not lose its surface states or
topological invariant for any perturbation that preserves the bulk
band gap and the protecting symmetry. This distinguishes strong
from weak topological insulators \cite{2007_FuKane_Z2Inversion}.
To apply this classification to SG~138 we consider arbitrary small
perturbations that break the symmetries except inversion. Thus,
the order of bands at the TRIMs does not change and all
$\delta_{\Gamma_i}$ will be preserved. Once the Dirac points on
the lines Z-R are slightly gapped, we can calculate the strong
invariant $\nu$ by $(-1)^\nu = \prod_{\Gamma_i} \delta_{\Gamma_i}
= \delta_\Gamma$. Hence, the band order at $\Gamma$ by itself
determines the stability of the topological phase.

To summarize the results for SG~138, we find besides four movable
Dirac points that the spatial symmetries enforce a nontrivial weak
topological invariant. We find that the overall robustness of the
topological phase depends on the order of bands at the point
$\Gamma$.

\subsubsection{Material example: AuBr}
\label{sec_AuBr}

\begin{figure}[t!]
\includegraphics[width = 0.89\columnwidth]{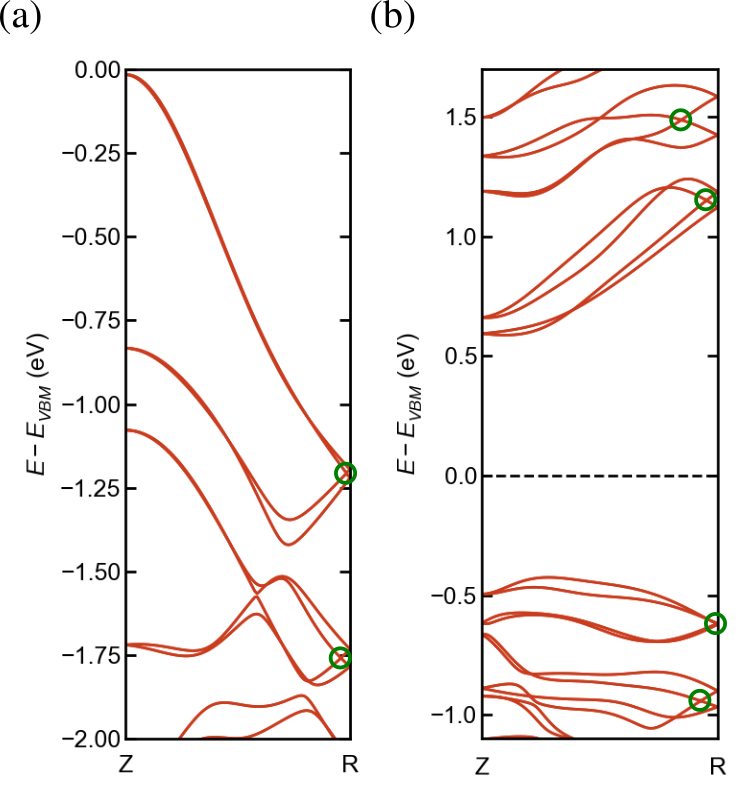}
\caption{ \label{Dirac_points_AuBr_Tl4PbSe3}
Movable Dirac points in AuBr and Tl$_4$PbSe$_3$. Bands along the
Z-R line for (a) AuBr (SG~138) and (b) Tl$_4$PbSe$_3$ (SG~130).
Green circles highlight the enforced Dirac points, see
Sec.~\ref{maintext_130}.
}
\end{figure}

The primitive tetragonal polymorph of AuBr, crystallizing in SG 138~\cite{AuBr_struct_ref_78}, is an example of a material
with Dirac points along the Z-R line, as discussed in
Sec.~\ref{sec_theo_Dirac_points}. It consists of layers of polymeric 
Au$-$Br zig-zag chains, and can be grown as yellow-brown crystals 
by vapor transport. In Fig.~\ref{Dirac_points_AuBr_Tl4PbSe3}(a) we show the
first-principles band dispersions of AuBr along the Z-R direction.
All bands are twofold Kramers degenerate due to time-reversal and
inversion symmetry. Along Z-R we observe groups of four connected
bands that form an odd number of Dirac crossings. AuBr is
insulating with a band gap of about 2~eV (see
Fig.~\ref{full_band_structures}(d)). This large gap makes it
difficult to experimentally measure the band structure using photoemission or scanning tunneling probes.

\subsection{SG 108}

SG 108 contains a fourfold crossing that is not pinned to a TRIM
but to the point P~\cite{MultipleDiracFermions}. Due to the
presence of mirror symmetries it does not carry a chirality as
topological charge. Yet, since a fourfold crossing away from
TRIMs is a unique feature of some tetragonal space groups, we will discuss its
origin.

The fourfold crossing in SG 108 can be understood with the
eigenvalues $\exp(\mathrm{i} (1 + 2p_{2_{001} } )\frac{\pi}{2})$ of
$2_{001}(0, 0, 0)$. At P the Kramers theorem is applicable to the
combined symmetry $\mathcal{T} M_{010}(0, 0, \frac{1}{2})$, which
pairs identical $p_{2_{001}}$ values, $(0,0)$ and $(1,1)$.
Furthermore the point P is left invariant by $M_{1\bar{1}0}(0, 0,
\frac{1}{2})$, which relates different $p_{2_{001}}$ yielding the
pair $(0,1)$. Therefore four bands are degenerate at P. The
fourfold crossing splits into nondegenerate bands away from P
except along the lines P-X and P-N.

\subsection{SG 142}
\label{maintext_142}

SG 142 is the only tetragonal SG with a Dirac crossing at the
point P. A different argument as for SG~108, which contains a
fourfold crossing at P, must be used because its corresponding
twofold rotation $2_{001}(\frac{1}{2}, 0, \frac{1}{2})$ is
off-centered, time-reversal with mirror or inversion symmetries
pair already identical eigenvalues. 

SG 142 is similar to SG 110, cf. Sec.~\ref{section_on_SG110} for
more details, because also in SG~142 three non-symmorphic
operations $2_{001}(\frac12, 0,\frac12), M_{110}(\frac34,
\frac14,\frac14),$ and $M_{1\bar{1}0}(\frac34, \frac34,\frac34)$
commute on the line X-P. The product of the corresponding
eigenvalues must fulfill $\alpha_{2_{001}}
\alpha_{M_{110}}\alpha_{M_{1\bar{1}0}} = \exp(\mathrm{i} 3 k_z/2)$, leading
to four possible eigenvalue configurations $\ket{0,0,1},
\ket{0,1,0}, \ket{1,0,0}, \ket{1,1,1}$ along the X-P line, where
the parameters $p$ defined analogously to before label the states
$\ket{p_{2_{001}},p_{M_{110}},p_{M_{1\bar{1}0}}}$. Time reversal
with inversion, $\mathcal{PT}$, pairs them in groups of two,
$(\ket{0,0,1}, \ket{0,1,0})$ and $(\ket{1,0,0}, \ket{1,1,1})$. 

There are irreducible representations of dimensions 2 and 4 at the
P point, due to an eigenvalue dependent Kramers theorem, see in
Sec.~\ref{eigval_Kramers}. The combination $\mathcal{T}
4_{001}(\frac{1}{4} , \frac{3}{4} , \frac{1}{4})$ leads to
orthogonal new states at P, if applied to a state with
$p_{2_{001}} = 1$. In this process the values of $p_{2_{001}}$
remain unchanged and the twofold degeneracy $(\ket{1,0,0},
\ket{1,1,1})$ is doubled and results in the Dirac point at P.

\section{Twofold Weyl lines}
\label{sec_two_fold_weyl_lines}
\label{weyl_lines}

Nodal lines pinned to rotation axes or equivalently the
intersection of two mirror planes are a common feature. These
pinned degeneracies can be understood from the structure of the
little group and are
tabulated~\cite{miller_love_irreps,elcoro_aroyo_JAC_17}. The
identification of movable nodal lines on the other hand poses more
intricacies, which we discuss in the following sections. Here, we
begin by considering the concept of glide mirror symmetries in the
presence of time reversal, before we move to more complex nodal
lines. In the simplest case so-called hourglass nodal lines are
found, which were classified on the grounds of compatibility
relations~\cite{hourglass_230SGclassification_2020_Wu}. Our
approach uses the explicit symmetry eigenvalues instead of
representations and facilitates the derivation of other movable
nodal lines between TRIMs, Sec.~\ref{sec_movableTRIMlines}.

In the absence of inversion any glide mirror symmetry naturally
leads to twofold nodal lines. 
Applying a glide mirror symmetry with translation $t_\perp$ within
the mirror plane twice results in a lattice translation by
$2t_\perp$. Consequently, the eigenvalues need to be the square
roots of this lattice translation and are $k$-dependent,
\begin{equation}
\alpha_p(\vk) = \exp(\mathrm{i} (2p+1)\frac{\pi}{2}) \exp(\mathrm{i} \mathbf{k}\cdot t_\perp),
\label{mirror_eigvals}
\end{equation}
where $t_\perp$ is the projection of the translational part to the
mirror plane and we use $p\in\{0,1\}$ as label.
When traversing the BZ, these eigenvalues exchange at least once.

In the presence of other symmetries, degeneracies between states
with different eigenvalues or, through Kramers theorem, identical
eigenvalues can be enforced.
For example, at TRIMs in the BZ
where $\vk\cdot t_\perp = 0 \mod 2\pi$
the values of $p$ are paired as $(0,1)$, while for 
$\vk \cdot t_\perp=\pi \mod 2\pi$
they are paired to be $(0,0), (1,1)$. On any path within the
mirror plane connecting two such points with different eigenvalue
pairing the standard argument for hourglass dispersions can be
invoked and thus each such path must contain an odd number of band
crossings. These enforced crossings form movable Weyl nodal lines,
away from high-symmetry points. 

We specify movable nodal lines in column 3 of Table~\ref{mTab1} by
grouping points according to their eigenvalue pairing. The left
hand side of a tuple always contains points with pairs of
alternating signs, whereas identical eigenvalues are paired at all
points on the right hand side. The nodal line then has to be in
between these sets and in the plane containing all points of the
tuple.
For example in SG~100 we denote with
($\Gamma$-Z ; X, R)(4) that at any point on the line $\Gamma$-Z
the value of $p$ is paired as $(0,1)$, whereas at X and R the
pairs $(0,0)$ and $(1,1)$ are formed. Hence, four bands are connected
on any path between $\Gamma$-Z to X or R each forming an hourglass
structure. Such a path exhibits a band connectivity as
depicted in Fig.~\ref{accordion_figure}(a) for the movable Weyl
points on a high-symmetry line.
Within the tetragonal space groups fourfold crossings do appear in
planes with movable nodal lines. It turns out that
on a path containing such a fourfold crossing, no additional
crossing is enforced.
Applied to SG~102 with the nodal line denoted by
\mbox{($\Gamma$-$Z$-R ; X)(4)} this means that 
a path from Z to X does not necessarily cross a nodal line.
We note this by writing a
cursive~$Z$ instead of Z on the line of identical paired
eigenvalues $\Gamma$-$Z$-R. One can confirm from the irreducible
representations that there are no fourfold crossings with only one
value of the mirror eigenvalue in the tetragonal SGs
\cite{elcoro_aroyo_JAC_17}.

In conclusion, we find that SGs~100, 102, 104, 106, 108, 109, 110,
117, 118, 120, and 122 enforce on at least one mirror plane a
movable twofold nodal line.

Using the previous discussion, it can be understood that
intersecting mirror planes may give rise to nodal chain metals
\cite{bzdusek_soluyanov_nature_16}. We will briefly discuss the
relevance to tetragonal space groups and give a material example.
Then, we focus on SG~110, although it does not contain a nodal
chain extended in reciprocal space, it exhibits intersecting nodal
lines and a band connectivity allowing in principle for a simple Fermi surface.
But before we move to these complex configurations of mirror
planes, we discuss a type of nodal lines which is easily missed
despite its conceptual simplicity. These nodal lines are 
also movable, except that they are pinned to certain TRIMs.

\subsection{Almost movable nodal lines}
\label{sec_movableTRIMlines}

In this section we will discuss twofold almost movable nodal
lines, which are enforced features pinned at a high-symmetry point
but which may be moved freely everywhere else in the mirror plane.
Before discussing the affected SG, we derive the existence of this
type of nodal line in body-centered SGs.

To begin, suppose there is a single (glide) mirror symmetry,
with eigenvalues $\pm \mathrm{i}$ at a TRIM on the mirror plane. Time
reversal pairs the symmetry eigenvalues $+\mathrm{i}$ and $-\mathrm{i}$ yielding
twofold degenerate bands at this TRIM. 
Let $\mathbf{q}$ be the coordinate of a point in the mirror plane
relative to the TRIM.
Time-reversal symmetry relates $\mathbf{q}$ to $-\mathbf{q}$,
while exchanging the mirror eigenvalues of opposite phase, $+$
with $-$. If one chooses a path leading from $\mathbf{q}$ to
$-\mathbf{q}$ within the mirror plane and without crossing the
TRIM, we know that the mirror eigenvalues must change. As the
eigenvalues may only change if bands cross, we can conclude from
the generality of the path that the twofold degeneracy at the
TRIM is part of a nodal line. We refer to these nodal lines as
almost movable nodal lines, because they are unpinned everywhere
except for a finite set of points, i.e. the TRIMs which enforce
them. 

Almost movable nodal lines are unlike pinned and movable nodal
lines. 
Although the above argument holds for all TRIMs in mirror planes
with eigenvalue pairing of opposite signs, i.e. $p=0$ and $1$ in
Eq.~\eqref{mirror_eigvals}, often there are further symmetries
present which pin the nodal line to a rotation axis.
Pinned nodal lines arrange for the exchange of eigenvalues in the
same way, but they are completely fixed to a straight line by the
anticommutation of symmetries or by Kramers theorem applied to the
combination of a glide mirror symmetry with time reversal. Their
appearance can therefore be understood from the group structure of
the involved symmetries alone and they can be found in tabulated
irreducible representations of the corresponding little
groups~\cite{elcoro_aroyo_JAC_17}. 
The other extreme is movable nodal lines, which are
only constrained in the sense that they mediate an exchange of
mirror eigenvalues within the mirror plane, as we have discussed
at the beginning of Section~\ref{weyl_lines}. This differs
from the almost movable nodal lines in that the latter are not
responsible for exchanging symmetry eigenvalues between different
TRIMs. Studying compatibility relations along high-symmetry paths
would not explain almost movable nodal lines
\cite{elcoro_aroyo_JAC_17, hourglass_230SGclassification_2020_Wu}.
Thus they are not just movable nodal lines pinned at a finite
number of points in the BZ, but they exist because nodal lines
have no endpoints.

\begin{figure}[th]
\centering
\includegraphics[width=0.45\textwidth]{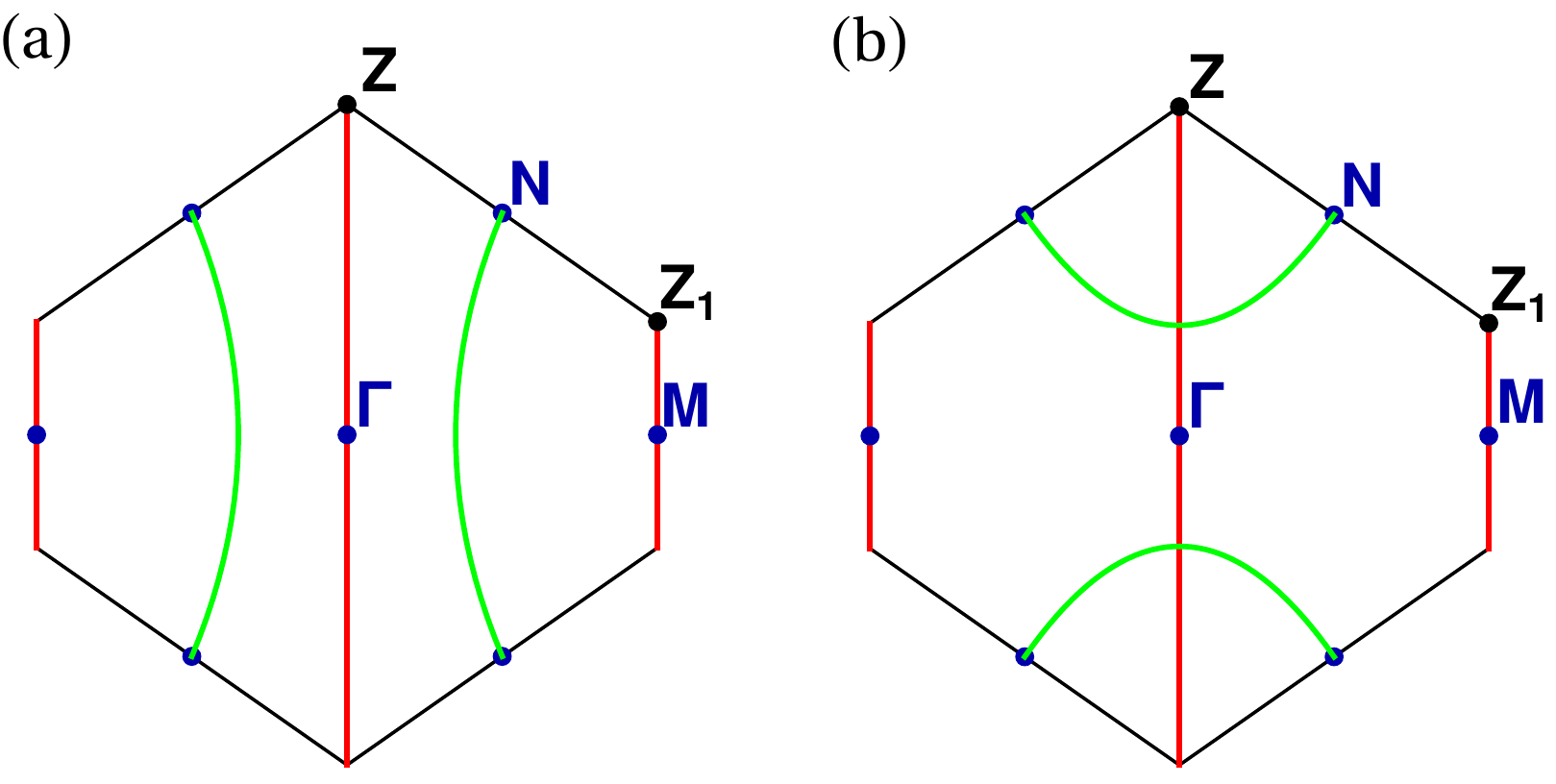}
\caption{\label{Fig_AlmostMovableSketch}
Almost movable nodal lines (green) and pinned nodal lines (red) 
in mirror planes of SGs~107, 109, and 119.
The connectivity of almost movable nodal lines is not determined
by symmetry alone.
(a) Nodal line connecting distinct N points. 
(b) Same as (a), but intersecting the pinned nodal line.
}
\end{figure}

A SG with almost movable lines needs to have a TRIM in a mirror
plane that is not part of a rotation axis, i.e. the little group
of the TRIM contains only the reflection. In the tetragonal crystal system, this
is only possible at N in body centered space groups with
crystallographic point group $4mm$ or $\bar{4}2m$. Of those, only
SGs 107, 109 and 119 pair $\pm \mathrm{i}$ at N. 
For these SGs we illustrate the possible connectivity of almost
movable lines qualitatively in Fig.~\ref{Fig_AlmostMovableSketch}.
Finally, there is a fourth case in the tetragonal SGs of an almost
movable line found at P in SG 110. P is not invariant under time
reversal alone, but under the antiunitary symmetry
$M_{010}(0,0,\frac12)\mathcal{T}$.
It creates a pair of
eigenvalues with opposite signs, $(0,1)$ for $p$ in the definition
of the 110-mirror eigenvalues, c.f. Eq.~\eqref{mirror_eigvals},
and creates an almost movable nodal line in the corresponding
invariant plane. A detailed discussion and derivation can be found
in Appendix~\ref{App_110P}.

The almost movable lines are indicated in the same column as the
movable lines, i.e. the third column in Table~\ref{mTab1}.
We use the same notation as for movable nodal lines, e.g. for
SG~107 we write (N, $\Gamma$-Z, M-Z$_1$ ; $-$)(2). This expression
is understood as before. Bands at the points and lines N,
$\Gamma$-Z, and M-Z$_1$ are twofold degenerate and comprise both
mirror eigenvalues. But here the second entry, where we would
denote points in the mirror plane of identical pairing remains
empty "$-$". Note that $\Gamma$-Z and M-Z$_1$ exhibit pinned nodal
lines, whereas the point N 
is crossed by an almost movable nodal line. The nodal line passing
through N may cross but cannot end at the pinned lines like
$\Gamma$-Z, cf. Appendix~\ref{Appendix_119} 
with SG~119 as an example. SG~110 is special in that its almost
movable nodal lines do not pass a TRIM but rather the point P, and they do
not exist for all bands (see Sec.~\ref{section_on_SG110}).

\subsection{Nodal chain metals}
\label{sec_nodal_chain_metals}

Nodal chain materials are characterized by enforced, movable nodal
lines that touch on the intersections of their mirror planes
~\cite{bzdusek_soluyanov_nature_16}. Several tetragonal SGs i.e.
102, 104, 109, 118, and 122, fall into this category. Nodal chain
metals appear, when there are two intersecting mirror planes,
whose eigenvalues are each paired differently for the two
high-symmetry points on the intersecting line. A detailed
discussion can be found in~\cite{bzdusek_soluyanov_nature_16}.
Note that systems with nodal chains always have other Fermi
surfaces besides the nodal chain.

\subsubsection{Material example: Ba$_5$In$_4$Bi$_5$}
\label{sec_Ba5In4Bi5}

\begin{figure}
\centering
\includegraphics[width=.9\linewidth]{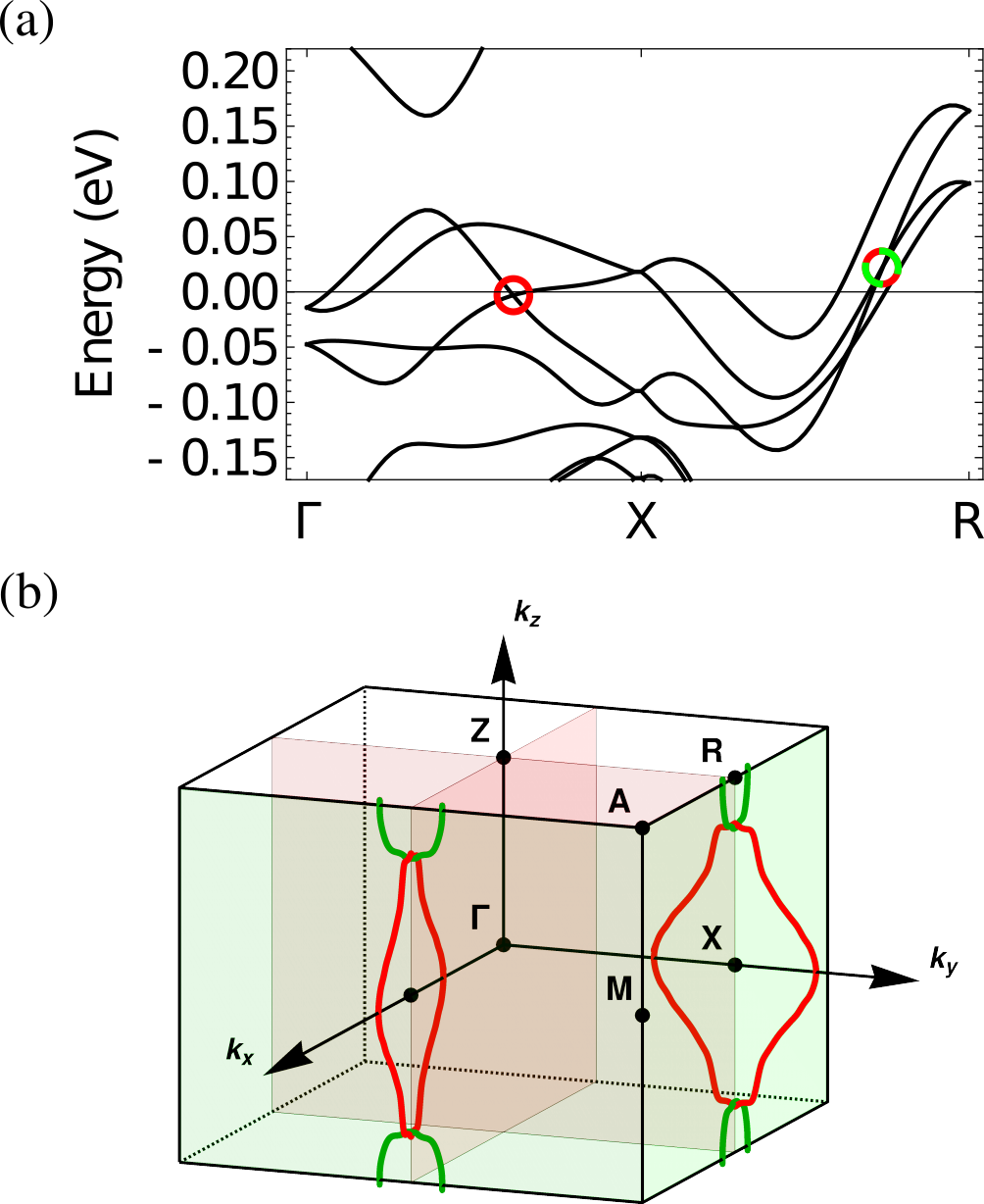}
\caption{ \label{Ba5In4Bi5_figure}
Nodal chain metal Ba$_5$In$_4$Bi$_5$ in SG 104. 
(a) Electron bands close to the Fermi energy exhibiting nodal
lines. Circles mark points on the nodal chain and their color
refers to the color code in (b). On the line X-R the red and green
sections of the nodal chain cross. 
(b) Nodal chain formed by the bands in (a) represented as lines in
the Brillouin zone. Red: Nodal lines in a mirror plane defined by
$k_x = 0$ or $k_y = 0$. Green: Nodal lines in a
mirror plane defined by $k_x = \pi$ or $k_y = \pi$.
}
\end{figure}

Ba$_5$In$_4$Bi$_5$, which crystallizes in SG 104, is an
electron-deficient polar intermetallic formed by the reduction of
In and Bi with Ba~\cite{ponou_Ba5In4Bi5}. Its DFT band structure
is presented in Fig.~\ref{Ba5In4Bi5_figure}
and~\ref{full_band_structures}(c). The nodal chain of
Ba$_5$In$_4$Bi$_5$ is in close proximity to the Fermi energy
presenting an experimentally accessible platform for the study of
nodal chains, see Fig.~\ref{Ba5In4Bi5_figure}(a). We show the
shape of the nodal chain within reciprocal space for the bands
close to the Fermi energy in Fig.~\ref{Ba5In4Bi5_figure}(b). 

\subsection{Intersecting nodal lines of SG 110}
\label{section_on_SG110}

\begin{figure}[th]
\centering
\includegraphics[width = 0.49 \textwidth]{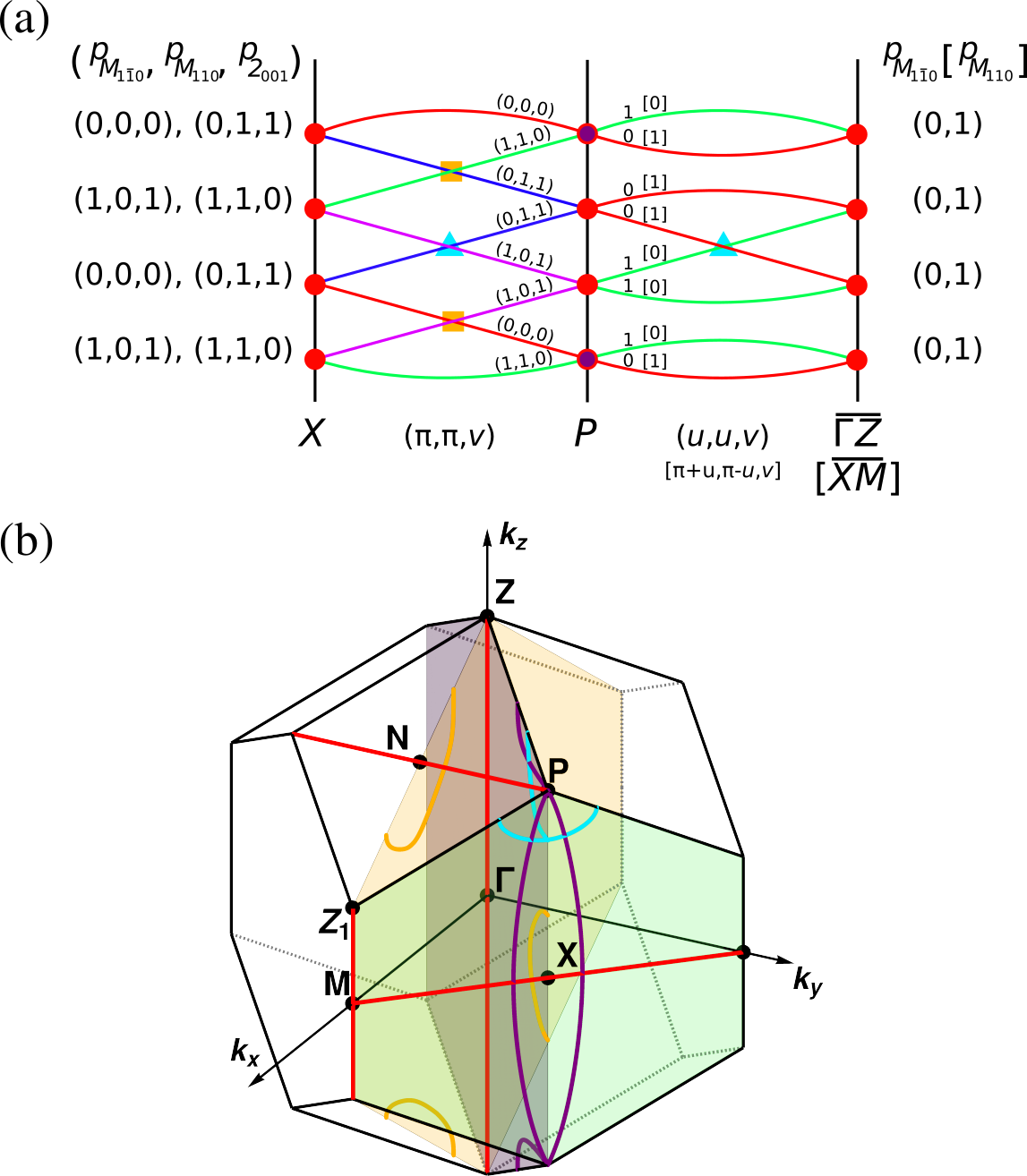}
\caption{
Band crossings in SG 110 including the nodal armillary sphere.
(a) Band connectivity, sets of symmetry eigenvalues and enforced
crossings on the line $(\pi,\pi,v)$ and on the planes $(u,u,v)$
and $(\pi+u, \pi-u,v)$, which includes $\Gamma$-Z and X-M,
respectively. Eigenvalues for the plane $(\pi+u, \pi-u,v)$ are
given in square brackets. Turquoise triangles: Nodal
armillary sphere. Yellow squares: movable nodal
lines forming an hourglass structure on a single mirror plane.
Red circles: Pinned nodal lines, which can be overlapped by purple circles indicating additional almost movable nodal lines.
(b) Qualitative arrangement of enforced nodal
lines in the 3D Brillouin zone with the same color code as the
crossings in (a). Additional copies due to fourfold rotation are
omitted.
}
\label{Fig:SG110_BZ_NodalLines}
\end{figure}

We now want to discuss the special case of the body-centered
SG~110 ($I4_1cd$) \cite{bzdusek_soluyanov_nature_16}. It combines
all types of nodal lines we have encountered so far, movable,
almost movable, and pinned nodal lines. We focus on the (almost) movable
nodal lines, which intersect without forming a nodal chain.
The qualitative results are summarized in Fig.~\ref{Fig:SG110_BZ_NodalLines}.

The movable nodal lines on mirror planes appear between
high-symmetry points and pinned nodal lines. To enforce the pinned
nodal lines, the twofold rotation
$2_{001}(\frac12,\frac12,\frac12)$ may relate eigenvalues of
mirror symmetries, e.g. $M_{010}(0,0,\frac12)$. 
As defined in Eq.~\eqref{mirror_eigvals},
$p_{M_{010}}$ labels the eigenvalue of $M_{010}(0,0,\frac12)$,
i.e. 
\mbox{$\exp(\mathrm{i} (2p_{M_{010}} +1) \frac{\pi}{2}) \exp(\mathrm{i} \frac{k_z}{2})$}.
Both symmetries commute on \mbox{X-P} and anticommute on
\mbox{$\Gamma$-Z-M}. Thus only \mbox{$\Gamma$-Z-M} is twofold
degenerate and $p_{M_{010}}$ values are paired (0,1). 

Another 2-fold degenerate line with $(0,1)$ pairing results from
Kramers theorem applied to the combined operation $\mathcal{T}
M_{110}(0,\frac12,\frac34)$ on the line \mbox{X-M}, which can be
parameterized by $\vk=(\pi+u,\pi-u,0)$ with 
\mbox{$u \in [ -\pi,\pi ]$}.
Kramers theorem applies, because the condition
$(\mathcal{T} M_{110}(0,\frac12,\frac34))^2 = \exp(-\mathrm{i} u) \neq 1 $ 
holds almost everywhere (see
Sec.~\ref{eigenvalue_dependent_Kramers_theorem}).
The third pinned nodal line is enforced by $M_{010}(0,0,\frac12)
\mathcal{T}$ on N-P, where
$(M_{010}(0,0,\frac12) \mathcal{T})^2=-1$.
All pinned nodal lines are shown as red lines in
Fig.~\ref{Fig:SG110_BZ_NodalLines}.

We can readily assess the plane $\Gamma$NM, because it is bounded
by the twofold degenerate rotation axis \mbox{$\Gamma$-Z-M}.
Therefore $M_{010}(0,0,\frac12)$ and time-reversal symmetry lead
to an hourglass structure between any point on \mbox{$\Gamma$-Z-M}
with $(0,1)$ pairing and the TRIM $N$ which exhibits the eigenvalue
pairs $(0,0)$ and $(1,1)$. The resulting movable nodal lines in the
$\Gamma$NM plane connect four bands and are denoted in
Table~\ref{mTab1} as \mbox{($\Gamma$-Z, M-Z$_1$; N)(4)}. 

A more intricate argument is needed to understand the enforced
band structure on the remaining two distinct planes, because they
share the rotation axis \mbox{X-P} with nontrivial pairing. On
the line \mbox{X-P} three symmetries commute:
$M_{1\bar{1}0}(\frac12,0,\frac14)$, $M_{110}(0,\frac12,\frac34)$,
and $2_{001}(\frac12,\frac12,\frac12)$. Their eigenvalues on the
line \mbox{X-P}, $(\pi,\pi,k_z)$ are 
\begin{align}
\alpha_{M_{1\bar{1}0} } 
&\overset{\text{X-P}}{=} \exp(\mathrm{i} ( 2p_{M_{1\bar{1}0} }+ 2 )\pi/2 ) \exp(\mathrm{i} k_z/4) 
\label{eq_m1-10_EV} \\
\alpha_{M_{110} } 
&\overset{\text{X-P}}{=} \exp(\mathrm{i} (2p_{M_{110} } + 1)\pi /2) \exp(\mathrm{i} 3 k_z/4) 
\label{eq_m110_EV}\\
\alpha_{2_{001} } 
&\overset{\phantom{\text{X-P}}}{=} \exp(\mathrm{i} (2p_{2_{001} }\; + 1 )\pi/2 ) \exp(\mathrm{i} k_z/2)
\label{eq_2001_EV}
\end{align}
Each band shall be labeled as
$\ket{p_{M_{1\bar{1}0}},p_{M_{110}},p_{2_{001}}}$, where each
$p \in \{0,1\}$. 
 
Before we can relate the states on \mbox{X-P} we have to determine
the possible combinations
$(p_{M_{1\bar{1}0}},p_{M_{110}},p_{2_{001}})$. The
states are restricted because the product of all 
symmetries is a translation $ M_{1\bar{1}0}(\frac12,0,\frac14) \,
M_{110}(0,\frac12,\frac34) \, 2_{001}(\frac12, \frac12, \frac12) =
t(\frac12,-\frac12,\frac32)$.
As the bands are generally non-degenerate along \mbox{X-P}, the
product of eigenvalues must fulfill
$ \alpha_{M_{1\bar{1}0} } \alpha_{M_{110} } \alpha_{2_{001} }
=\exp(\mathrm{i} \frac{3k_z}{2}) $~\cite{furusaki_non_symmorphic_17}.
One concludes that
$p_{M_{1\bar{1}0}} + p_{M_{110}} + p_{2_{001} } \mod 2 = 0$ and
thereby the only possible bands on \mbox{X-P} are $\ket{0,0,0},
\ket{1,1,0}, \ket{1,0,1},$ and $\ket{0,1,1}$. 

On the line \mbox{X-P} no further symmetries apply, but
time-reversal symmetry $\mathcal{T}$ at X and $\mathcal{T}
M_{010}(0,0,\frac12)$ at P each invoke Kramers theorem. At X time
reversal yields two different twofold bands
$(\ket{0,0,0},\ket{0,1,1})$ and $(\ket{1,1,0},\ket{1,0,1})$.
Notice that the pairing of $M_{110}(0,\frac12,\frac34)$
eigenvalues at X, i.e. $(0,1)$ is consistent with the action of
$\mathcal{T} M_{110}(0,\frac12,\frac34)$ on the line \mbox{X-M}
discussed before. For P we have to consider how the three
eigenvalues in Eqs.~\eqref{eq_m1-10_EV} -~\eqref{eq_2001_EV} are
related by $\mathcal{T} M_{010}(0,0,\frac12)$. When acting on a
generic state on the right, one obtains the following relations
\begin{align} &M_{110}(0,\tfrac12,\tfrac34) \; T
M_{010}(0,0,\tfrac12) \nonumber\\ &= \exp(\mathrm{i} \pi (-
2p_{M_{1\bar{1}0} }-1)/2 ) \exp(\mathrm{i} 3 k_z/4) \; T M_{010}(0, 0,
\tfrac12) , \label{110P_eigvals1} \\ &2_{001}(\tfrac12, \tfrac12,
\tfrac12) \; T M_{010}(0, 0, \tfrac12) \nonumber\\ &= \exp(\mathrm{i} \pi
(- 2 p_{2_{001} } -3 )/2 ) \exp(\mathrm{i} k_z/2) \; T M_{010}(0, 0,
\tfrac12). \label{110P_eigvals2} \end{align}
Thus, at P the band doublets $(\ket{0,0,0},\ket{1,1,0})$,
$(\ket{1,0,1},\ket{1,0,1})$, and $(\ket{0,1,1},\ket{0,1,1})$ are
independent and degenerate. 
If the representations are ordered in energy such that the minimal
number of crossings appear, then nodal lines on different mirror
planes cross on the \mbox{X-P} axis, as shown by the turquoise lines in
the visualization of SG~110 in
Fig.~\ref{Fig:SG110_BZ_NodalLines}(b). 
Unlike for nodal chain metals the nodal lines do not need to
extend over the full height of the Brillouin zone. 
To our knowledge, a similar arrangement of bands 
has only been discussed in systems without spin-orbit
coupling. For the cubic SG~221 three intersecting planes with
joined nodal lines were
considered~\cite{noSOC_intersectingNodeLines_Yu,noSoc_intersectingNodalLine_Du},
whereas in the orthorhombic SG~61 two intersecting nodal lines
were referred to as a nodal armillary sphere~\cite{shao2019composite}.
These nodal lines gap out once spin-orbit is relevant, unlike the
feature of SG~110 presented here.
There are no enforced crossings at the same energy as the
connected nodal lines around P. If no accidental crossings or band
pockets occur and the filling is $8n+4$ ($n \in \mathbb{N}_0$)
electrons per unit cell, materials in SG~110 with a weak
dispersing nodal line are not just metals with nodal points but
genuine enforced semimetals. 

At different band number we find additional nodal lines appearing
in the same mirror plane around different high-symmetry points. We
denote this by labeling the nodal lines as ($\Gamma$-Z, P ; X)(4)
and ($\Gamma$-Z, X ; P)(4) in Table~\ref{mTab1} and show them as
orange lines in Fig.~\ref{Fig:SG110_BZ_NodalLines}(b). Therefore
the combination of different connectivities raises the number of
connected bands to $8 \mathbb{N}$. 

Furthermore, SG~110 exhibits almost movable nodal lines as
introduced in Sec.~\ref{sec_movableTRIMlines}. They can be
inferred from the pairing of symmetry eigenvalues at the point P
we already derived. Out of the possible representations only the
paired states $(\ket{0,0,0},\ket{1,1,0})$ contain different mirror
eigenvalues. Hence, only twofold degeneracies of this type are
part of almost movable nodal lines. A possible arrangement of
these nodal lines is displayed by purple lines in
Fig.~\ref{Fig:SG110_BZ_NodalLines}(b), where the relevant
crossings at P are highlighted in
Fig.~\ref{Fig:SG110_BZ_NodalLines}(a) by a purple dot. These
almost movable nodal lines may intersect the pinned nodal lines
shown in red, whereas they are at a different band index, i.e.
different energy, than the turquoise colored nodal armillary
sphere and thus do not cross them in the simplest case. See
Appendix~\ref{App_110P} for a detailed analysis of bands in the
vicinity of the point P.

\section{Fourfold Weyl nodal lines}
\label{4fold_lines}
\label{sec_four_fold_weyl_lines}

Fourfold degenerate Weyl nodal lines are symmetry enforced
degeneracies which split into non-degenerate bands when moving
away from the nodal line in almost all directions. This
distinguishes them from Dirac lines, which are also fourfold
degenerate, but split into twofold degenerate states in all
perpendicular directions. Fourfold degenerate nodal lines are
enforced in band structures of SG 113 and 114.
The generators of these two SGs are translations, a fourfold
rotoinversion 
$\overline{4}_{001} \equiv 4_{001}(0,0,0) \mathcal{P}$ and a
twofold screw rotation $2_{010}(\frac12,\frac12,c)$ around an axis
perpendicular to it, where the factor $c=0,\frac{1}{2}$
distinguishes between SG~113 and SG~114 and is not relevant in the
following discussion.

The fourfold degenerate line follows from the two antiunitary
symmetries created by combining the generators with time reversal
symmetry, $\overline{4}_{001}\mathcal{T}$ and
$2_{010}(\frac12,\frac12,c)\mathcal{T}$.
As mentioned before, $2_{010}(\frac12,\frac12,c)\mathcal{T}$
leaves points in the two planes $k_y=0$ and $k_y=\pi$ invariant
and squares to $-1$ on the latter plane, therefore enforcing
twofold degenerate nodal planes.
$\overline{4}_{001}\mathcal{T}$ squares to $-2_{001}$ and leaves
the lines \mbox{$\Gamma$-Z} and \mbox{M-A} invariant. It further
fulfills the conditions of Kramers theorem~\eqref{Kramers_def} on
these lines, since eigenvalues of $2_{001}$ are $\pm\mathrm{i}$.

On the M-A axis, pairing conditions of both symmetries are
present, resulting in a pinned fourfold degenerate line. To track
the pairing from each of these two anti\-unitary symmetries, we
label the states with the sign of their $2_{001}$-eigenvalues $\pm
\mathrm{i}$. Since $2_{001}$ and $\overline{4}_{001}\mathcal{T}$ naturally
commute, the eigenvalues are related only via complex conjugation,
i.e. the pairing is $(+,-)$.
$2_{010}(\frac12,\frac12,c)\mathcal{T}$, on the other hand,
anticommutes with $2_{001}$, adding another sign change to the
conjugated eigenvalue. Hence it pairs $(+,+)$ and $(-,-)$. In
combination, we find the fourfold degeneracy of the nodal line
with the pairing $(+,+,-,-)$. This feature has recently been
reported as a Dirac line~\cite{D0CP03686B}, however, apart from the
planes $k_y=\pi$ and $k_x=\pi$, the bands are non-degenerate. The
dispersion perpendicular to the nodal line is linear and remains
twofold degenerate in the nodal planes $k_x=\pi$ and $k_y=\pi$.
For constant $k_z$, this is the same structure seen in the
fourfold Weyl point shown in Fig.~\ref{higher_weyl_fig}(b).

\begin{figure}[hbt]
\centering
\includegraphics[width=0.9\linewidth]{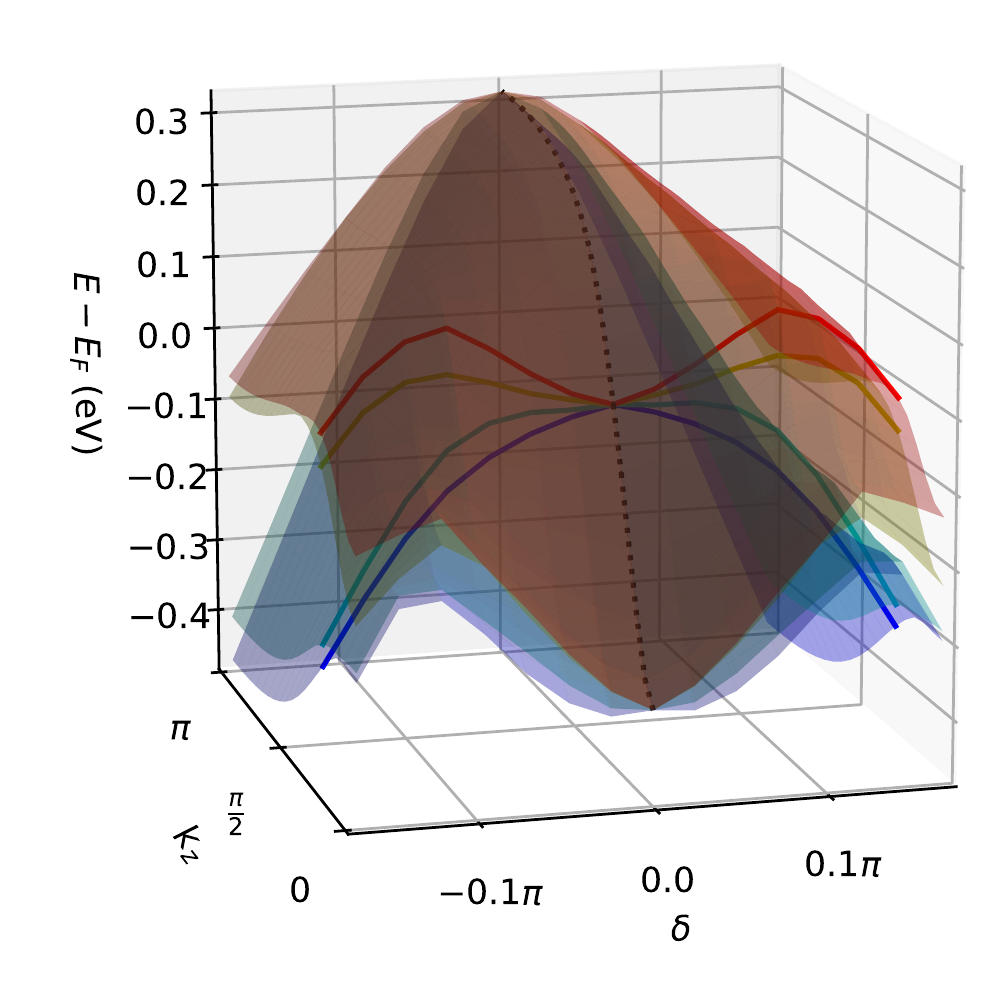}
\caption{
        Dispersion around the fourfold nodal line in NaSn$_5$ in
        the plane defined by 
        \mbox{$\vk=(\pi+\delta,\pi+\delta,k_z)$} 
        in the vicinity of the line M-A.
        The dashed line highlights the fourfold degeneracy on the
        high-symmetry axis. The colored solid lines show the
        dispersion perpendicular to the nodal line at constant
        $k_z$, where the nodal line	crosses the Fermi level.
        }
\label{NaSn5}
\end{figure}

The topological protection of this feature can best be understood
from the linearized Hamiltonian around the nodal line, see
Appendix~\ref{low_energy_4fold_line}. It can be transformed into
block diagonal form via a $\vk$-independent, unitary matrix. Each
of the blocks describe one independent Weyl line, each protected
by a $\pi$ Berry phase.

\subsubsection{Material example: NaSn$_5$}
\label{sec_NaSn5}

NaSn$_5$ crystallizes in SG 113~\cite{fassler1998nasn5} and shows
the fourfold Weyl line along M-A and the twofold degeneracy at the
BZ boundary, which can be seen as fourfold and twofold
degeneracies along high-symmetry lines in
Fig.~\ref{full_band_structures}(g). It is metallic and the
fourfold nodal line crosses the Fermi energy. A closeup of bands
calculated from first principles in the plane parameterized by
$\vk=(\pi+\delta,\pi+\delta,k_z)$ 
is shown in Fig.~\ref{NaSn5}. It shows the dispersion of the nodal
line along $k_z$ as well as the separation into individual bands
for non-vanishing $\delta$, with a linear dispersion only for very
small values. The same features have been reported in band
structure calculations for Pd$_4$S, which crystallizes in
SG~114~\cite{D0CP03686B}.


\section{Non-symmorphic nodal planes}
\label{weyl_planes}

Band degeneracies may not only appear as points and lines but are
known to occur also as two-dimensional
manifolds~\cite{herring1937accidental}, among them nodal 
planes~\cite{xiao2017topologically,turker2018weyl,bzduvsek2017robust,PhysRevB.99.235423,Xiaoeaav2360, chang2018topological}.
As we have seen above, the combination of a twofold screw rotation
with time reversal symmetry is one mechanism to enforce nodal
planes via a generalized Kramers theorem.
For example a nodal plane is enforced in SG~76, where the combined
antiunitary symmetry $2_{001}(0,0,\frac12)\mathcal{T}$ leaves
points in the $k_z=0$ and $k_z=\pi$ plane invariant and squares to
a full lattice translation $t(0,0,1)$. In the $k_z=\pi$ plane,
the translation eigenvalue is $-1$ and the condition for Kramers
theorem is fulfilled, c.f. Eq.~\eqref{Kramers_def}. All such nodal
planes are listed in Table~\ref{mTab1} in the column 'nodal
planes'.

Recently, nodal planes have been discussed in the context of
topology~\cite{bzduvsek2017robust,turker2018weyl,CicumventingNoGo_yuxin},
as they can also act as sources and sinks for Berry curvature.
Within the tetragonal space groups there are three cases, SGs~92,
94, and 96, of enforced nodal planes, which additionally must have
a topological charge, i.e. they act as source or sink of Berry
flux~\cite{CicumventingNoGo_yuxin} to balance the topological
charges of symmetry enforced Weyl points in the interior of the
BZ.

\subsection{Space groups 92 and 96}

The SGs~92 and 96 form an enantiomorphic pair, i.e. they are
mirrored versions of each other. They are generated from
translations of the primitive lattice and the screw rotations
$4_{001}(\frac12,\frac12,\frac{c}{4})$ and
$2_{010}(\frac12,\frac12,\frac{c}4)$, $c=1$ and 3, respectively.
The fourfold screw rotation squares to a twofold screw rotation,
which means these two space groups have twofold screw rotations
along 100, 010 and 001. With time-reversal symmetry present, all
planes on the surface of the BZ are therefore Kramers degenerate,
and the nodal planes form a closed box. Formally, there is a
twofold degeneracy for every $\vk$ with at least one component
$k_i=\pi$, $i \in \{x,y,z\}$. To calculate the topological charge
of this box, it can be enclosed by the surface of a slightly
smaller box within, because bands are non-degenerate in the
interior of the BZ. This smaller box is simultaneously enclosing
all band degeneracies in the interior when inverting the surface
normal. Thus the topological charge of the nodal planes equals the
negative sum of all topological charges in the interior, which
close the same gap as the nodal surface in question. Consequently,
we have to consider the band connectivity across the whole BZ.
Bands in these band structures appear always as multiples of eight
with a certain minimum number of crossings, see
Fig.~\ref{SG96_bands}.
As shown above, the non-symmorphic rotations enforce accordion and
hourglass states, i.e. Weyl points at $\Gamma$ and movable
topological crossings along the high-symmetry lines $\Gamma$-X and
$\Gamma$-Z, cf. Sec.~\ref{Sec_NonSymWeyl}. 
Although various symmetry enforced band crossings appear in SGs~92
and 96, all topological charges in the interior can not add up to
zero for any odd number of bands above (or equivalently below) the
gap. In this case, there is always an enforced Weyl point at
$\Gamma$ with a charge of $\pm1$, whereas the movable band
crossings on the rotation axis close the gap in between even sets
of bands, see Fig.~\ref{accordion_figure}(a) and (c). 
Additional accidental band crossings might exist on the rotation
axes, but they will have a multiplicity of two if they are on the
line $\Gamma$-Z or four on the other two high-symmetry axes
$\Gamma$-X and $\Gamma$-M. Since Weyl points with higher
multiplicity are always symmetry related by rotations in these
SGs, symmetry related copies will always have the same charge.
Thus, the total charge in the interior of the BZ remains odd and
the nodal plane carries an odd non-zero topological
charge, which cannot be removed without breaking
symmetries~\cite{CicumventingNoGo_yuxin}. 
Despite this non-zero Chern number, any surface termination will overlap the nodal plane with its corresponding Weyl points such that, in this case, no Fermi arcs are associated with the topological nodal plane. 

\subsubsection{Material example: Ba$_3$Sn$_2$}
\label{Ba3Sn2_chapter_zwei}

As discussed in Section~\ref{Ba3Sn2_chapter}, Ba$_3$Sn$_2$
crystallizes in SG 96 (or its enantiomorphic pair, SG 92). The
highest occupied bands are clearly spin split, due to strong
spin-orbit coupling, see Fig.~\ref{Ba3Sn2_full}.
Figure~\ref{SG96_bands} compares \emph{ab initio} calculations
with the sketch made from connectivity of irreducible
representations. The enforced nodal plane can be seen as twofold
degeneracy on all paths on the BZ surface for any band index. Due
to the multiplicity of accidental crossings the enforced
Kramers-Weyl point at $\Gamma$ is necessarily compensated by the
topological charge of the nodal plane. 

\subsection{Space group 94}

Space group 94 is special, since only two surfaces are nodal
planes and their charge is even and non-vanishing, i.e. at least
$\pm2$. In contrast to the above, the fourfold screw rotation
$4_{001}(\frac12,\frac12,\frac12)$ in SG~94 squares to a symmorphic
twofold rotation. Thus the plane $k_z=\pi$ is not a nodal plane
and the nodal manifold is restricted to the $k_x=\pi$ and
$k_y=\pi$ planes. There is a conceptual difference to SGs
discussed in~\cite{CicumventingNoGo_yuxin}, because SG~94 hosts
Weyl points at the two TRIMs $\Gamma$ and Z. However, they cannot
balance their charges as without additional crossings, the
connectivity of bands requires that the Weyl point at Z has the
same charge as the one at $\Gamma$, as in 
Fig.~\ref{accordion_figure}(b)~\cite{tsirkin_vanderbilt_PRB_17}.
Accidental crossings along $\Gamma$-Z may occur in time reversal related pairs but
they cannot compensate the chirality, because the total phase picked up with band exchanges is restricted by the periodicity of the BZ~\cite{MnSi}.
As a result the smallest possible change from these additional crossings on the
rotation axis $\Gamma$-Z is $\pm4$. Away from the rotation axis and the
nodal planes, any Weyl point appears at least four times with
equal chirality, such that the nodal charge of the surface needs
to be $\mathcal{C} = 2 +4 z$, $z\in \mathds{Z}$. On a (001)
surface in SG~94, two Fermi arcs must connect the projected Weyl
points from Gamma and Z to the sides of the surface BZ.

\section{Accidental band crossings/ Off-centered Symmetries}
\label{Sec_AccBandCross_OffCenter}

Notwithstanding crossings protected by topological charges,
symmetry allowed perturbations to a given Hamiltonian may
generally gap any crossing between bands unless they can be
labeled by different symmetry eigenvalues. A specific realization
of topological semimetallic phases may also contain accidental
band crossings, and we briefly discuss such cases here.
The dimensions of irreducible representations are
comprehensively addressed in the
literature~\cite{elcoro_aroyo_JAC_17, miller_love_irreps}.
Therefore we restrict this discussion to a selection of these results
and short remarks for the tetragonal space groups. 
 
In space groups without inversion (Table~\ref{mTab1}), 
eigenvalues are unpaired by default and crossings may appear at any
$k$ with a symmetry beside translations in its little group.
Exceptions can be the result of the nodal planes and certain
rotation axes in mirror planes. Conversely, for inversion
$\mathcal{P}$ and time reversal $\mathcal{T}$ the combination
$\mathcal{PT}$ pairs eigenvalues, such that twofold symmetries
may not lead to accidental band crossings anymore. The only
exceptions to this 
are off-centered symmetries~\cite{yang_furusaki_PRB_2017}. 

Off-centered rotation (mirror) symmetries are characterized by
translations perpendicular to the rotation axis (mirror plane). In
real space they can be understood as having a different center
than inversion, i.e. their sets of invariant points do not overlap.
When the off-centered symmetry is a screw or glide operation it will
enforce an odd number of point or line crossings, respectively.
Hereby, the defining off-centered translation parts of the
symmetries lead to a pairing of identical eigenvalues. Whereas
nodal lines enforced by off-centered mirror symmetries
are generally movable on the mirror plane, in the tetragonal space groups
they are pinned to high-symmetry lines. They explain the existence
of `Dirac lines' listed in Table~\ref{mTab2}. Note that point
crossings due to off-centered symmetries are always pinned by
time-reversal symmetry to TRIMs. They are included with other
Dirac points in the column `Dirac points.'

Finally, we list all possible sets of k-points with accidental
crossings in the columns `accidental points' and `accidental
lines,' which follow from the off-centered symmetries. With
knowledge of accidental nodal points and lines, it is evident
whether a crossing along a high-symmetry path is a solely a Dirac point or
part of a nodal line.

\section{Conclusions}
\label{sec_conclusion}
 
In conclusion, we classified all possible symmetry-enforced band
crossings in tetragonal materials with strong spin-orbit coupling.
We considered both movable and pinned band degeneracies. We
uncovered a rich variety of topological band crossings, which
arise due to the intricate interplay of symmetry and topology
(Tables~\ref{mTab1} and~\ref{mTab2}). This includes different
types of pinned and movable Weyl points, specifically,
single/double Weyl points and fourfold double/quadruple Weyl
points (Sec.~\ref{Sec_NonSymWeyl} and Fig.~\ref{higher_weyl_fig}).
Usually, these Weyl points come in multiple copies that are
related by symmetry (i.e. they have large multiplicity).
However, for SGs~119 and 120 with body-centered unit cells we
found that there can exist Weyl points that have only \emph{one}
symmetry-related partner (i.e. multiplicity two). This
could be important for applications, since materials with fewer
Weyl points exhibit simpler and more interesting (magneto-)optical
responses and transport properties. We also classified all
possible symmetry-enforced Dirac points, which can be movable or
pinned, in inversion symmetric tetragonal systems
(Sec.~\ref{sec_theo_Dirac_points}). Interestingly, SGs 130, 133,
and 138 have movable Dirac points with an hourglass dispersion
(Fig.~\ref{movable_Dirac_points}). SG 138 also has a symmetry-enforced weak $\mathbb{Z}_2$ invariant
leading to Dirac surface states. Similar features are 
expected in orthorhombic
systems~\cite{Orthorhombic_to_be_published}.

Nodal lines also exist in different varieties. We catalogued all
possible symmetry-enforced Dirac nodal lines, and twofold and
fourfold Weyl nodal lines (Secs.~\ref{sec_two_fold_weyl_lines}
and~\ref{sec_four_fold_weyl_lines}). In SGs 102, 104, 109, 118,
and 122 the Weyl nodal lines form chains off connected rings,
i.e. nodal chains (Fig.~\ref{Ba5In4Bi5_figure}), while in SG 110
they form an armillary sphere
(Fig.~\ref{Fig:SG110_BZ_NodalLines}). Interestingly, in SG 110 the
global band topology allows, in principle, for a band structure
with only the armillary sphere at the Fermi energy. The low-energy
physics of a material with this property would be dominated by the
nontrivial topology of the armillary sphere, which could prove
useful for applications.
 
Finally, we investigated nodal planes whose existence is enforced
by the combination of screw rotations with time reversal
(Sec.~\ref{weyl_planes}). Remarkably, in SGs 92, 94, and 96 the
symmetries enforce a nontrivial topological charge of the nodal
planes. This charge is compensated by a \emph{single} Weyl point
at $\Gamma$ for SGs 92 and 96. SG 94 is particularly interesting
since it has a nodal plane duo (as opposed to a nodal plane trio
in SGs 92 and 96) with arc surface states connecting two Weyl
points with equal chirality at $\Gamma$ and Z to the nodal planes.

We emphasize that all of the aforementioned band degeneracies are
purely symmetry enforced. That is, they occur in all bands of all
materials crystallizing in the given SG, regardless of the
chemical composition and other material details. These symmetry
enforced band degeneracies cannot be annihilated by any
symmetry-preserving perturbation. This is in contrast to
accidental band degeneracies which can be pair annihilated and
which are only perturbatively stable.

Using our classification Tables~\ref{mTab1} and~\ref{mTab2} it is
now possible to specifically design (meta-)materials with the
desired band topologies. It is also possible to search for
existing compounds with these topological band crossings using
materials databases, such as the ICSD. We performed such a search and
found seven candidate materials (Fig.~\ref{heat_map}).
Particularly interesting are the compounds where the band
degeneracies cross the Fermi level, which is the case for the
twofold and fourfold Weyl nodal lines in Ba$_5$In$_4$Bi$_5$ and
NaSn$_5$, respectively. We hope that our findings will stimulate
experimentalists to synthesize and characterize these materials and
to design new (meta-) materials based on our classifications.

\acknowledgments
The authors thank C.~Ast, C.-K. Chiu, L.~M.~Schoop,
M.~G.~Vergniory, and A.~Yaresko for useful discussions. This
research was supported in part by the National Science Foundation
under Grant No.\ NSF PHY-1748958. D.H.F. gratefully acknowledges
financial support from the Alexander von Humboldt Foundation.

\vspace{1.0cm}
\appendix

\section{Additional band structure calculations}
\label{appendix_band_structure} 

\begin{figure*}[pt!]
\centering
\subfloat[Hf$_3$Sb]{
\label{Hf3Sb_full}
\includegraphics[width =.31\linewidth]{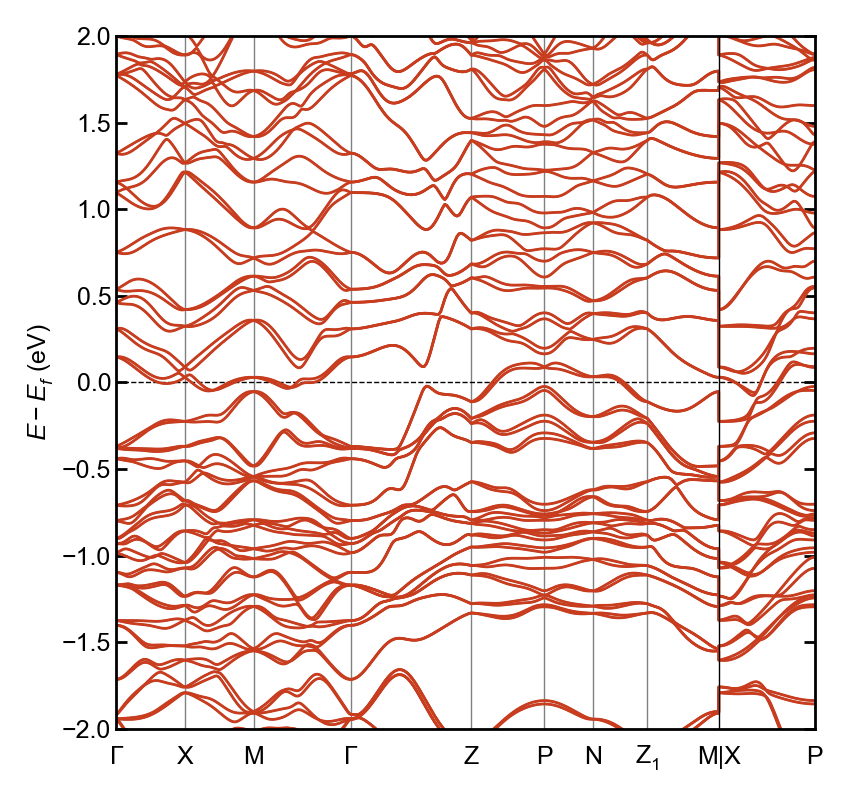}
}
\subfloat[Cs$_2$Tl$_3$]{
\includegraphics[width =.31\linewidth]{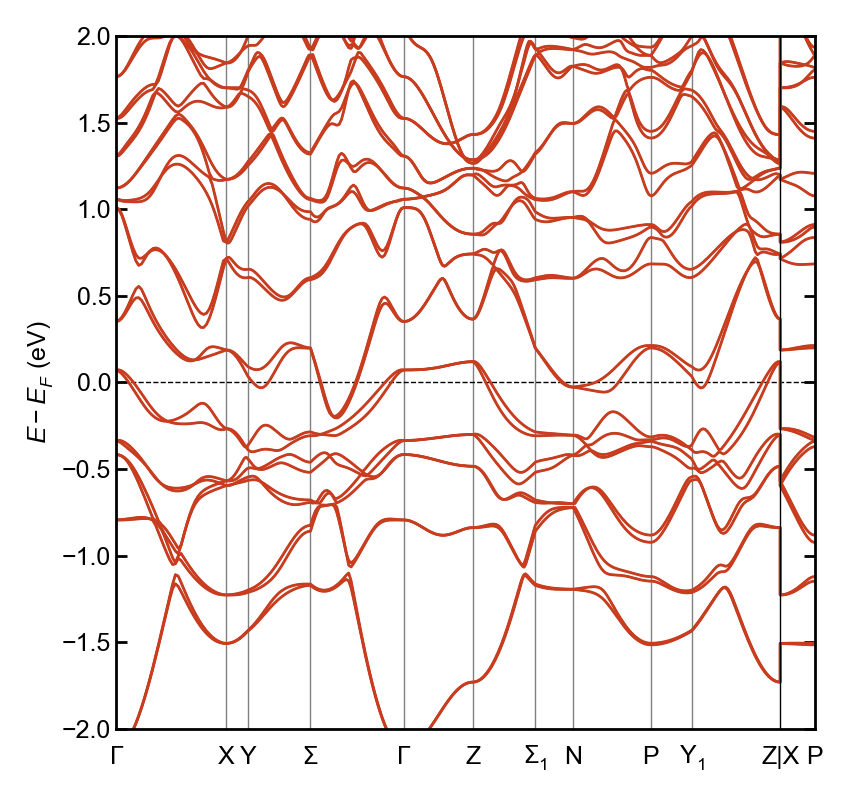}
\label{Cs2Tl3_full}
}
\subfloat[Ba$_5$In$_4$Bi$_5$]{
\includegraphics[width =.31\linewidth]{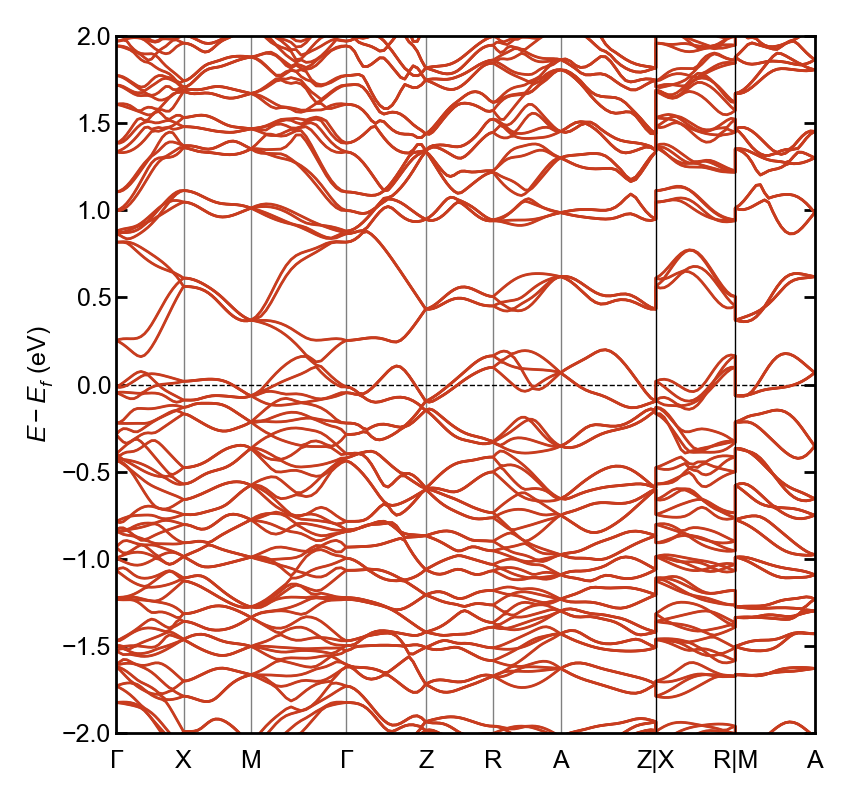}
\label{Ba5In4Bi5_full}
}
 
\subfloat[AuBr]{
\includegraphics[width =.31\linewidth]{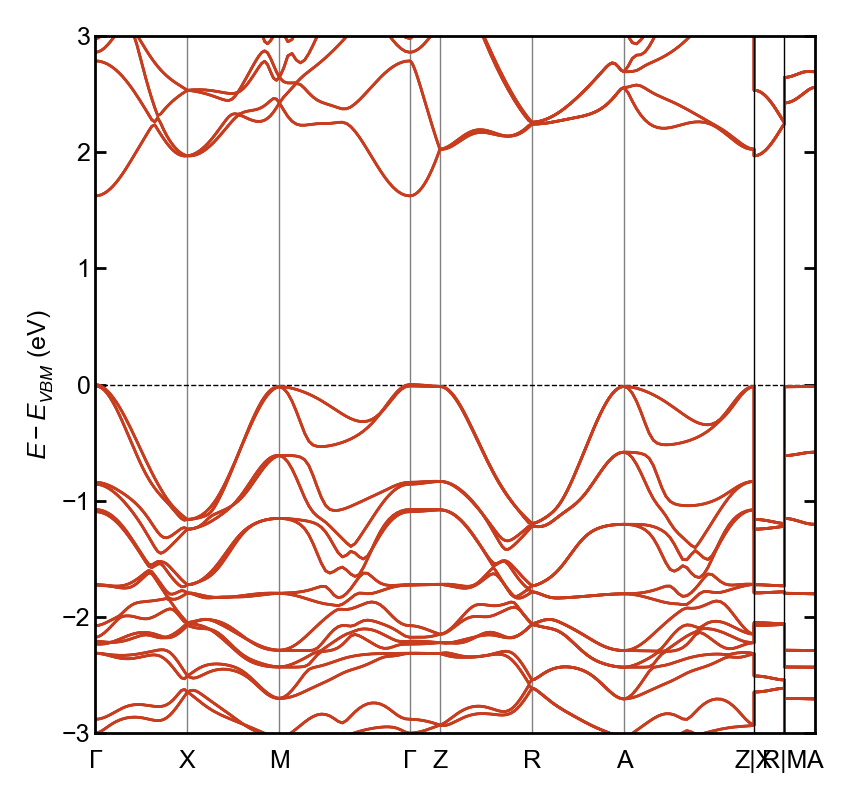}
\label{AuBr_full}
}
\subfloat[Tl$_4$PbSe$_3$]{
\includegraphics[width =.31\linewidth]{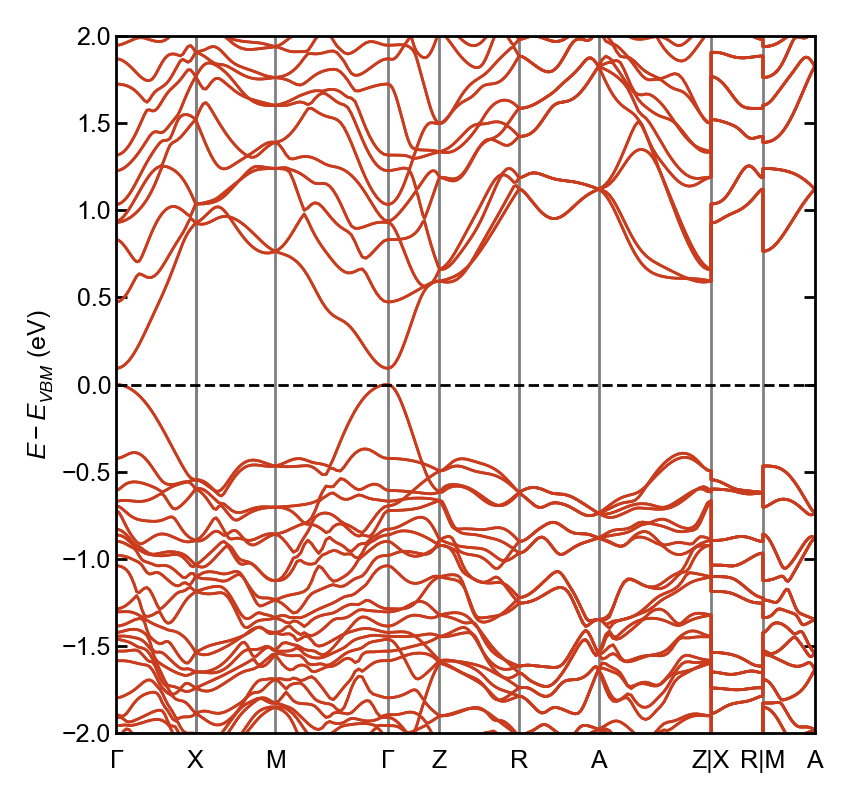}
\label{Tl4PbSe3_full}
}
\subfloat[Ba$_3$Sn$_2$]{
\includegraphics[width =.31\linewidth]{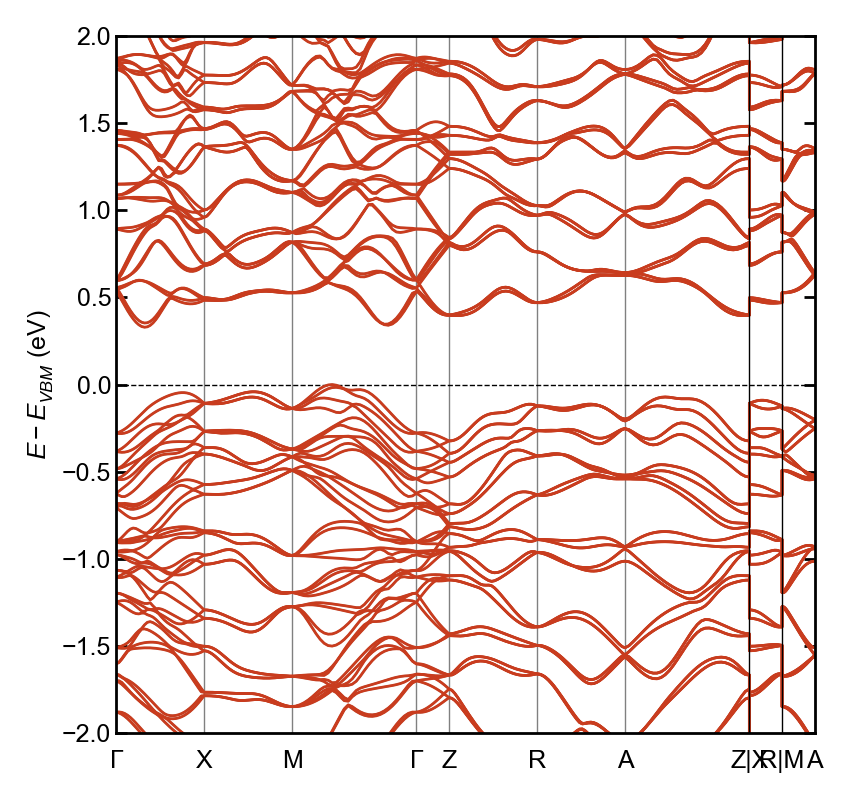}
\label{Ba3Sn2_full}
}

\subfloat[NaSn$_5$]{
\includegraphics[width =.31\linewidth]{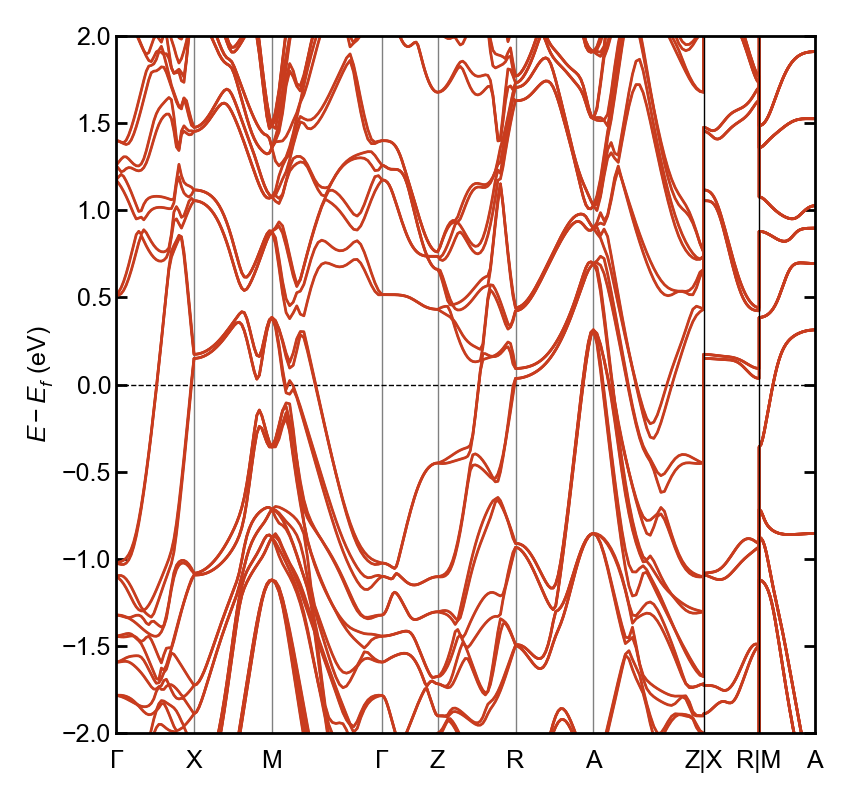}
\label{NaSn5_full}
}
 
\caption{ \label{full_band_structures}
    DFT band structures of the examples presented in the main
    text.
    Detail figures of the relevant sections are shown in the
    corresponding chapters.
}
\end{figure*}

In Fig.~\ref{full_band_structures} we present the bandstructure
calculations along standard high-symmetry
paths~\cite{setyawan2010high} for the example materials selected
according to the selection criteria mentioned in
Sec.~\ref{example_materials} and studied in the main text. Four
of these are metallic with the topological features close to the
Fermi level. Of the gapped examples, AuBr and Ba$_3$Sn$_2$ have a
large gap and serve only to illustrate the features of their
corresponding space groups. 

\section{Generic models and Qualitative description of surface states}
\label{generic_models}

The arguments of the main text yield the enforced features for any
material realization. Yet, in real materials the features can not
always be resolved and a tight-binding model exhibiting
sufficiently large spin-orbit coupling is more suitable to point
out the enforced behavior. 

We derive generic models by placing s-orbitals with a spin degree
of freedom on a Wyckoff position with lowest multiplicity,
considering hopping from one site to the others up to a
sufficiently large cutoff distance, and obtain all symmetry
related terms. Thereby symmetry forbidden terms vanish and all
exchange terms including spin are treated on the same footing.
Finally, random values for the remaining hopping amplitudes are
used to evaluate the model. Numeric calculations of Chern numbers
and chiralities are performed by the Wilson loop
method~\cite{fukui}.

In the following sections we denote the high-symmetry points of
the surface BZ by one of the respective bulk positions, which are
projected onto it. A bar is added to differentiate the labels,
e.g. $\Gamma$ becomes $\bar{\Gamma}$ for any surface termination.
The Figs.~\ref{Surface_SG119},~\ref{Surface_SG106_SG133},
and~\ref{Surface_SG130_138} represent the surface states of a slab
with 30 or more layers at a cut at constant energy through the
spectrum. At each $k$ point the color is determined by the
expectation value $\langle P_r \rangle_\psi$ of the projection
operator $P_r$ on the surface, where $\psi$ is the eigenstate that
gives the largest value of $\langle P_r \rangle_\psi$. The surface
is defined to be 10\% of the full slab thickness.

\subsection{Minimal number of Kramers points - SG 119}
\label{Appendix_119}

\begin{figure}
\includegraphics[width = 0.49 \textwidth]{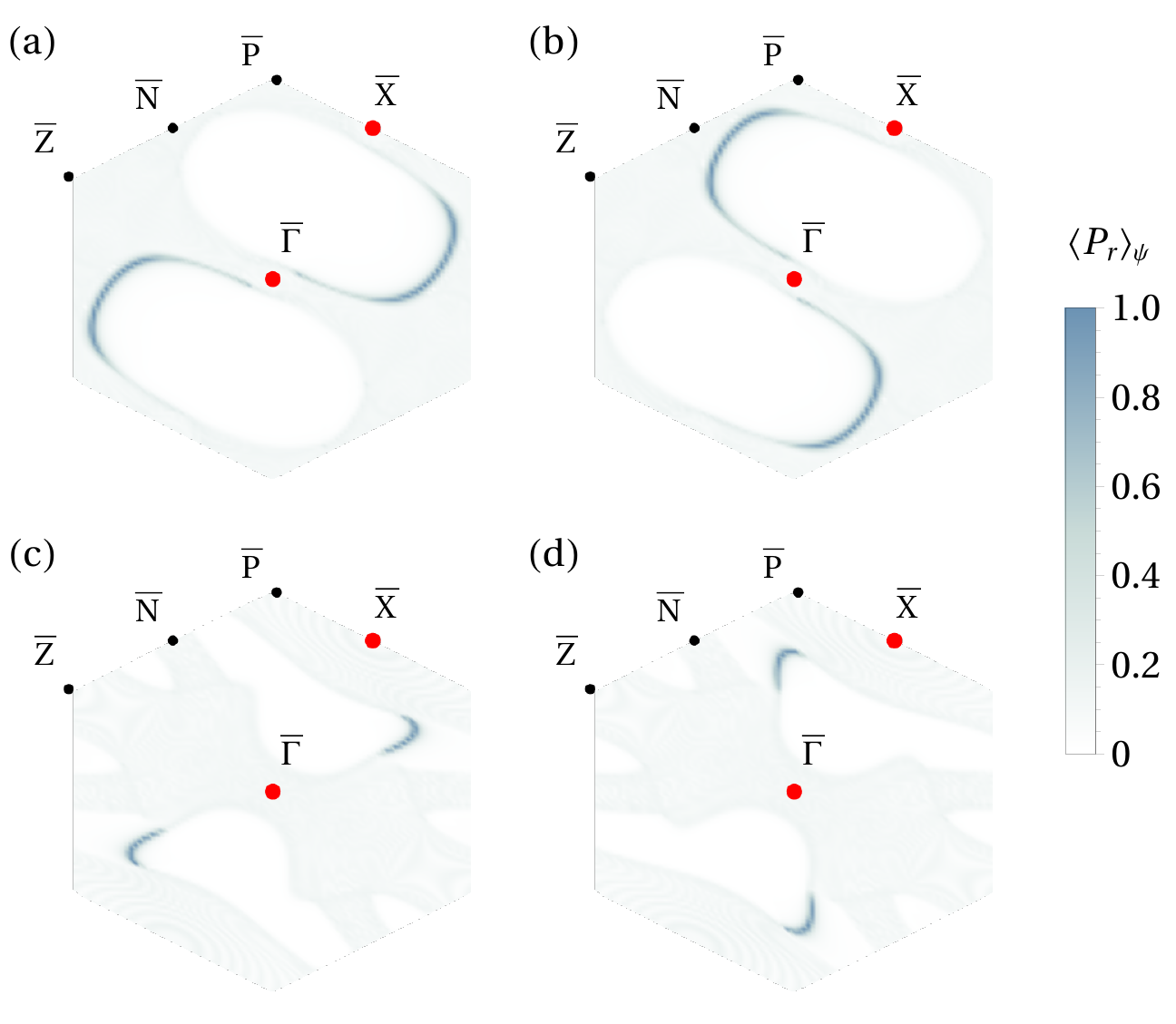}
\caption{
Surface states from SG~119 for two generic models. The slab is
$(1\bar{1}0)$ terminated with a chemical potential chosen in the
vicinity of the Kramers-Weyl points. Red discs mark the
projections of the only Weyl points of the respective model. In
(a,b) [(c,d)] the model contains up to 2nd [3rd] nearest neighbor
hopping. (a,c) show the top and (b,d) the bottom surfaces.
}
\label{Surface_SG119}
\end{figure}

SGs~119 (and 120) exhibit the minimal number of two Weyl points in
the absence of accidental crossings. To confirm this assessment
we devise generic models as described above with 2nd and 3rd
nearest neighbor hopping without accidental features. We then
confirm the chirality at the point X to be $\mathcal{C} = \pm 1$
and determine the surface states for the $(1\bar{1}0)$
termination, see Fig.~\ref{Surface_SG119}. Indeed, Fermi arcs
connect the projections of the Weyl points and attach to the bulk
bands close to the expected positions $\bar{\Gamma}$ and
$\bar{\mathrm{X}}$.

When calculating the Chern number on any plane spanned by [001] and a linear combination of
[100] and [010] one does not enclose one of the
Kramers-Weyl points and thus obtains zero. 
This is due to the connectivity of the body-centered BZ (here
BCT$_1$). 
We find that lines parallel to
$\bar{\Gamma}$-$\bar{\mathrm{Z}}$ cross both Fermi arcs (or the bulk
bands close the surface gap), see Fig.~\ref{Surface_SG119}.
If one assigns a direction to the arcs, e.g. both can be considered to start at $\bar{\Gamma}$ and end at $\bar{\mathrm{X}}$, then they cross the line parallel to $\bar{\Gamma}$-$\bar{\mathrm{Z}}$ in opposite directions in accordance with the vanishing Chern number on the corresponding bulk surface.
The structure of Fermi arcs thereby respects the twofold rotation symmetry $2_{1\bar{1}0}$ of the slab and the surfaces in Figs.~\ref{Surface_SG119}(a) and (b)
are related by $2_{001}$. 

The termination is chosen such that the two opposite chiralities
at X, which are related by the fourfold rotoinversion
$\bar{4}_{001}$ are projected onto two distinct points of the
surface BZ, which are labeled by $\bar{\Gamma}$ and
$\bar{\mathrm{X}}$. Simpler terminations like (100) or (001) map
both X points on top of each other and the Fermi arc could not be
observed.

Kramers theorem pairs opposite mirror eigenvalues at the TRIM N.
As we have described in Sec.~\ref{sec_movableTRIMlines}, the
crossing of mirror eigenvalues at N leads to almost movable nodal
lines in SG~119. Although we focus on surface states, their
existence is recognizable in bulk states shown in
Fig.~\ref{Surface_SG119}. For both models the nodal lines have
the qualitative shape illustrated in
Fig.~\ref{Fig_AlmostMovableSketch}(b). Whereas they appear as
lines through N at constant $k_z$ for the parameters chosen in
Fig.~\ref{Surface_SG119}(a,b), they cross roughly at the midpoint
of \mbox{$\Gamma$-Z} for the model used in in
Fig.~\ref{Surface_SG119}(c,d). Especially in the second case the
bulk bands crossing $\bar{\mathrm{N}}$ evidently belong to the
almost movable nodal line. Its drumhead surface state is not
visible at the chemical potential chosen in
Fig.~\ref{Surface_SG119}.

Challenges for a material realization are highlighted by the two
different models shown in Fig.~\ref{Surface_SG119}. The line
\mbox{$\Gamma$-Z-M} hosts twofold degenerate bulk bands, i.e. they
close the bulk gap in which the Fermi arcs appear. Thus, along the
lines $\bar{\Gamma}$-$\bar{\mathrm{Z}}$ and
$\bar{\mathrm{X}}$-$\bar{\mathrm{P}}$ bulk bands overlap the
projection of both Weyl points at X for the $(1\bar{1}0)$
termination. Beyond sufficient band splitting it is advantageous
if the twofold degeneracy at the $\Gamma$ and M points are at
similar energy compared to the X point.

\subsection{Movable fourfold points - SGs 106 and 133}
\label{Appendix_106_133}

For SGs 106 and 133 there is no chirality associated with the
fourfold crossing on the M-A line due to the mirror symmetries.
Yet, these semimetals generally have surface states, because they
consist of two superimposed Weyl semimetals. This can be
understood by Weyl semimetal precursors that are distinguished by
the action of the rotation symmetry $4_{001}(0, 0, \frac{1}{2})$.
The screw rotation alone yields two unrelated representations,
which each contain one Weyl point along the M-A line.
Adding time-reversal and the mirror symmetries yields the fourfold
crossings, which are Dirac points for SG~133. The mirror symmetry
relates the opposite chiralities and surface states. 

\begin{figure}
\includegraphics[width = 0.49 \textwidth]{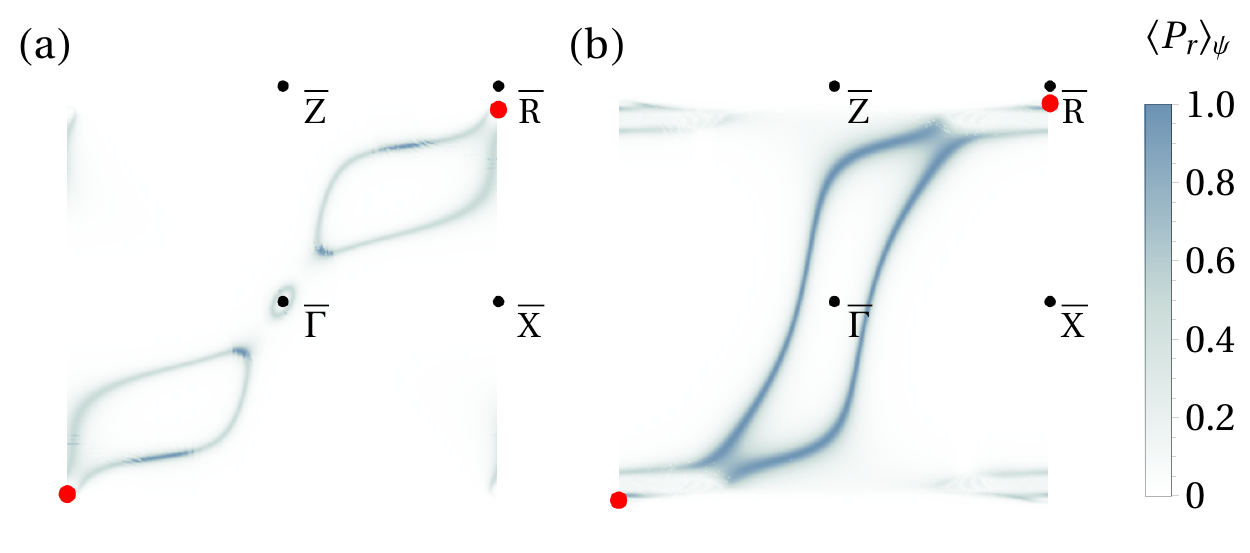}
\caption{
Surface states of SGs 106 (a) and 133 (b). The energy cut
is in the respective vicinity to the fourfold crossings. Both
slabs are (100) terminated and calculated for models with up to
4th nearest neighbor hopping. Red discs mark the projections of
the two movable fourfold crossings for each model. Vanishing Chern
numbers allow the hybridization of surface states.
}
\label{Surface_SG106_SG133}
\end{figure}

With our generic models we confirm the presence surface states in
SG 106 and 133, see Fig.~\ref{Surface_SG106_SG133}. As expected
for semimetals the surface states lie in the simplest case within
a bulk gap. Whereas in principle the surface states look similar
for SG~106 and 133, we display them in two different variations.
Figure~\ref{Surface_SG106_SG133}(a) shows intersecting surface
states, which hybridize close to $\bar{\Gamma}$ and split into
three disjoint Fermi surfaces. In
Fig.~\ref{Surface_SG106_SG133}(b) the surface states do not
intersect and only meet close to the surface projection of the
fourfold crossing marked in red. At a chemical potential closer
to the energy of the fourfold crossing the surface states gap out
as well. 

In conclusion, the surface arcs split into Fermi pockets which
attach either twice to a fourfold crossing or are disconnected
from it. Depending on the details of the model, intersections lead
to regions of energetically flat surface bands.

\subsection{Movable Dirac points - SG 130 and 138}
\label{Appendix_130_138}

\begin{figure*}[t!]
\centering
\includegraphics[width=\linewidth]{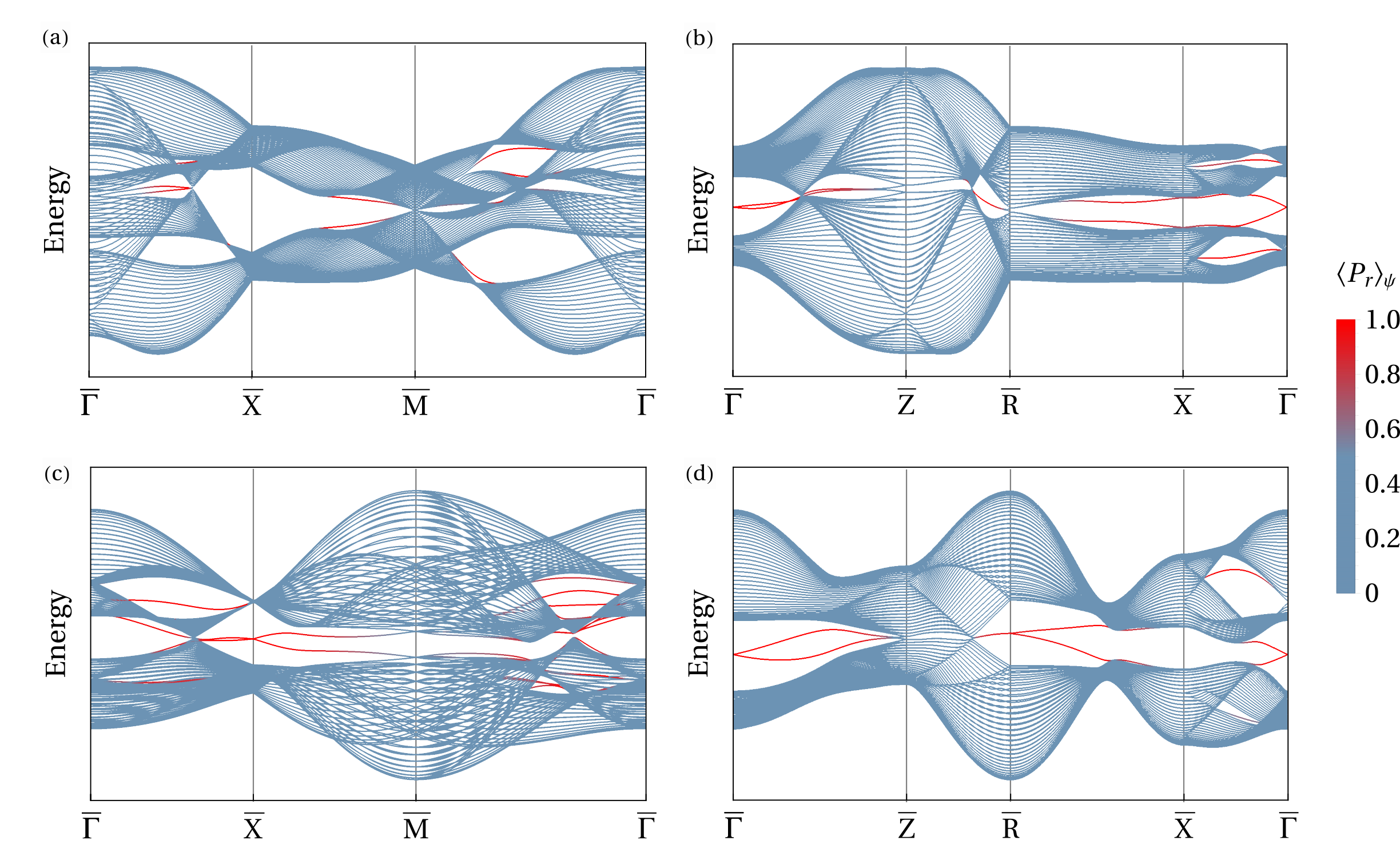}
\caption{
Bandstructure for SGs 130 (a,b) and SG 138 (c,d). Two generic
models with up to 7th nearest neighbor hopping are truncated and
the surface localization is color coded. In left column (a,c) the
termination (001) is chosen. For the right column (b,d) the model
is (100) terminated. 
}
\label{Surface_SG130_138}
\end{figure*}

SGs~130 and 138 exhibit four movable Dirac points in the bulk on
the line Z-R, whereas only SG~130 exhibits an additional eightfold
crossing at the point A. There are no other enforced crossings at
the considered filling of $4+ 8\mathbb{N}$, which translates into
a half filling for the slab calculations in
Fig.~\ref{Surface_SG130_138}. In the following we discuss the
arrangement of inversion eigenvalues in our models and their
surface states for (001) and (100) terminations.

As discussed in the main text the product $\delta_{\Gamma_i}$ of
every other occupied inversion eigenvalue at a TRIM $\Gamma_i$ is
determined by the symmetries except for the center of the BZ,
$\Gamma$. For our specific models of SGs~130 and 138 we obtain
$\delta_{\Gamma} = +1$, the same value that is enforced by SG~138
at A, i.e. $\delta_{\text{A}} = +1$. All other TRIMs yield
\mbox{
$\delta_{\text{X}} = \delta_{\text{R}} = \delta_{\text{M}} =
\delta_{\text{Z}} = -1 $}. We group the next arguments by the
chosen termination, which determines how the $\delta_{\Gamma_i}$
must be combined to obtain the relevant time-reversal polarization
\cite{2007_FuKane_Z2Inversion}.

If a bulk system with SG~130 or 138 is terminated in (001)
direction, the time-reversal polarizations are $\pi_{\bar{\Gamma}}
= \pi_{\bar{\text{M}}} = -1$ (if applicable) and
$\pi_{\bar{\text{X}}} = 1$, which we label by the TRIMs of the
surface BZ. For SG~130, see Fig.~\ref{Surface_SG130_138}(a), the
model exhibits a bulk band touching at the projection of the
eightfold crossing at $\bar{\text{M}}$. With this caveat in mind
the differences in the time-reversal polarization do not lead to
characteristic surface states on the line
$\bar{\text{X}}$-$\bar{\text{M}}$. Also no feature related to a
change in $\pi_a$ is expected along
$\bar{\Gamma}$-$\bar{\text{X}}$ because the bulk Dirac point on
Z-R closes the band gap. On this path we find states associated with
the movable Dirac point emerging from the bulk crossing. Due to
the vanishing Chern number these surface states are not protected.
Using a different parameter set an even number of gapped states
was achieved at $\bar{\text{X}}$ but the band connectivity is not
affected. For SG~138 a nontrivial enforced surface state appears
on $\bar{\text{X}}$-$\bar{\text{M}}$ shown in
Fig.~\ref{Surface_SG130_138}(c). The Dirac surface state at
$\bar{\text{X}}$ exhibits the characteristic connections to
valence and conduction bands. 

For the (100) termination we need to consider the time-reversal
polarizations $\pi_{\bar{\Gamma}} = \pi_{\bar{\text{R}}} = -1$ and
$\pi_{\bar{\text{X}}} = \pi_{\bar{\text{Z}}} = +1$. Note that in
the presence of the movable Dirac points on Z-R
$\pi_{\bar{\text{Z}}}$ cannot be rigorously used, whereas yet
again $\pi_{\bar{\text{R}}}$ is not defined for SG~130 due to its
eightfold crossing at A. For SGs~130 and 138 the values of $\pi_a$
change along $\bar{\Gamma}$-$\bar{\text{X}}$ and we find the
surface Dirac point with the expected connectivity at
$\bar{\Gamma}$, see Fig.~\ref{Surface_SG130_138}(b,d). The same
happens for SG~138 along $\bar{\text{X}}$-$\bar{\text{R}}$, with
the key difference that here this feature is enforced. Notably,
whereas all bands are twofold degenerate by the action of
$\mathcal{PT}$, the surface states on the line
$\bar{\mathrm{Z}}$-$\bar{\mathrm{R}}$ are Kramers degenerate due
to $M_{010}(0, \tfrac12, \tfrac12) \mathcal{T}$. The latter is
only true in the limit of large slabs, where the states localized
on opposing surfaces cannot interact with each other. 

In the discussion of surface states for SGs~106 and 133 as well as
SGs~130 and 138 one notices that the former tend to resemble Fermi
arcs more closely. We briefly try to rationalize this observation.
For the latter SGs the interpretation as two different
superimposed Weyl semimetals does not apply to the same extent as
to the former SGs. A possible explanation for this is that on the
line \mbox{Z-R} only one spinless representation exists for
SGs~130 and 138, whereas there are two on the fourfold rotation
axis M-A, where the fourfold crossing appears for SGs~106 and 133.
This and the absence of surface states due to the $\mathbb{Z}_2$
invariant underlines the varying extent to which Fermi arcs
vanish in our generic models.

\section{Effective low-energy Hamiltonians}
\label{appendix_low_energy_ham}

In this appendix we derive effective Hamiltonians, describing the
low-energy physics near different types of band crossings.
Throughout this chapter, we use coordinates 
$\mathbf{q} = \vk -\mathbf{K}_0$ 
relative to the point $\mathbf{K}_0$ of the degeneracy. The
Hamiltonian is then given up to the lowest order in $\mathbf{q}$
such that only enforced degeneracies remain.

For all the spatial symmetries in the little group of
$\mathbf{K}_0$, the Hamiltonian has to fulfill the condition 
\mbox{$U_i^\dagger H(\mathbf{q})U_i = H(R_i\mathbf{q})$},
where $U_i$ is the representation of the spatial symmetry and
$R_i$ the action of the symmetry in $k$-space, i.e. the spatial
symmetry without translational parts modulo reciprocal lattice
vectors.

If antiunitary symmetries like time reversal are present,
their representation can be given in terms of an unitary part
$U_{\tilde{\mathcal{T}}}$ and complex conjugation $\mathcal{K}$.
The condition on the local Hamiltonian is then
$U_{\tilde{\mathcal{T}}}^\dagger H^*(\mathbf{q})U
    =H(-R_{\tilde{\mathcal{T}}}\mathbf{q})$.

\subsection{Fourfold double Weyl point at R in SG 96}
\label{Appendix_RinSG96}

The little group of the TRIM $\mathrm{R}{=}(0,\pi,\pi)$ is generated 
from lattice translations and two of the three 
twofold screw rotations $2_{001}(0,0,\frac12)$,
$2_{010}(\frac12,\frac12,\frac34)$ and 
$2_{100}(\frac12,\frac12.\frac14)$.
The two-dimensional irreducible representation of these symmetries
can be given by $\sigma_z$, $\sigma_y$ and
$\mathrm{i}\sigma_x$ respectively.
Time-reversal symmetry couples two copies of these
representations and takes the form
\mbox{$U_\mathcal{T}\mathcal{K}=\mathrm{i}\tau_y\sigma_z\mathcal{K}$}
in this basis. 
Here we write $\sigma_i$, $i=0,x,y,z$ for the Pauli matrices
acting within a representation and $\tau_i$ for the Pauli matrices
acting in the space of the two representations.

This restricts the linearized Hamiltonian to the form 
(in $\tau$-space)
\begin{equation}
H(\mathbf{q}) =
    \left( \begin{array}{cc}
            \mathbf{d}\cdot\boldsymbol{\sigma} & \boldsymbol{\Lambda}\cdot\boldsymbol{\sigma} \\
            \boldsymbol{\Lambda}^*\cdot\boldsymbol{\sigma} & \mathbf{b}\cdot\boldsymbol{\sigma} \\
    \end{array} \right),
\end{equation}
with $d_i = v_i q_i$, $i=x,y,z$ and $b_x=d_x$, $b_y=-d_y$, $b_z =
-d_z$. Furthermore, $\Lambda_i = \lambda_i q_i$
for $i=y,z$ and 0 otherwise.

Using the unitary transformation
$U=\frac{1}{\sqrt{2}}(\tau_z\sigma_0 + \tau_y\sigma_x)$,
the Hamiltonian becomes block diagonal with two decoupled Weyl
points mentioned in the main text and shown in
Fig.~\ref{higher_weyl_fig}(b),
\begin{eqnarray}
H_\pm(\mathbf{q}) = 
    v_x q_x \sigma_x &
    +(\pm v_y q_y + \lambda_z q_z)\sigma_y
\nonumber \\ &
    +(\pm v_z q_z - \lambda_y q_y)\sigma_z .
\end{eqnarray}
The different signs have an influence on the dispersion only when
$q_z$ and $q_y$ are both non-zero in accordance with requirement
of the nodal planes $q_z=0$ and $q_y=0$.

\subsection{Low energy Hamiltonian for fourfold quadruple Weyl point}
\label{appendix_4fold_quadruple}

The point A is a TRIM and its little group consists of all the
symmetries of the SG. Among the spinful irreducible
representations, i.e. the ones with eigenvalue -1 for $2\pi$
rotations, four are one-dimensional and one is
two-dimensional~\cite{elcoro_aroyo_JAC_17}. The latter one is
paired with itself by time reversal symmetry and thus makes up the
fourfold degeneracy.

This representation is completely defined for the following choice of
matrices for the generators.
\begin{eqnarray}
4_{001}^+,(\tfrac12,\tfrac12,\tfrac34) :& U_4 = \tau_0\sigma_z  \\
2_{010}  ,(\tfrac12,\tfrac12,\tfrac34) :& U_2 = \tau_0\sigma_y
\end{eqnarray}
With this basis choice, the unitary part of the time reversal
symmetry takes the form 
$U_\mathcal{T} = \mathrm{i}\tau_y\sigma_z$.

Including all terms up to second order in
$\mathbf{q}$ and comparing coefficients leads to the Hamiltonian
\begin{eqnarray}
H(\mathbf{q}) 
    & = &
        \left(
        \begin{array}{cc}
     \mathbf{d          }   (\mathbf{q})\cdot\boldsymbol{\sigma}     &  
     \boldsymbol{\Lambda}   (\mathbf{q})\cdot\boldsymbol{\sigma}     \\
     \boldsymbol{\Lambda}^* (\mathbf{q})\cdot\boldsymbol{\sigma}     & 
    -\mathbf{d          }   (\mathbf{q})\cdot\boldsymbol{\sigma}^*
\end{array}
\right),
\end{eqnarray}
with the Hamiltonian of a double Weyl on the diagonal,
$d_x = v_x q_x q_y$,
$d_y = v_y (q_x^2 - q_y^2)$,
$d_z = v_z q_z$,
and the symmetry-allowed spin-orbit coupling terms
$\Lambda_x=\lambda_x q_x q_y$, $\Lambda_y=0$, 
$\Lambda_z=\lambda_z q_z$.
Note that $-\mathbf{d}\cdot\boldsymbol{\sigma}^*$
effectively inverts the signs of $v_x$
and $v_z$, leaving the Chern number of the block invariant.

The block diagonal form in Eq.~\eqref{Ham_4fold_quadruple} in the
main text can be achieved with the unitary transformation \mbox{$U
=\frac{1}{\sqrt{2}}(\tau_0\sigma_0 +\tau_x\sigma_y)$}.  This
matrix diagonalizes $\mathrm{i}\tau_y \sigma_y$, which commutes
with the initial Hamiltonian. Each block corresponds to a different
eigenvalue of this symmetry, which has been used to color
Fig.~\ref{higher_weyl_fig}(c).

\subsection{Low energy Hamiltonian for P in SG 110}
\label{App_110P}

The little group of P in SG 110 is generated by translations,
including $t(\frac12,\frac12,\frac12)$, the rotation
$2_{001}(\frac12,\frac12,\frac12)$, the glide reflection
$M_{110}(0,\frac12,\frac34)$, and the combination of time reversal
and another reflection, $M_{010}(0,0,\frac12)\mathcal{T}$.

From Eqs~\eqref{110P_eigvals1} and \eqref{110P_eigvals2} we know,
that different mirror eigenvalues are paired when both rotation
eigenvalues are $+1$, i.e. $p_{2_{001}}=0$ in
\eqref{eq_2001_EV}. The
representations can therefore readily be given as 
\begin{eqnarray}
U_{2_{001}} &=& \sigma_0 \\
U_{M_{110}} &=& \mathrm{i}\mathrm{e}^{\mathrm{i}3\pi/ 4}\sigma_z
\end{eqnarray}
In this basis, $M_{010}(0,0,\frac12)\mathcal{T}$ takes the form
$\mathrm{i}\sigma_y\mathcal{K}$.

To fully recreate the gap in a low-energy model, it is necessary
to go to third order in $\mathbf{q}$. 
The Hamiltonian is restricted to the form 
\begin{equation}
H(\mathbf{q}) =
a (q_x^2-q_y^2)q_z \sigma_+ 
+ (c_1 q_z + 2 c_2 q_x q_y )\sigma_z +\mathrm{h.c.},
\end{equation}
with $\sigma_+ = \sigma_x + \mathrm{i}\sigma_y$, $a\in\mathds{C}$
and $c_i \in \mathds{R}$.

To better show the behavior in the mirror planes, we introduce rotated 
coordinates
$p_- = \frac{1}{\sqrt{2}}( q_x - q_y )$ and
$p_+ = \frac{1}{\sqrt{2}}( q_x + q_y )$,
which span together with $q_z$ the mirror planes
$M_{110}$ and 
$M_{1\bar10}$, respectively.
With this, the Hamiltonian reads
\begin{equation}
H(\mathbf{q}) =
2 a p_+ p_- q_z \sigma_+ 
+ (c_1 q_z + c_2\left( p_+^2 - p_-^2 ) \right)\sigma_z +\mathrm{h.c.}.
\end{equation}
This shows that within each mirror plane, only the prefactor of
$\sigma_z$ is non-zero. Within this term, there is a $q_z$ for
every $p_\pm$ such that this term vanishes as well, leading to the
almost movable nodal line in each mirror plane as required by the
exchange of mirror eigenvalues at P. 
Because of the restrictions for odd and even contributions for $q_z$
and $p_\pm$ respectively, this still holds when including
arbitrary higher order terms. The nodal lines in both planes are
furthermore symmetry related because of the constraint on the
coefficient $c_2$ of $p_+$ and $p_-$.

\begin{figure}
\centering
\includegraphics[width=.7\linewidth]{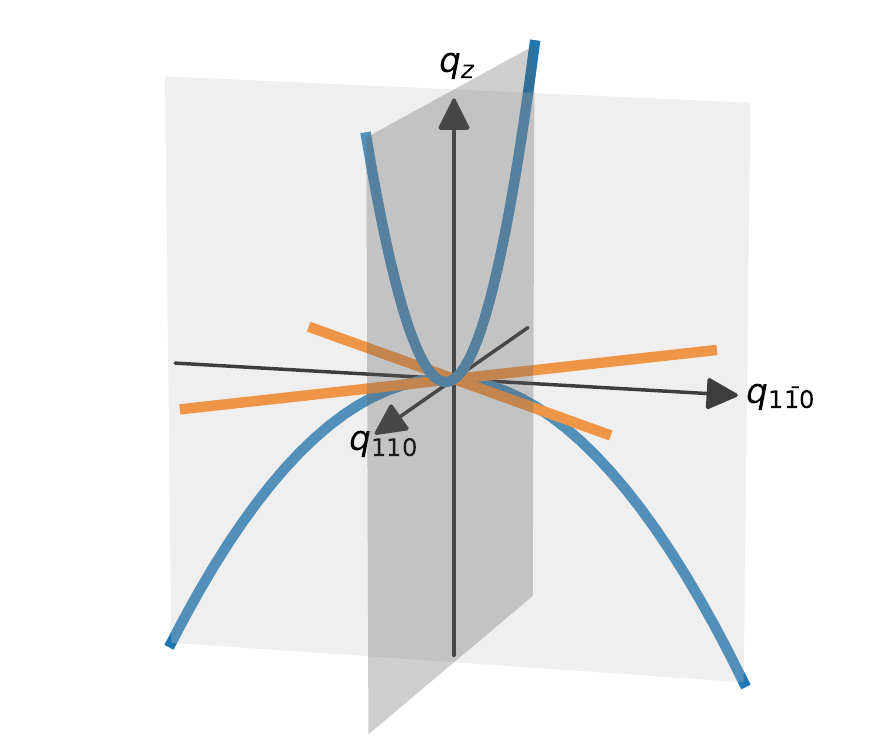}
\caption{
Touching nodal lines at P in SG 110. The orange lines are pinned
to the path P-N, the blue lines are almost movable nodal lines
each within a mirror plane. The latter result from the fact that
the bands at P are formed by mirror eigenvalues of different
signs, see Sec.~\ref{sec_movableTRIMlines}.
}
\end{figure}

\subsection{Linearized Hamiltonian for fourfold Weyl line}
\label{low_energy_4fold_line}

In this section we construct a linearized Hamiltonian for the
fourfold nodal line on the M-A line in SGs 113 and 114.
We follow the procedure of~\cite{PhysRevB.98.155145}, where a
linearized Hamiltonian for fourfold double Weyl points can be found.
Note that only the directions perpendicular to the nodal line are
expanded in relative coordinates $q_x$ and $q_y$, as $k_z$ spans
the full range from $-\pi$ to $\pi$ along the nodal line.

The little group of a vector $\vk = (\pi,\pi,k_z)$ is $C_{2v}$,
generated by $2_{001}$ and $M_{110}(\frac{1}{2},\frac{1}{2},c)$
and has two-dimensional spinful
irreducible representations~\cite{elcoro_aroyo_JAC_17}. With
time-reversal symmetry present, an additional generator
$\overline{4}^+_{001}\mathcal{T}$ is included, which again relates
two copies of the double group representation of $C_{2v}$ and
squares to $-2_{001}$. The other antiunitary symmetry mentioned
in the main text, $2_{010}(\frac12,\frac12,c)\mathcal{T}$ can be
written as product of the above generators.

A matrix representation can be given by
\begin{eqnarray}
U_{2_{001}} &=& 
    \tau_0 (-\mathrm{i}\sigma_y) \\
U_{M_{110}} &=& 
    \tau_0 
        (\mathrm{i}\mathrm{e}^{\mathrm{i}k_zc} \sigma_x) \\
U_{\bar{4}^+_{001}\mathcal{T}}\mathcal{K} &=& 
    \mathrm{-i}\tau_y 
        \frac{\sigma_0 - \mathrm{i}\sigma_y}{\sqrt{2}}\mathcal{K}.
\end{eqnarray}
The linearized Hamiltonian close to the nodal line at
\mbox{$\mathbf{q} = (\pi+q_x,\pi+q_y,k_z)$}
is restricted by these symmetries to the 
form 
\begin{eqnarray}
H(\mathbf{q}) &=& \left( \begin{array}{cc}
    \alpha q_+ \sigma_z +\beta q_-\sigma_x &
    \lambda(q_+ \sigma_z -q_- \sigma_x) \\
    \lambda(q_+ \sigma_z -q_- \sigma_x) &
    \beta q_+ \sigma_z +\alpha q_-\sigma_x 
    \end{array} \right) \nonumber \\
    && \quad + \varepsilon_0 \mathds{1},
\end{eqnarray}
with the perpendicular momentum components $q_\pm = q_x \pm q_y$
and the parameters $\alpha,\beta, \lambda$ and $\varepsilon_0$ 
all being $k_z$-dependent without any restrictions.

This Hamiltonian can be block diagonalized via 
the unitary, $q_x$- and $q_y$-independent matrix
\mbox{
$T = \cos\phi\tau_0\sigma_0
    +\sin\phi\mathrm{i}\tau_y\sigma_0$},
where $\tan(2\phi) = \frac{2\lambda}{\alpha-\beta}$.
The block diagonal Hamiltonian reads
\begin{eqnarray}
H(\mathbf{q}) &=& \left( \begin{array}{cc}
    \tilde{\alpha} q_+ \sigma_z -\tilde{\beta} q_-\sigma_x &
    0 \\
    0 &
    \tilde{\beta} q_+ \sigma_z -\tilde{\alpha} q_-\sigma_x 
    \end{array} \right) \nonumber \\
    && \quad + \varepsilon_0 \mathds{1},
\end{eqnarray}
with two decoupled Weyl lines with the modified velocities
\begin{eqnarray}
\tilde{\alpha} 
    = &
        \frac{\alpha+\beta}{2}+\mathrm{sgn}(\alpha-\beta)
            \sqrt{\left(\frac{\alpha-\beta}{2}\right)+\lambda^2},
\nonumber \\
\tilde{\beta} 
    = &
        \frac{\alpha+\beta}{2}-\mathrm{sgn}(\alpha-\beta)
            \sqrt{\left(\frac{\alpha-\beta}{2}\right)+\lambda^2}.
\end{eqnarray}
In the nodal planes we find $|q_+|=|q_-|$, leading to identical
eigenvalues as demanded by symmetry. Furthermore, a
$\frac{\pi}{2}$ rotation relates the eigenvalues of the upper
block to the ones in the lower one and vice versa. The nodal line
in each subspace is protected by a Berry phase of $\pi$. Including
symmetry allowed terms of higher order in $\mathbf{q}$ adds an
identical quadratic contribution to both bands, seen in the
collective bending of bands in the dispersion of NaSn$_5$.

\bibliographystyle{apsrev4-1}
\bibliography{tetragonals_literature}

\appendix

\end{document}